\documentclass[a4paper,fleqn,usenatbib]{mnras}

\usepackage{txfonts}

\usepackage[T1]{fontenc}
\usepackage{ae,aecompl}

\usepackage{psfig}   
\usepackage{graphicx}
\usepackage{amssymb}
\usepackage{subfigure}

\title[Disc destruction in star-forming regions]{Far and extreme ultraviolet radiation fields and consequent disc destruction in star-forming regions}

\author[R.~J.~Parker et al.]{Richard  J. Parker\thanks{E-mail: R.Parker@sheffield.ac.uk}\thanks{Royal Society Dorothy Hodgkin Fellow}, Rhana B. Nicholson  and Hayley L. Alcock \vspace*{0.1cm}\\
Department of Physics and Astronomy, The University of Sheffield, Hicks Building, Hounsfield Road, Sheffield, S3 7RH, UK}

\begin{document}

\date{}
                             
\pagerange{\pageref{firstpage}--\pageref{lastpage}} \pubyear{2020}

\maketitle

\label{firstpage}

\begin{abstract}
The first stages of planet formation usually occur when the host star is still in a (relatively) dense star-forming region, where the effects of the external environment may be important for understanding the outcome of the planet formation process. In particular, star-forming regions that contain massive stars have strong far ultraviolet (FUV) and extreme ultraviolet (EUV) radiation fields, which can induce mass-loss from protoplanetary discs due to photoevaporation. In this paper we present a parameter-space study of the expected FUV and EUV fields in $N$-body simulations of star-forming regions with a range of initial conditions. We then use recently published models to determine the mass-loss due to photoevaporation from protoplanetary discs.  In particular, we focus on the effects of changing the initial degree of spatial structure and initial virial ratio in the star-forming regions, as well as the initial stellar density. We find that the FUV fields in star-forming regions are much higher than in the interstellar medium, even when the regions have stellar densities as low as in the Galactic field, due to the presence of intermediate-mass, and massive, stars ($>$5\,M$_\odot$). These strong radiation fields lead to the destruction of the gas component in protoplanetary discs within 1\,Myr, implying that gas giant planets must either form extremely rapidly ($<$1\,Myr), or that they exclusively form in star-forming regions like Taurus, which contain no intermediate-mass or massive stars. The latter scenario is in direct tension with meteoritic evidence from the Solar system that suggests the Sun and its protoplanetary disc was born in close proximity to massive stars.  
\end{abstract}


\section{Introduction}

Most stars form in regions with tens to thousands of other stars where the stellar density of these groups exceeds the density of the Galactic field by at least several orders of magnitude \citep{Korchagin03,Lada03,Bressert10}. The majority of star-forming regions are short-lived, tending to disperse after around 10\,Myr \citep[e.g.][]{Lada10,Chevance20}.

On similar timescales, young stars in these regions are observed to host protoplanetary discs \citep{Andre94,Odell94,Mann14,Brogan15,Ansdell17,vanTerwisga19}, which are far more abundant at ages less than 5\,Myr than at older ages \citep{Haisch01,Richert18}. The reason for this observed depletion is likely to be a combination of rapid planet formation \citep{Johansen07}, accretion on to the central star \citep{Hartmann98,Stamatellos11}, internally driven winds from the host star \citep[e.g.][]{Ercolano17}, as well as destruction from external processes, such as truncation due to encounters with passing stars \citep{Scally01,Olczak08,Rosotti14,Vincke15,Vincke16,Zwart16,Winter18a,Winter18b}.

As well as truncation due to encounters with passing stars, discs can be destroyed by photoionising radiation from massive stars ($>$10\,M$_\odot$). The Initial Mass Function predicts far more low-mass ($<$1\,M$_\odot$) stars than massive stars \citep{Kroupa02,Chabrier05,Maschberger13}, but if a star-forming region contains 100s to 1000s of low-mass stars, then the formation of at least one or more massive stars is likely \citep{Parker07,Nicholson17}. 

Massive stars produce both extreme ultraviolet (EUV) radiation, where the individual photon energies are $h\nu > 13.6$\,eV, and far ultraviolet (FUV) radiation, where the individual photon energies are in the range $6 < h\nu \leq 13.6$\,eV. Many authors have demonstrated that both forms of radiation are extremely destructive to the gaseous component of protoplanetary discs \citep{Johnstone98,Henney99,Storzer99,Armitage00,Hollenbach00,Scally01,Adams04,Winter18b,Nicholson19a,ConchaRamirez19,Haworth20}. Unless the dust particles are particularly small, in which case they can be entrained in the photoevaporative wind \citep{Miotello12}, the dust content of the disc is largely unscathed by photoionising radiation \citep{Haworth18a,Sellek20}.

Recently, \citet{Haworth18b} introduced a new set of models for calculating the mass loss from protoplanetary discs due to photoevaporation caused by FUV radiation. The \texttt{FRIED} grid \citep{Haworth18b} requires as an input the stellar mass, disc mass, disc radius and ambient FUV radiation field, expressed in terms of the \citet{Habing68} unit, $G_0 = 1.8 \times 10^{-3}$erg\,s\,$^{-1}$\,cm$^{-2}$, which is the underlying FUV flux in the interstellar medium. The output is a mass-loss, which can be used in a post-processing analysis of an $N$-body simulation to determine the impact of photoionising radiation on discs in simulated star-forming regions.

Whilst several authors have used the \texttt{FRIED} grid to determine mass-loss due to photoevaporation in specific star-forming regions or planetary systems, \citep{Haworth18a,Winter18b,Winter19,Winter19b,ConchaRamirez19}, to our knowledge no comprehensive parameter space study has yet been carried out to calculate both the EUV and FUV flux in star-forming regions with realistic initial conditions \citep[spatial and kinematic substructure, e.g.][]{Cartwright04,Goodwin04c,Parker14b,Lomax18}, and the effects of these radiation fields on protoplanetary discs in such star-forming regions. \citet{Fatuzzo08} determined the EUV and FUV fluxes in nearby star-forming regions, but these are largely devoid of massive stars \citep[and may not be representative of all star formation,][]{Kruijssen12b}. \citet{Winter18b} considered more distant star-forming regions, but tailored the stellar content of their simulations to match those regions. 

In this paper, we take a more general approach and calculate the EUV and FUV fluxes in star-forming regions with different initial densities, spatial and kinematic substructure, virial ratios and stellar mass. The paper is organised as follows. In Section~2 we describe our simulations, including the post-processing analysis we use to calculate the effects of the photoionising radiation on protoplanetary discs. We present our results in Section~3, we provide a discussion in Section~4 and we conclude in Section~5.








\section{Method}

In this Section we describe the set-up of $N$-body simulations used to model evolution of the star-forming regions, and we then describe the post-processing routine used to model photoevaporative mass-loss from protoplanetary discs.

\subsection{Star-forming regions}

For each set of initial conditions, we create twenty versions of the same simulation to gauge the effects of stochasticity on the results. Our default models contain $N = 1500$ stars, with masses drawn from a \citet{Maschberger13} IMF, which has a probability distribution of the form 
\begin{equation}
p(m) \propto \left(\frac{m}{\mu}\right)^{-\alpha}\left(1 + \left(\frac{m}{\mu}\right)^{1 - \alpha}\right)^{-\beta}.
\label{maschberger_imf}
\end{equation}
Here, $\mu = 0.2$\,M$_\odot$ is the average stellar mass, $\alpha = 2.3$ is the \citet{Salpeter55} power-law exponent for higher mass stars, and $\beta = 1.4$ describes the slope of the IMF for low-mass objects \citep*[which also deviates from the log-normal form;][]{Bastian10}. We randomly sample this distribution in the mass range 0.1 -- 50\,M$_\odot$, i.e.\,\,we do not include brown dwarfs in the simulations. Typically, for 1500 stars we draw between one and five massive ($>20$\,M$_\odot$) stars for each realisation of the simulation. We do not include primordial binary stars in the simulations; although binary stars are a common outcome of star formation \citep{Raghavan10,Duchene13b}, they complicate the formation, evolution and stability of protoplanetary discs and so we defer their inclusion to a future paper. 

In one set of simulations we keep the IMF constant so that we can isolate the effects of stochastically sampling the IMF from the stochastic dynamical evolution of the star-forming region. These simulations contain a 31\,M$_\odot$ star, an 18\,M$_\odot$ star, as well as around 10 stars with masses in the range 5 -- 15\,M$_\odot$.

In another set of simulations, we draw $N = 150$ stars from the IMF. Statistically, we expect fewer massive stars (both O-type and lower-mass B-type stars) in these regions, and star-forming regions with this number of stars are much more common in the vicinity of the Sun. With fewer or no massive stars, these regions will have lower ionising radiation fluxes and so we expect protoplanetary discs to be less affected by photoevaporation in these low-mass star-forming regions. 

The star-forming regions are set up as fractals in an attempt to mimic the spatial and kinematic substructure observed in young star-forming regions \citep{Gomez93,Larson95,Cartwright04,Sanchez09,Andre14,Hacar16}. We refer the interested reader to \citet{Goodwin04a} and \citet{Parker14b} for a comprehensive description of the fractal distributions we use here. In brief, we use the box fractal method, which proceeds by defining a `parent' in the centre of a cube which has sides of length $N_{\rm div} = 2$, which then spawns  $N_{\rm div}^3$ subcubes. Each of the subcubes contains a `child' at its centre, and the construction of the fractal proceeds by determining which of the children become parents themselves. The probability that a child becomes a parent is given by $N_{\rm div}^{D - 3}$, where $D$ is the fractal dimension. In this scheme, the lower the fractal dimension, the fewer children becomes parents and so there is more substructure. 

The velocities of the parent particles are drawn from a Gaussian distribution with mean zero, and the children inherit these velocities plus a small random component (the size of which scales as $N_{\rm div}^{D - 3}$) that decreases with each subsequent generation of the fractal. This results in a kinematic distribution in which the stars on local scales have very similar velocities, whereas on larger scales the velocities between stars can be very different. In the box fractal method we adopt here,  on scales of size $L$ the velocities scale as $v(L) \propto L^{3 - D}$, so for $D = 1.6$, $v(L) \propto L^{1.4}$ and for $D = 2.0$ $v(L) \propto L$. (Note that the \citet{Larson81} linewidth relation roughly scales as $v(L) \propto L^{0.38}$.)

We create fractals with three different amounts of substructure. In the first, the fractal dimension $D = 1.6$, which results in a high degree of substructure, and the stellar velocities are strongly correlated on local scales. Most of our simulations have $D = 2.0$, which is a moderate amount of spatial substructure with some correlation in the velocities of nearby stars. Finally, we run models with $D = 3.0$, which is a uniform sphere with minimal correlation in the stellar velocities. 

Once the fractal star-forming regions have been created, we scale the velocities of the individual stars to a bulk virial ratio, $\alpha_{\rm vir} = \mathcal{K}/|\Omega|$, where  $\mathcal{K}$ is the total kinetic energy and $|\Omega|$ is the total potential energy. Most young stars are observed to have subvirial velocities, so most of our simulations are scaled to $\alpha_{\rm vir} = 0.3$. This initiates a global collapse, although the timescale on which this occurs depends on both the fractal dimension and the local stellar density. We also run simulations with $\alpha_{\rm vir} = 0.5$ (virial equilibrium) and $\alpha_{\rm vir} = 1.5$ (supervirial) to gauge the effects of the bulk motion of a star-forming region on the FUV and EUV fields, and subsequent photoevaporative mass-loss. 

Finally, we vary the initial median stellar density in each star-forming region, by altering the radius of the fractal. We mostly use simulations with moderate substructure ($D = 2.0$) and subvirial velocities ($\alpha_{\rm vir} = 0.3$), but for the comparison of the effect of changing the initial degree of substructure, we keep the median density constant and change the fractal dimension. In order to keep the stellar density constant, simulations with a high degree of substructure ($D = 1.6$) have a larger radius, $r_F$ than simulations with no substructure ($D = 3.0$), because a high degree of substructure skews the median local density to higher values. We adopt four different initial local stellar densities, 1000\,M$_\odot$\,pc$^{-3}$, 100\,M$_\odot$\,pc$^{-3}$, 10\,M$_\odot$\,pc$^{-3}$ and 0.2\,M$_\odot$\,pc$^{-3}$. The highest density is thought to be commensurate with the initial densities of regions such as the Orion Nebula Cluster \citep{Parker14e}, whereas many star-forming regions are consistent with having lower densities \citep{Parker17a}. Very diffuse stellar associations (such as Taurus or Cyg~OB2) may have been low-density at birth \citep[e.g.\,\,$\sim$10\,M$_\odot$\,pc$^{-3}$][]{Wright14,Wright16}, and for completeness  we run simulations where the stellar densities are similar to those in the Sun's local neighborhood today \citep{Korchagin03}. 

We summarise the different combinations of parameters used as initial conditions for the simulations in Table~\ref{simulations}. 

The simulations are evolved for 10\,Myr using the \texttt{kira} integrator within the \texttt{Starlab} environment \citep{Zwart99,Zwart01}. We do not include stellar evolution in the simulations.

\begin{table}
  \caption[bf]{A. summary of the different initial conditions of our simulated star-forming regions. The columns show the number of stars, $N_{\rm stars}$, the initial radius of the star-forming region, $r_F$, the initial median local stellar density, $\tilde{\rho}$, the fractal dimension $D$, the initial virial ratio $\alpha_{\rm vir}$, and the variation of the \citet[][M13]{Maschberger13} IMF (either stochastic between the different realisations of the same simulation, or constant across all realisations). }
  \begin{center}
    \begin{tabular}{|c|c|c|c|c|c|}
      \hline

$N_{\rm stars}$ & $r_F$ & $\tilde{\rho}$ & $D$ & $\alpha_{\rm vir}$ & IMF \\
\hline
1500 & 1\,pc & $1000$\,M$_\odot$\,pc$^{-3}$ & 2.0 & 0.3 & M13, stochastic \\
1500 & 2.5\,pc & $100$\,M$_\odot$\,pc$^{-3}$ & 2.0 & 0.3 & M13, stochastic \\
1500 & 5.5\,pc & $10$\,M$_\odot$\,pc$^{-3}$ & 2.0 & 0.3 & M13, stochastic \\
1500 & 20\,pc & $0.2$\,M$_\odot$\,pc$^{-3}$ & 2.0 & 0.3 & M13, stochastic \\ 
\hline
1500 & 5\,pc & $100$\,M$_\odot$\,pc$^{-3}$ & 1.6 & 0.3 & M13, stochastic \\
1500 & 1.1\,pc & $100$\,M$_\odot$\,pc$^{-3}$ & 3.0 & 0.3 & M13, stochastic \\
\hline
1500 & 2.5\,pc & $100$\,M$_\odot$\,pc$^{-3}$ & 2.0 & 0.5 & M13, stochastic \\
1500 & 2.5\,pc & $100$\,M$_\odot$\,pc$^{-3}$ & 2.0 & 1.5 & M13, stochastic \\
\hline
1500 & 2.5\,pc & $100$\,M$_\odot$\,pc$^{-3}$ & 2.0 & 0.3 & M13, constant \\
150 & 0.75\,pc & $100$\,M$_\odot$\,pc$^{-3}$ & 2.0 & 0.3 & M13, stochastic \\ 

      \hline
    \end{tabular}
  \end{center}
  \label{simulations}
\end{table}
 
\subsection{Disc photoevaporation and internal evolution}
  
  Directly including discs around stars in $N$-body simulations is too computationally prohibitive, especially in the non-equilibrium initial conditions we adopt for our star-forming regions \citep[some attempts have been made to include discs in simulations using hybrid codes, e.g.][]{Rosotti14}. In our simulations we model the discs and the mass-loss due to photoevaporation  in a semi-analytic post-processing routine after the simulations have run \citep[see also][for a similar approach]{Scally01,Adams04,Winter18b,ConchaRamirez19,Nicholson19a}.
  
 We set the initial disc mass to be 10\,per cent of the host star's mass 
 \begin{equation}
 M_{\rm disc} = 0.1\,M_\star,
 \end{equation}
 which is higher than the minimum mass Solar Nebula \citep{Hayashi81}, but comfortably lower than the regime where the disc could become gravitationally unstable and fragment \citep{Toomre64,Mayer02,Meru15}. We do not allow stars more massive than 3\,M$_\odot$ to host discs. Typically, out of 1500 stars in a simulation, around 1460 will host a disc, though this is subject to some scatter due to stochastic sampling of the IMF and the direct proportionality of disc mass to host star mass. 
 
 At each snapshot output of the simulation, we determine the distance of each disc-hosting star $d$ to all stars more massive than 5\,M$_\odot$, which we adopt as the lowest mass star that emits far ultraviolet (FUV) radiation. We then take the FUV and EUV luminosities, $L_{\rm FUV}$, $L_{\rm EUV}$, in \citet{Armitage00}, which are calculated from stellar atmosphere models \citep{Buser92,Schaller92} and calculate the FUV and EUV flux for each star:
 \begin{equation}
 F_{\rm FUV} = \frac{L_{\rm FUV}}{4\pi d^2},
 \end{equation}
 and
 \begin{equation}
 F_{\rm EUV} = \frac{L_{\rm EUV}}{4\pi d^2}.
 \end{equation}
 
 In the subsequent analysis we retain the EUV flux in cgs units (erg\,s$^{-1}$\,cm$^{-2}$), but present the FUV flux in terms of the \citet{Habing68} unit, $G_0 = 1.8 \times 10^{-3}$\,erg\,s$^{-1}$\,cm$^{-2}$, which is the background FUV flux in the interstellar medium. 

In most simulations there is more than one star that emits FUV radiation (and often more than one star emitting EUV radiation), so we sum the fluxes from all of the emitting stars to obtain a total flux. 

In addition to presenting the FUV and EUV fluxes for our different initial conditions for star-forming regions, we will also calculate the likely mass-loss due to photoevaporation in these FUV and EUV radiation fields. For EUV radiation, we adopt the mass loss from \citet{Johnstone98}:
\begin{equation}
\dot{M}_{\rm EUV} \simeq 8 \times 10^{-12} r^{3/2}_{\rm disc}\sqrt{\frac{\Phi_i}{d^2}}\,\,{\rm M_\odot \,yr}^{-1}.
\label{euv_equation}
\end{equation}
Here, $\Phi_i$ is the  ionizing EUV photon luminosity from each massive star in units of $10^{49}$\,s$^{-1}$ and is dependent on the stellar mass according to the observations of \citet{Vacca96} and \citet{Sternberg03}. For example, a 41\,M$_\odot$ star has $\Phi = 10^{49}$\,s$^{-1}$ and a 23\,M$_\odot$ star has $\Phi = 10^{48}$\,s$^{-1}$. The disc radius $r_{\rm disc}$ is expressed in units of au and the distance to the massive star $d$ is in pc.

To determine the photoevaporative mass loss due to FUV radiation, we utilise the \texttt{FRIED} grid from \citet{Haworth18b}, which consists of a grid of mass-loss rates for given combinations of stellar mass, $G_0$, disc mass, disc radius and disc surface density. We interpolate over the \texttt{FRIED} grid to choose most appropriate mass-loss value given an input of stellar mass, $G_0$, disc mass and disc radius.

  We subtract mass from the discs according to the FUV-induced mass-loss rate in the \texttt{FRIED} grid and the EUV-induced mass-loss rate from Equation~\ref{euv_equation}. Models of mass loss in discs usually assume the mass is removed from the edge of the disc (where the surface density is lowest) and we would expect the radius of the disc to decrease in this scenario. We employ a very simple way of reducing the radius by assuming the surface density of the disc at 1\,au, $\Sigma_{\rm 1\,au}$, from the host star remains constant during mass-loss \citep[see also][]{Haworth18b,Haworth19}. If
  \begin{equation}
\Sigma_{\rm 1\,au} = \frac{M_{\rm disc}}{2\pi r_{\rm disc} [{\rm 1\,au}]},
  \end{equation}
  where $M_{\rm disc}$ is the disc mass, and $r_{\rm disc}$ is the radius of the disc, then if the surface density at 1\,au remains constant, a reduction in mass due to photoevaporation will result in the disc radius decreasing by a factor equal to the disc mass decrease:
\begin{equation}
r_{\rm disc}(t_k) = \frac{M_{\rm disc}(t_k)}{M_{\rm disc}(t_{k-1})}r_{\rm disc}(t_{k-1}).
\label{rescale_disc}
\end{equation}  
  
Here, the radius after mass loss, {\bf $r_{\rm disc}(t_k)$}, is then a function of the radius before mass loss, {\bf $r_{\rm disc}(t_{k-1})$}, multiplied by the new disc mass ({\bf $M_{\rm disc}(t_k)$}) divided by the previous disc mass ({\bf $M_{\rm disc}(t_{k-1})$}).  

The decrease in disc radius due to photoevaporation will be countered to some degree by expansion due to the internal viscous evolution of the disc. We implement viscous expansion by utilising the diffusion equation \citep{LyndenBell74,Pringle81}, with the parameterisation given in \citet{Hartmann98} and \citet{Hartmann09}. In brief, the surface density $\Sigma$ at a given radius $R$ is 
\begin{equation}
  \Sigma = 1.4 \times 10^3 \frac{{\rm e}^{-R/(R_1t_d)}}{(R/R_1)t_d^{3/2}}\left(\frac{M_{\rm disc}(0)}{\rm 0.1 M_\odot}\right)\left(\frac{R_1}{\rm 10\,au}\right)^{-2}{\rm g\,cm}^{-2},
  \label{diffusion}
\end{equation}
where $M_{\rm disc}(0)$ is the disc mass before viscous evolution and $R_1$ is a radial scaling factor, which we set as $R_1 = 10$\,au. $t_d$ is a non-dimensional time, such that at a given physical time $t$
\begin{equation}
t_d = 1 + \frac{t}{t_s},
\end{equation}
and the viscous timescale, $t_s$ is given by
\begin{equation}
  t_s = 8 \times 10^4\left(\frac{R_1}{\rm 10\,au}\right)\left(\frac{\alpha}{10^{-2}}\right)^{-1}\left(\frac{M_\star}{\rm 0.5\,M_\odot}\right)^{1/2}\left(\frac{T_{100\,{\rm au}}}{\rm 10\,K}\right)^{-1}{\rm yr}.
  \label{time_scale}
\end{equation}
Here, $\alpha$ is the disc viscosity parameter \citep{Shakura73} and $T_{\rm 100\,au}$ is the temperature of the disc at a distance of 100\,au from the star. We assume the temperature profile of the disc has the form 
\begin{equation}
T(R) = T_{\rm 1\,au} R^{-q},
\end{equation}
where $T_{\rm 1\,au}$ is the temperature at 1\,au from the host star and is derived from the stellar luminosity for pre-main sequence stars by \citet{Luhman03b,Luhman04a} and \citet{Kirk10}.  We adopt $q = 0.5$ and $\alpha = 0.01$ \citep{Hartmann98}.

Given the mass of the star, we calculate the temperature at 100\,au and then calculate the viscous timescale $t_s$. We then use this to calculate the surface density $\Sigma$ as a function of radius $R$ (Eqn.~\ref{diffusion}), to determine the outer radius of the disc, $r_{\rm disc}$. We set the surface density threshold below which we consider the disc to be truncated, $R_{\rm trunc}$ to be  $10^{-6}$\,g\,cm$^{-2}$. 

Following mass-loss due to photoevaporation and the subsequent inward movement of the disc radius according to Equation~\ref{rescale_disc}, we calculate the change in truncation radius $R_{\rm trunc}(t_n)/R_{\rm trunc}(t_{n-1})$
and scale the disc radius $r_{\rm disc}$ accordingly:
 \begin{equation}
 r_{\rm disc}(t_n) = r_{\rm disc}(t_{n - 1})\frac{R_{\rm trunc}(t_n)}{R_{\rm trunc}(t_{n-1})}.
 \end{equation} 
(Note that the subscripts differ from those in Equation~\ref{rescale_disc} as they refer to different stages in the process; subscript $_k$ refers to the mass and radius before and after mass-loss due to photoevaporation, whereas subscript $_n$ refers to the mass and radius before and after viscous spreading and accretion onto the star.) 
 
 Finally, with viscous spreading in the disc we would expect some disc material to be accreted onto the host star. As our disc evolution occurs in a post-processing analysis, we do not add any extra mass to the star, but instead assume this extra mass is negligible compared to the stellar mass and merely subtract that mass from the disc itself, according to 
 \begin{equation}
 M_{\rm disc}(t) = M_{\rm disc}(0)\left( 1 + \frac{t}{t_s}\right)^{\frac{1}{2\gamma - 4}},
 \end{equation}
 where $M_{\rm disc}(0)$ is the initial disc mass and  $M_{\rm disc}(t)$ is the disc mass at time $t$, following viscous evolution,  and the viscosity exponent $\gamma$ is unity \citep{Andrews10}.

 As an example, for a 1\,M$_\odot$ star  with a $M_{\rm disc} = 0.1$\,M$_\star$ disc with initial radius $r_{\rm disc} = 100$\,au, after 0.1\,Myr the disc will have a new radius $r_{\rm disc} = 173$\,au  and mass $M_{\rm disc} = 0.073$\,M$_\star$. After 1\,Myr the disc radius will be $r_{\rm disc} = 693$\,au  and mass $M_{\rm disc} = 0.032$\,M$_\star$, and after 5\,Myr the disc radius will be $r_{\rm disc} = 2314$\,au  and mass $M_{\rm disc} = 0.015$\,M$_\star$. After 10\,Myr of viscous evolution, the disc radius will be $r_{\rm disc} = 3826$\,au  and mass $M_{\rm disc} = 0.011$\,M$_\star$. These values are similar to other analytic estimates \citep{Hartmann09,Lichtenberg16b,ConchaRamirez19a} as well as numerical simulations \citep{Krumholz15}.
  
If the disc mass falls below zero, the disc is assumed to be destroyed and the star is denoted `disc-less' (though in reality a significant amount of dust may still be present \citep{Haworth18a}). The dynamical information (i.e.\,\,masses, positions and velocities of the stars) is outputted every 0.1\,Myr. However, in order to capture as much of the disc physics as possible we implement a much smaller timestep ($10^{-3}$\,Myr) for the disc mass-loss due to photoevaporation, and the internal viscous evolution. In Appendix~\ref{appendix:resolution} we demonstrate the effects of decreasing the timestep in the disc calculations.

\section{Results}

In this section we will  describe the evolution of the FUV and EUV fluxes in star-forming regions with different initial conditions and then describe the mass-loss in discs due to photoevaporation in these radiation fields. Where we plot the evolution of the disc fraction in simulations, we show observed disc fractions in star-forming regions from \citet{Richert18} for comparison (and these are shown by the dark grey points in the relevant figure panels). Note that we only implement viscous spreading in the simulation described in Section~3.3; all other models include inward evolution of the disc radius only.

\subsection{FUV and EUV flux in a single star-forming region}

\begin{figure}
\begin{center}
\rotatebox{270}{\includegraphics[scale=0.4]{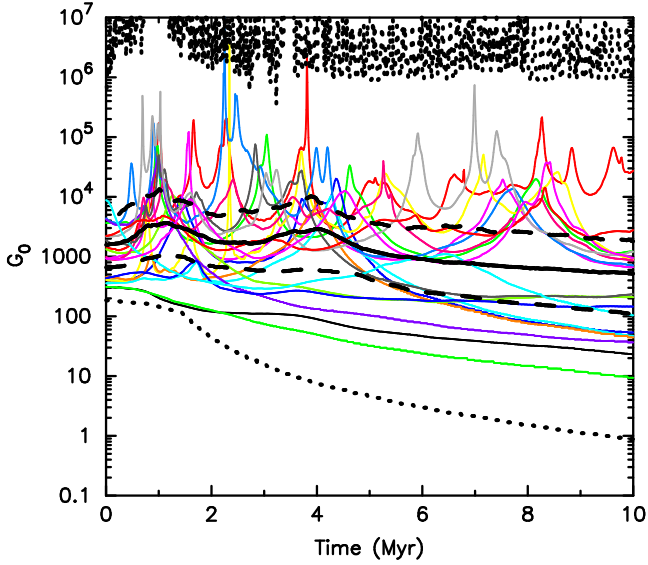}}
\caption[bf]{Evolution of the FUV flux in one simulated star-forming region where the initial local stellar density is 100\,M$_\odot$\,pc$^{-3}$. The solid black line shows the median $G_0$ value for all stars as a function of time, with the dashed lines indicating the interquartile range, and the dotted lines showing the full range. The $G_0$ values for 20 individual stars are shown by the coloured lines. }
\label{G0_singular}
  \end{center}
\end{figure}

In Fig.~\ref{G0_singular} we show the evolution of the FUV flux ($G_0$) in a subvirial ($\alpha_{\rm vir} = 0.3$), moderately substructured ($D = 2.0$) star-forming region with initial local stellar density $\tilde{\rho} = 100$\,M$_\odot$\,pc$^{-3}$. The median $G_0$ field (the solid black line) is initially $G_0 \sim 2000$, and this increases slightly as the star-forming region becomes more compact, before slowly decreasing (though it remains well above $G_0 = 100$, so the radiation field is more than 100 times that in the interstellar medium). 

Across the simulation, there is a huge range in possible values (indicated by the dotted lines, which show the full range $G_0 = 100 - 10^7$ at the start of the simulation, and $G_0 = 1 - 10^6$ after 10\,Myr). Interestingly, the $G_0$ field experienced by an individual star can hugely fluctuate, as shown by the coloured lines in Fig.~\ref{G0_singular}. 

\subsection{Different initial conditions and ensembles of simulations} 

\subsubsection{Stellar density}

\begin{figure*}
  \begin{center}
\setlength{\subfigcapskip}{10pt}
\vspace*{-0.7cm}
\hspace*{-1.0cm}\subfigure[$\tilde{\rho} \sim 1000$\,M$_\odot$\,pc$^{-3}$]{\label{G0_densities-a}\rotatebox{270}{\includegraphics[scale=0.19]{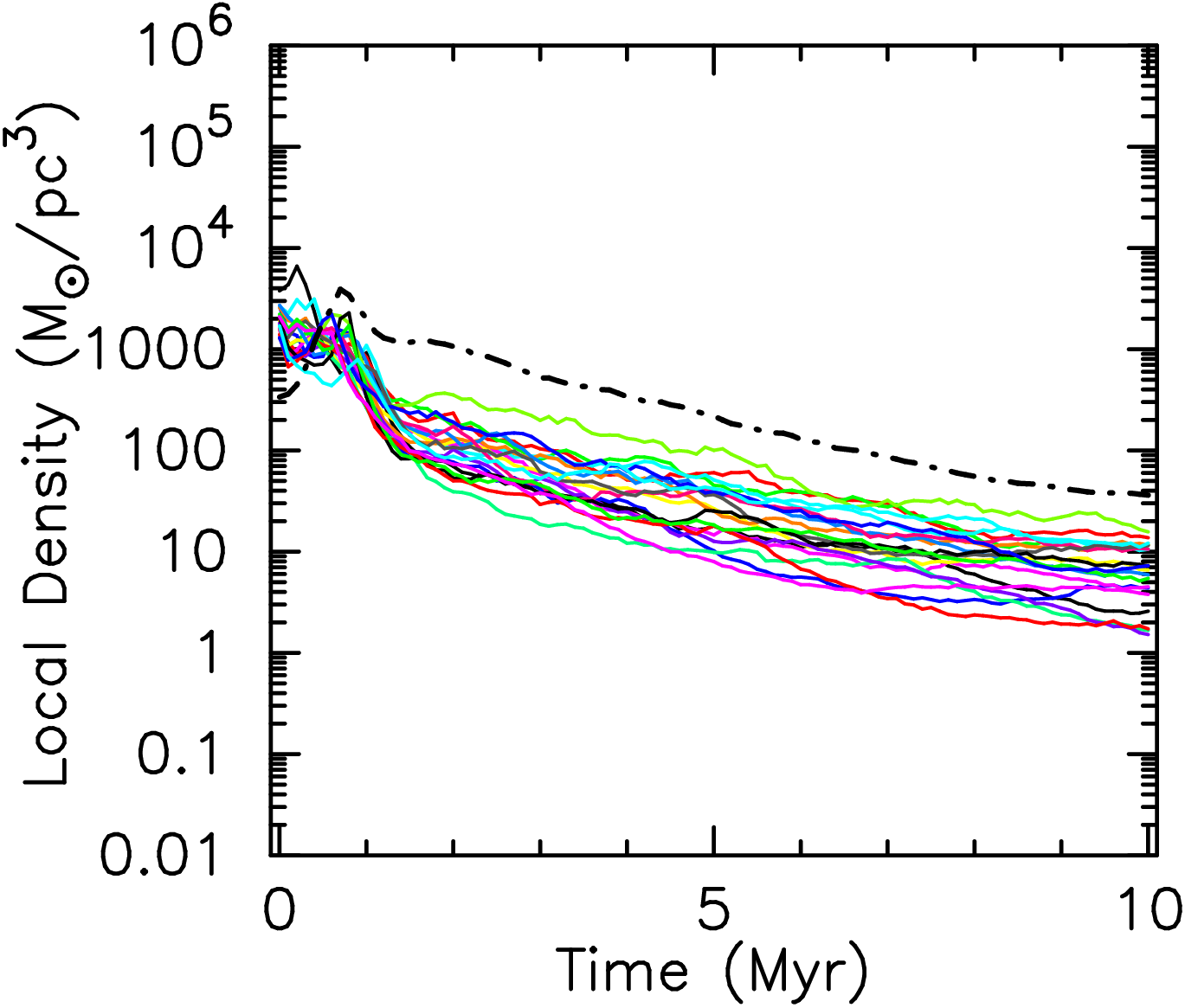}}}
\hspace*{0.2cm}\subfigure[$\tilde{\rho} \sim 100$\,M$_\odot$\,pc$^{-3}$]{\label{G0_densities-b}\rotatebox{270}{\includegraphics[scale=0.19]{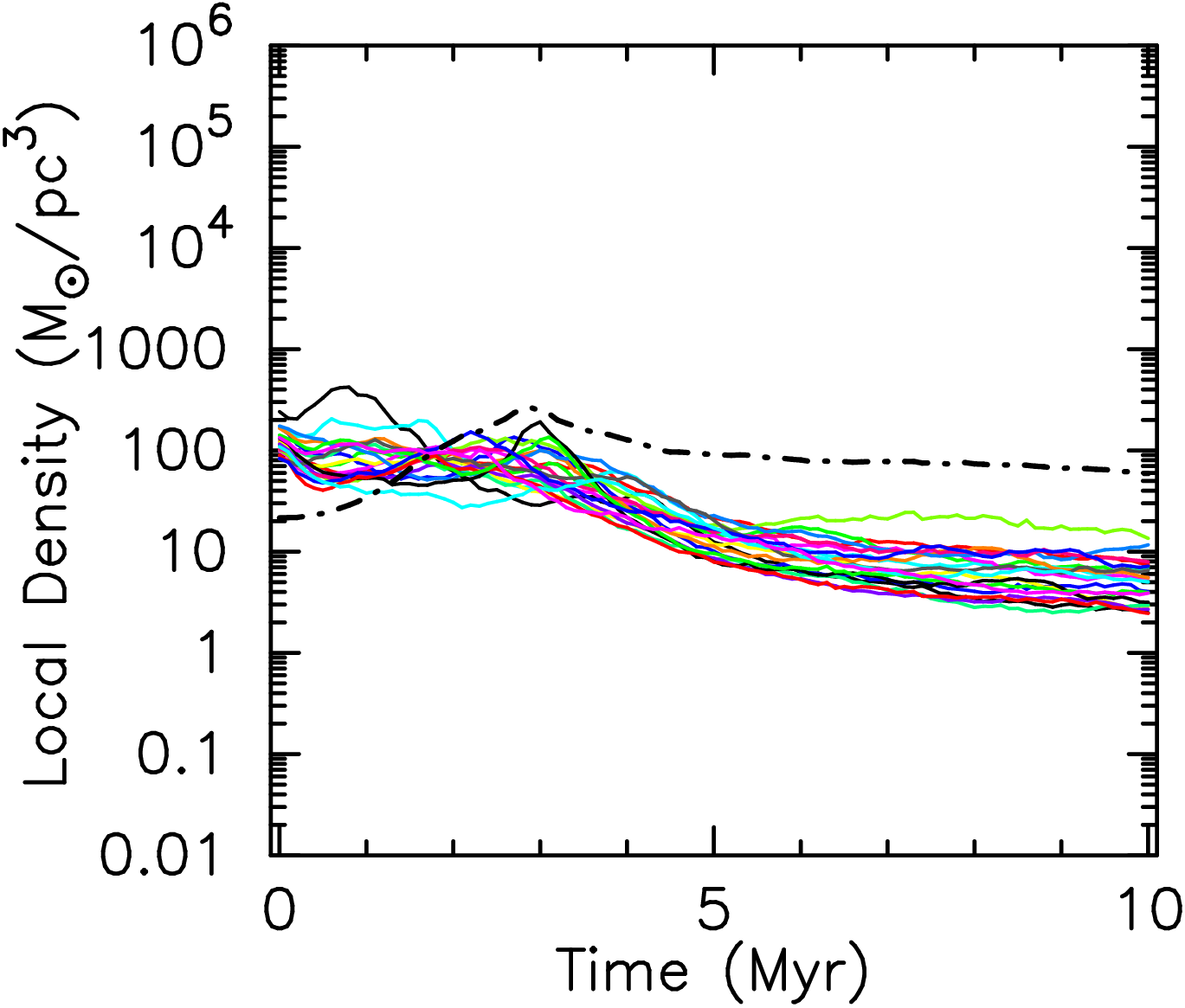}}}
\hspace*{0.2cm}\subfigure[$\tilde{\rho} \sim 10$\,M$_\odot$\,pc$^{-3}$]{\label{G0_densities-c}\rotatebox{270}{\includegraphics[scale=0.19]{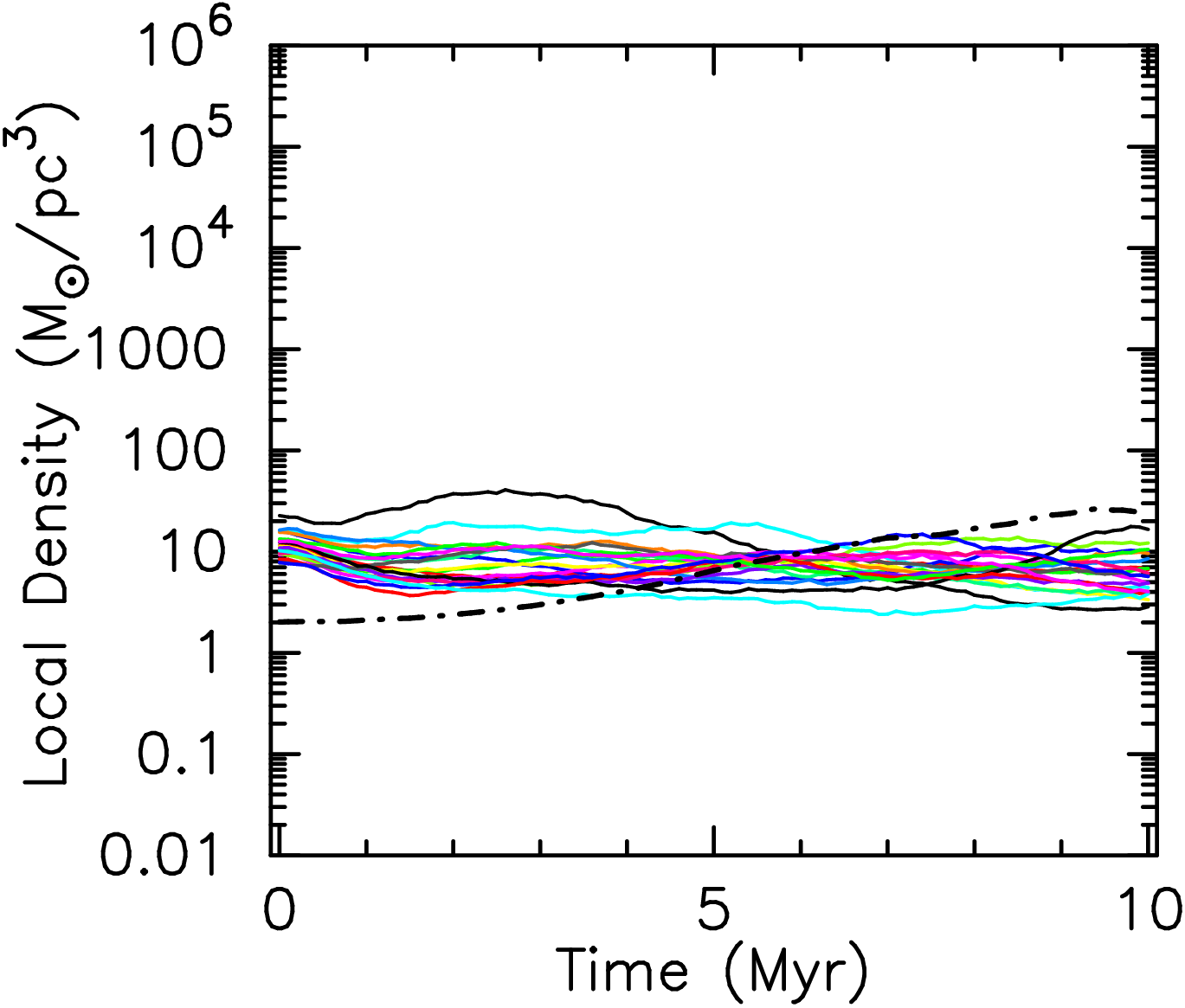}}}
\hspace*{0.2cm}\subfigure[$\tilde{\rho} \sim 0.2$\,M$_\odot$\,pc$^{-3}$]{\label{G0_densities-d}\rotatebox{270}{\includegraphics[scale=0.19]{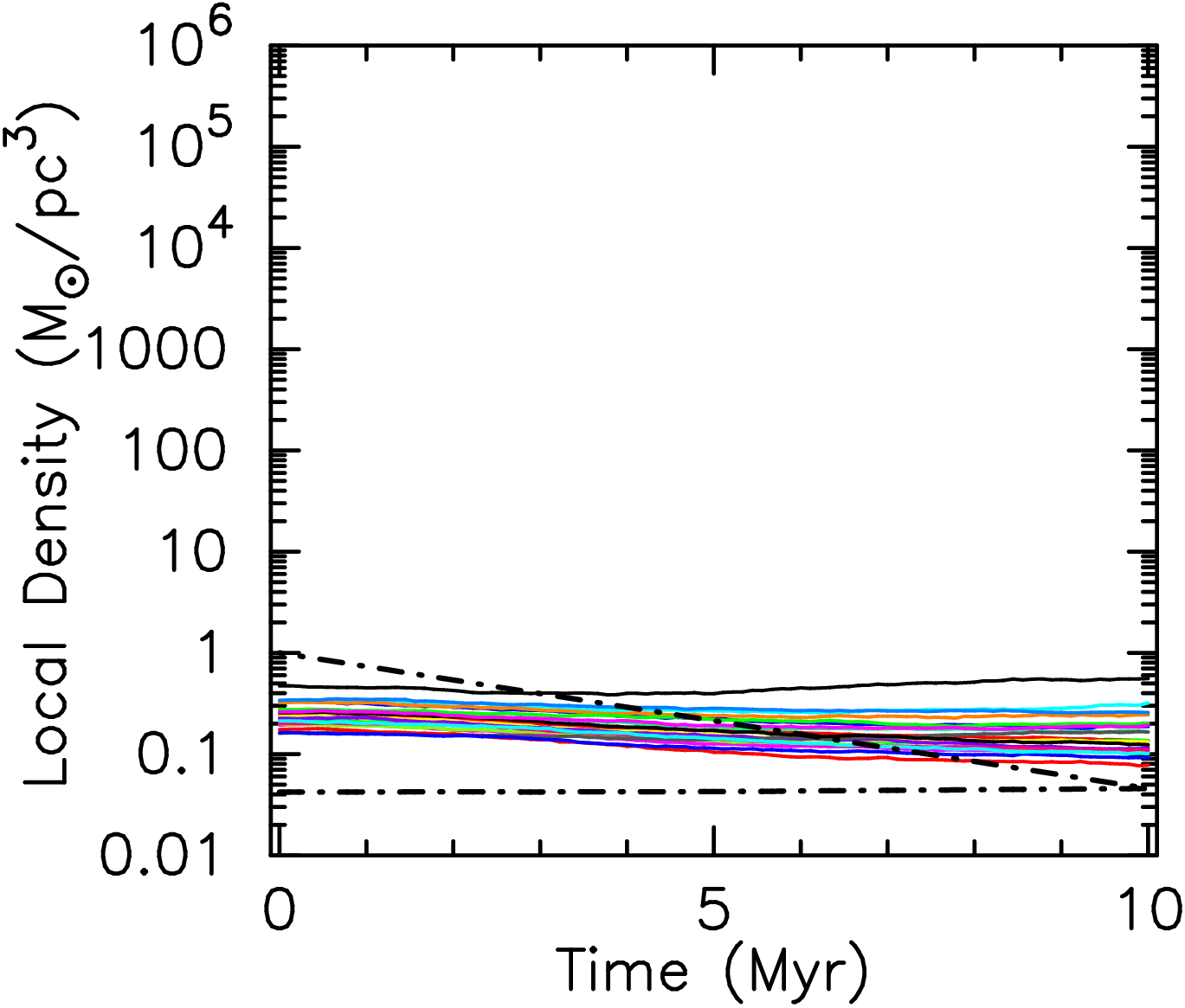}}}

\hspace*{-1.0cm}\subfigure[$\tilde{\rho} \sim 1000$\,M$_\odot$\,pc$^{-3}$]{\label{G0_densities-e}\rotatebox{270}{\includegraphics[scale=0.19]{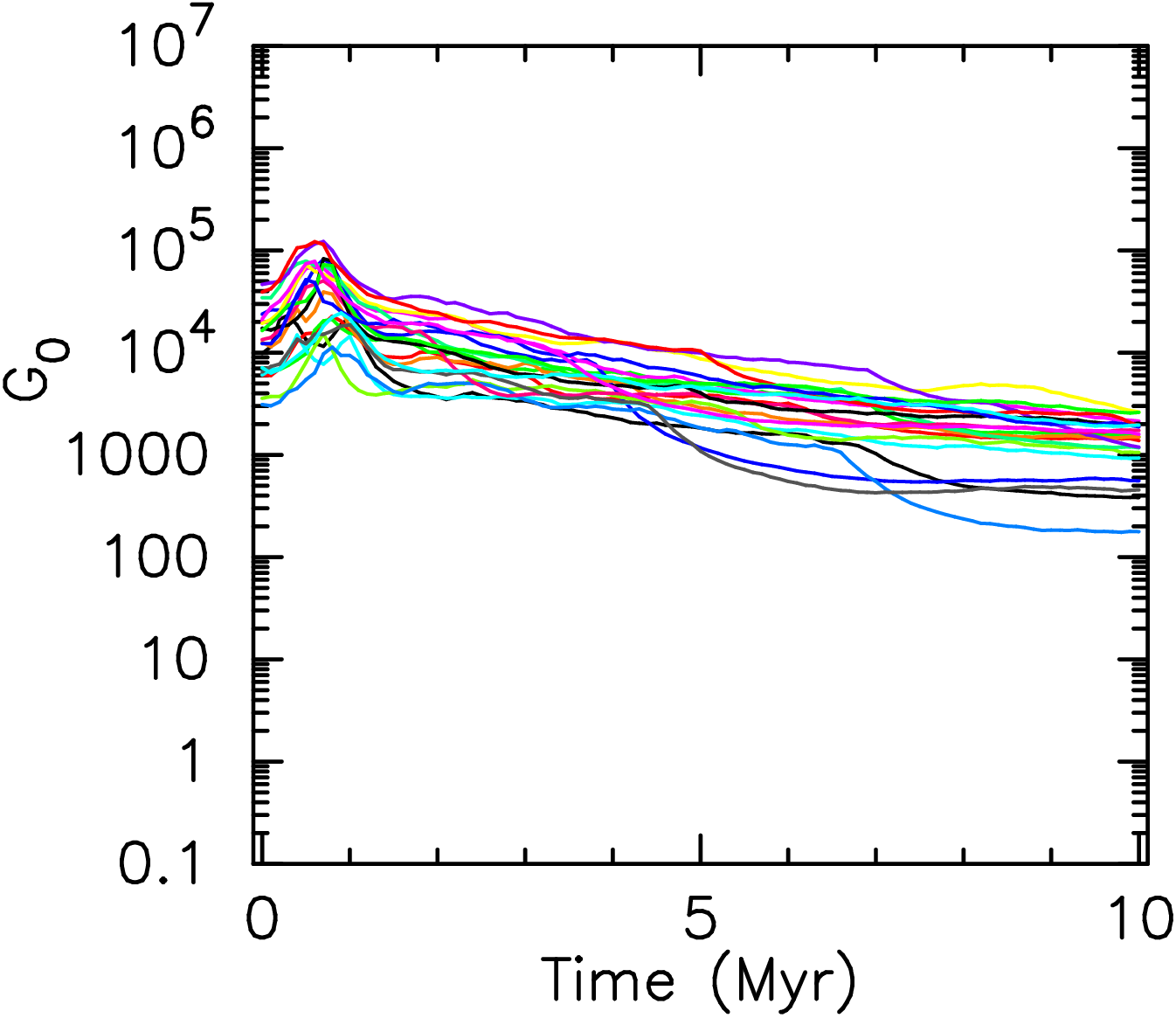}}}
\hspace*{0.3cm}\subfigure[$\tilde{\rho} \sim 100$\,M$_\odot$\,pc$^{-3}$]{\label{G0_densities-f}\rotatebox{270}{\includegraphics[scale=0.19]{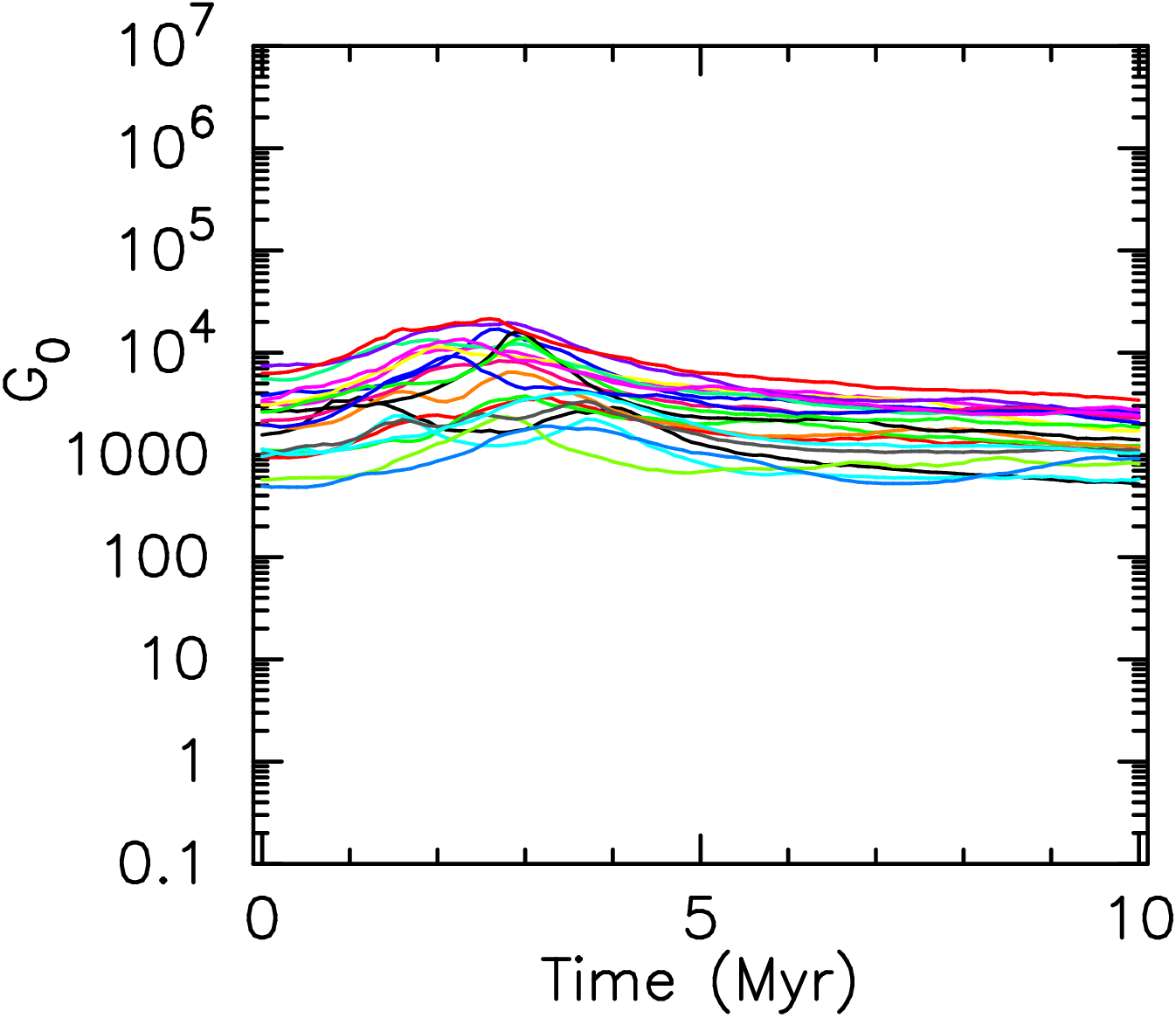}}}
\hspace*{0.3cm}\subfigure[$\tilde{\rho} \sim 10$\,M$_\odot$\,pc$^{-3}$]{\label{G0_densities-g}\rotatebox{270}{\includegraphics[scale=0.19]{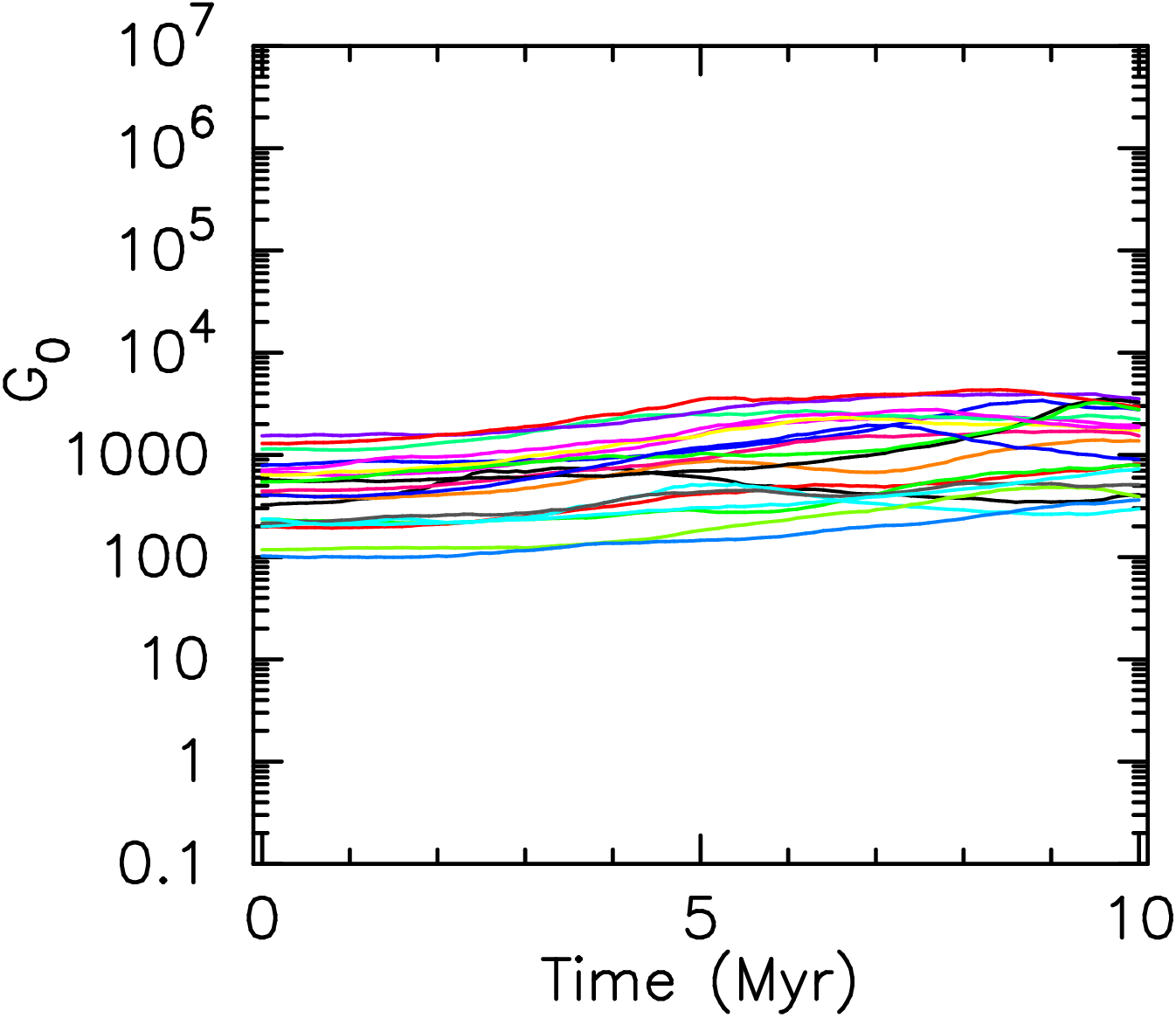}}}
\hspace*{0.3cm}\subfigure[$\tilde{\rho} \sim 0.2$\,M$_\odot$\,pc$^{-3}$]{\label{G0_densities-h}\rotatebox{270}{\includegraphics[scale=0.19]{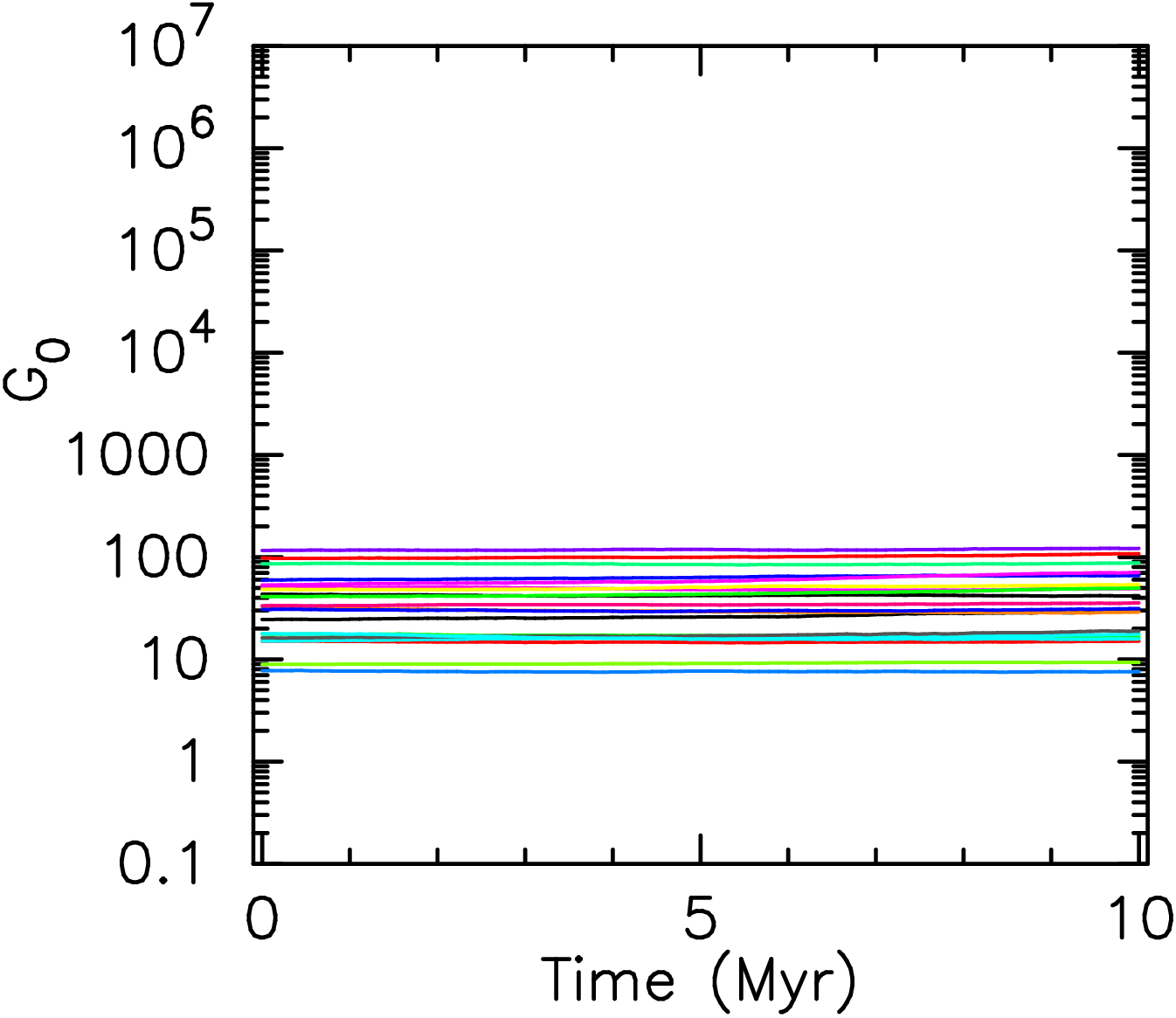}}}

\hspace*{-1.0cm}\subfigure[$\tilde{\rho} \sim 1000$\,M$_\odot$\,pc$^{-3}$]{\label{G0_densities-i}\rotatebox{270}{\includegraphics[scale=0.19]{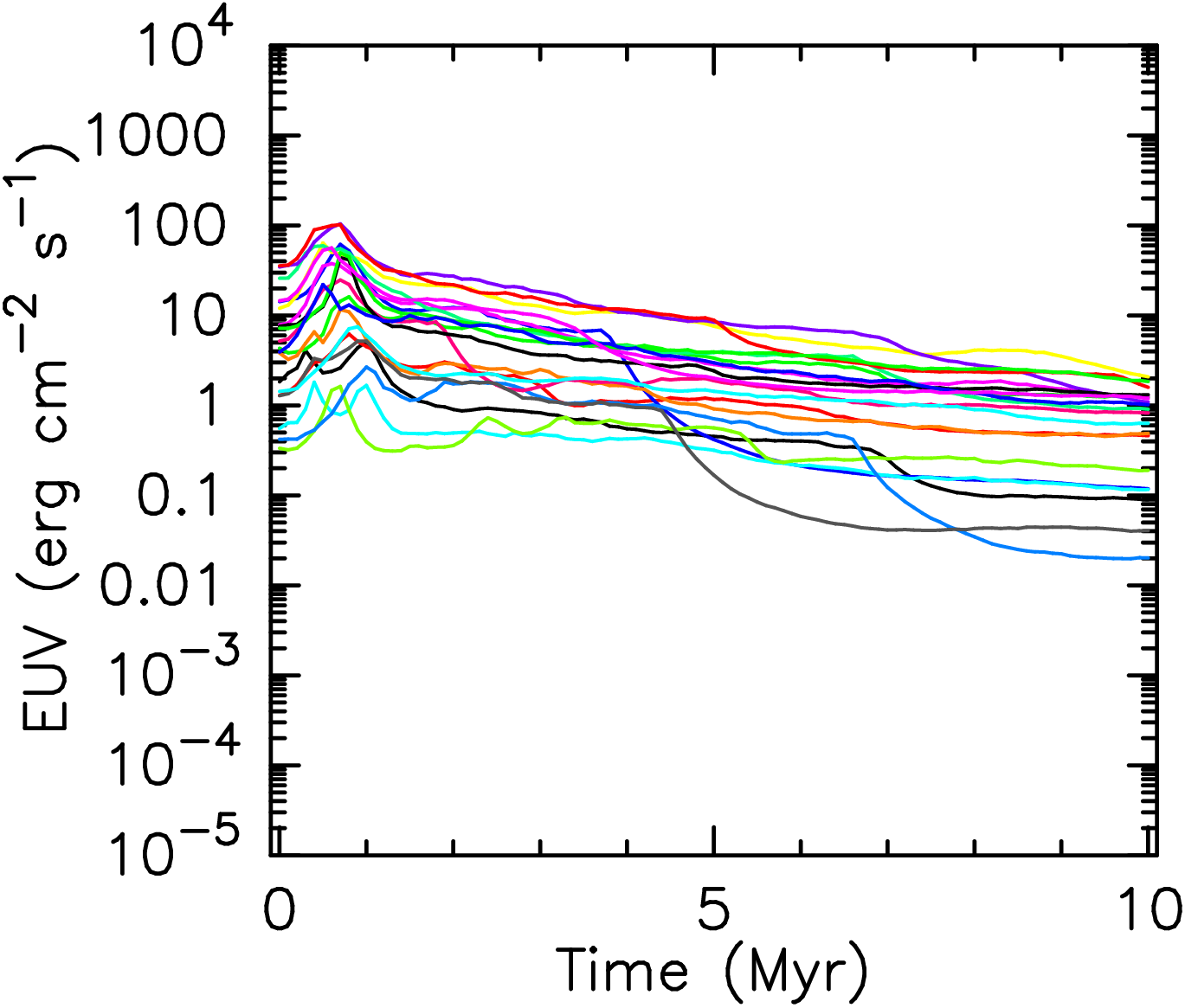}}}
\hspace*{0.2cm}\subfigure[$\tilde{\rho} \sim 100$\,M$_\odot$\,pc$^{-3}$]{\label{G0_densities-j}\rotatebox{270}{\includegraphics[scale=0.19]{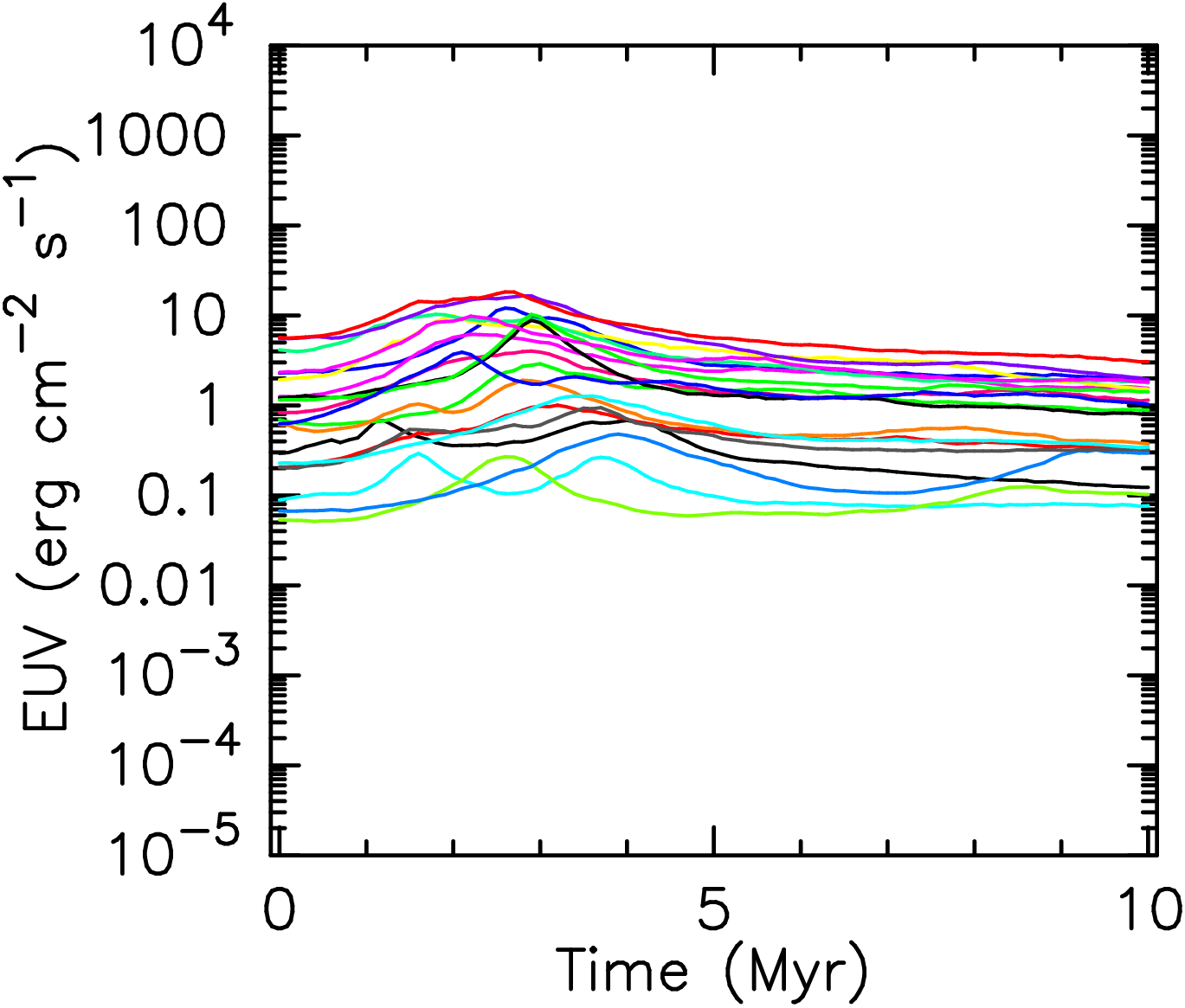}}}
\hspace*{0.2cm}\subfigure[$\tilde{\rho} \sim 10$\,M$_\odot$\,pc$^{-3}$]{\label{G0_densities-k}\rotatebox{270}{\includegraphics[scale=0.19]{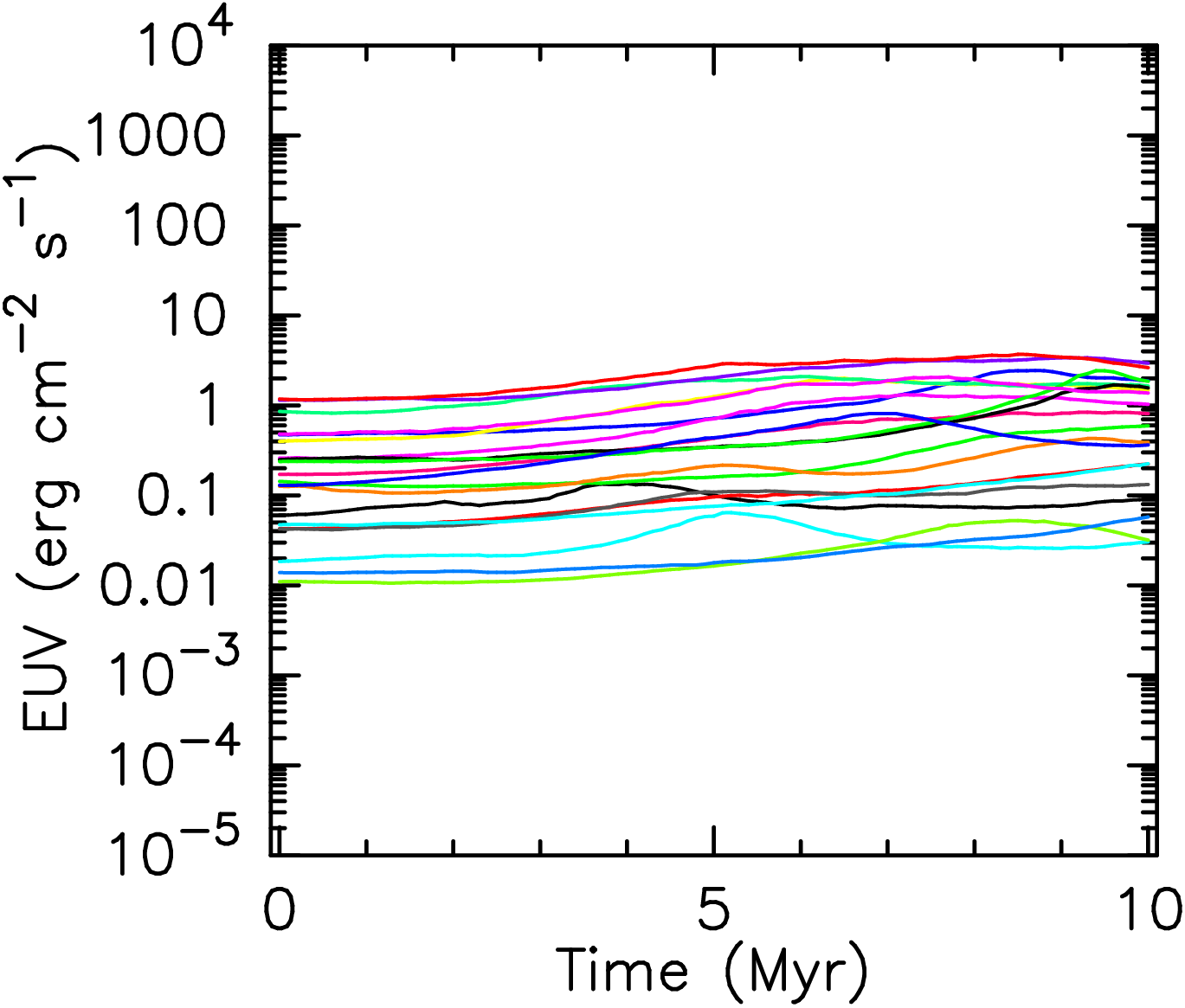}}}
\hspace*{0.2cm}\subfigure[$\tilde{\rho} \sim 0.2$\,M$_\odot$\,pc$^{-3}$]{\label{G0_densities-l}\rotatebox{270}{\includegraphics[scale=0.19]{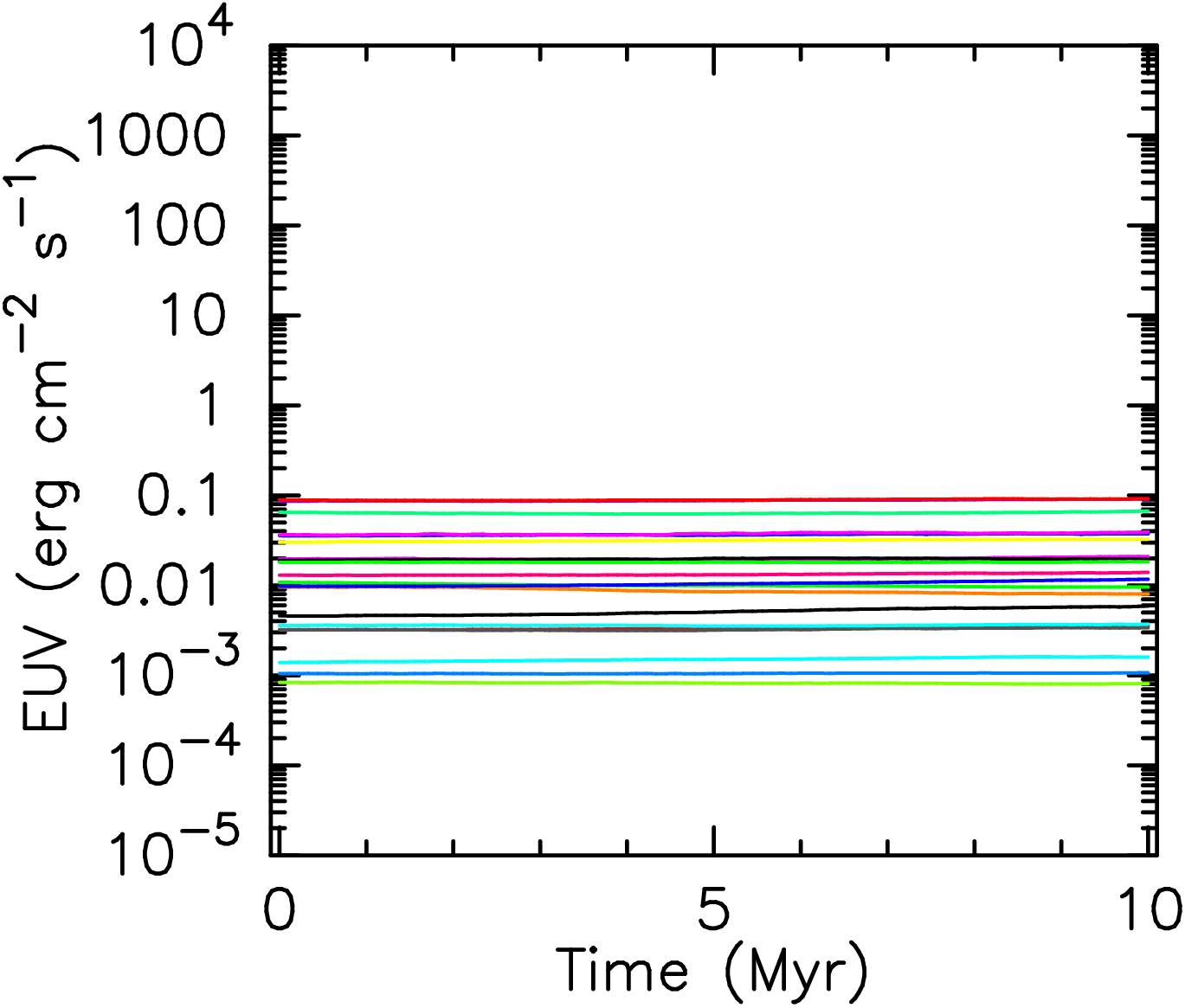}}}
\hspace*{-1.cm}\subfigure[$\tilde{\rho} \sim 1000$\,M$_\odot$\,pc$^{-3}$]{\label{G0_densities-m}\rotatebox{270}{\includegraphics[scale=0.19]{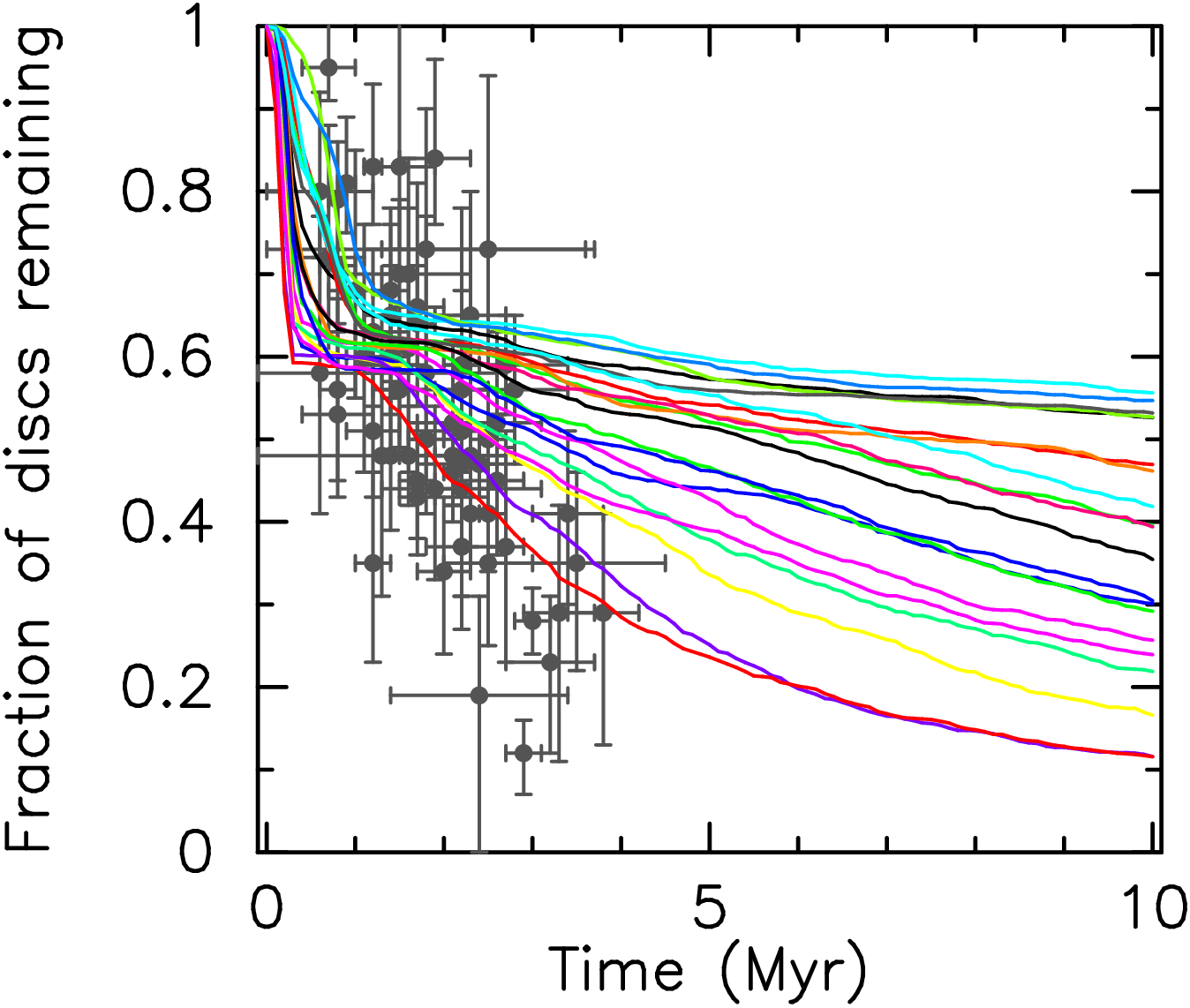}}}
\hspace*{0.2cm}\subfigure[$\tilde{\rho} \sim 100$\,M$_\odot$\,pc$^{-3}$]{\label{G0_densities-n}\rotatebox{270}{\includegraphics[scale=0.19]{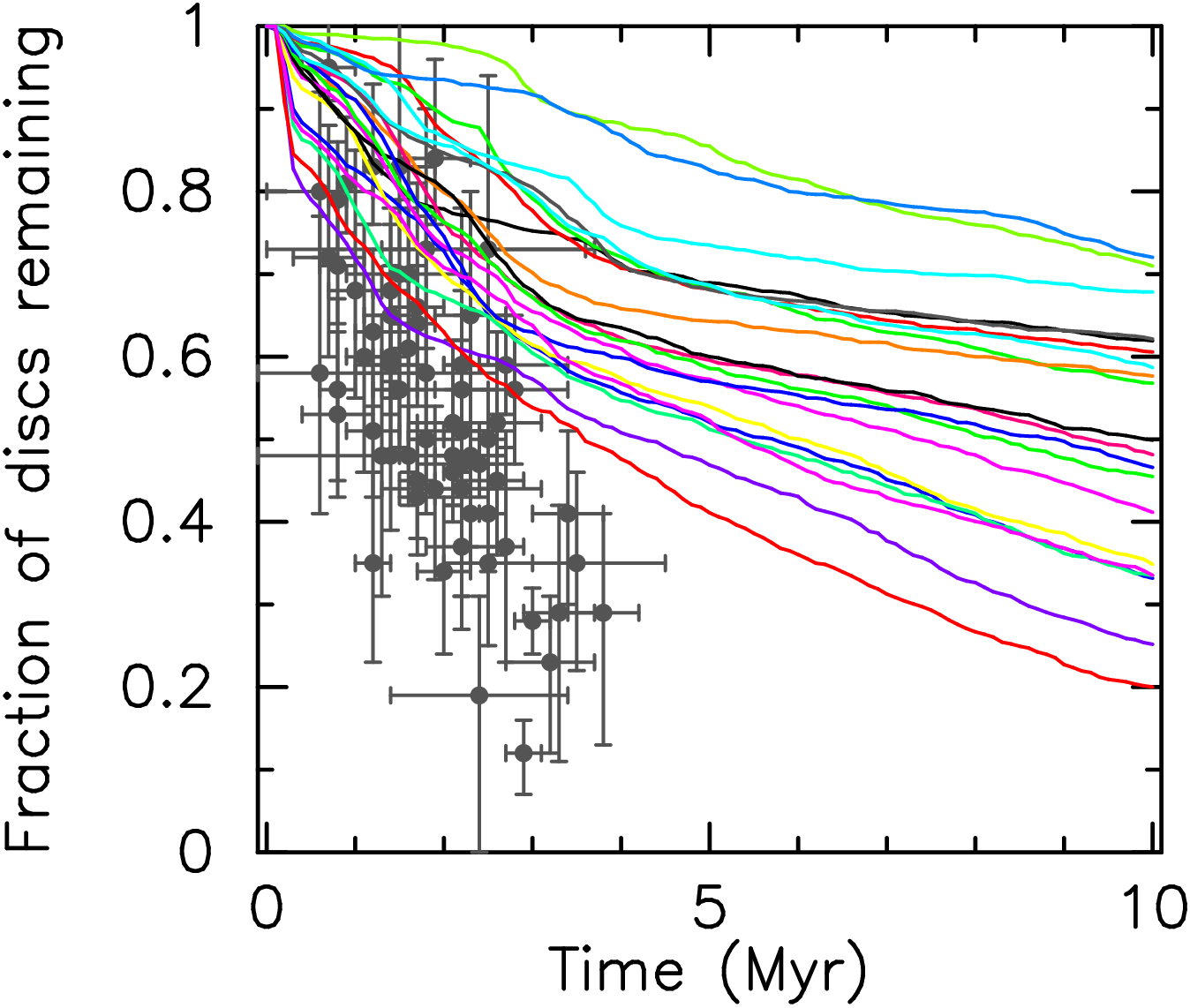}}}
\hspace*{0.2cm}\subfigure[$\tilde{\rho} \sim 10$\,M$_\odot$\,pc$^{-3}$]{\label{G0_densities-o}\rotatebox{270}{\includegraphics[scale=0.19]{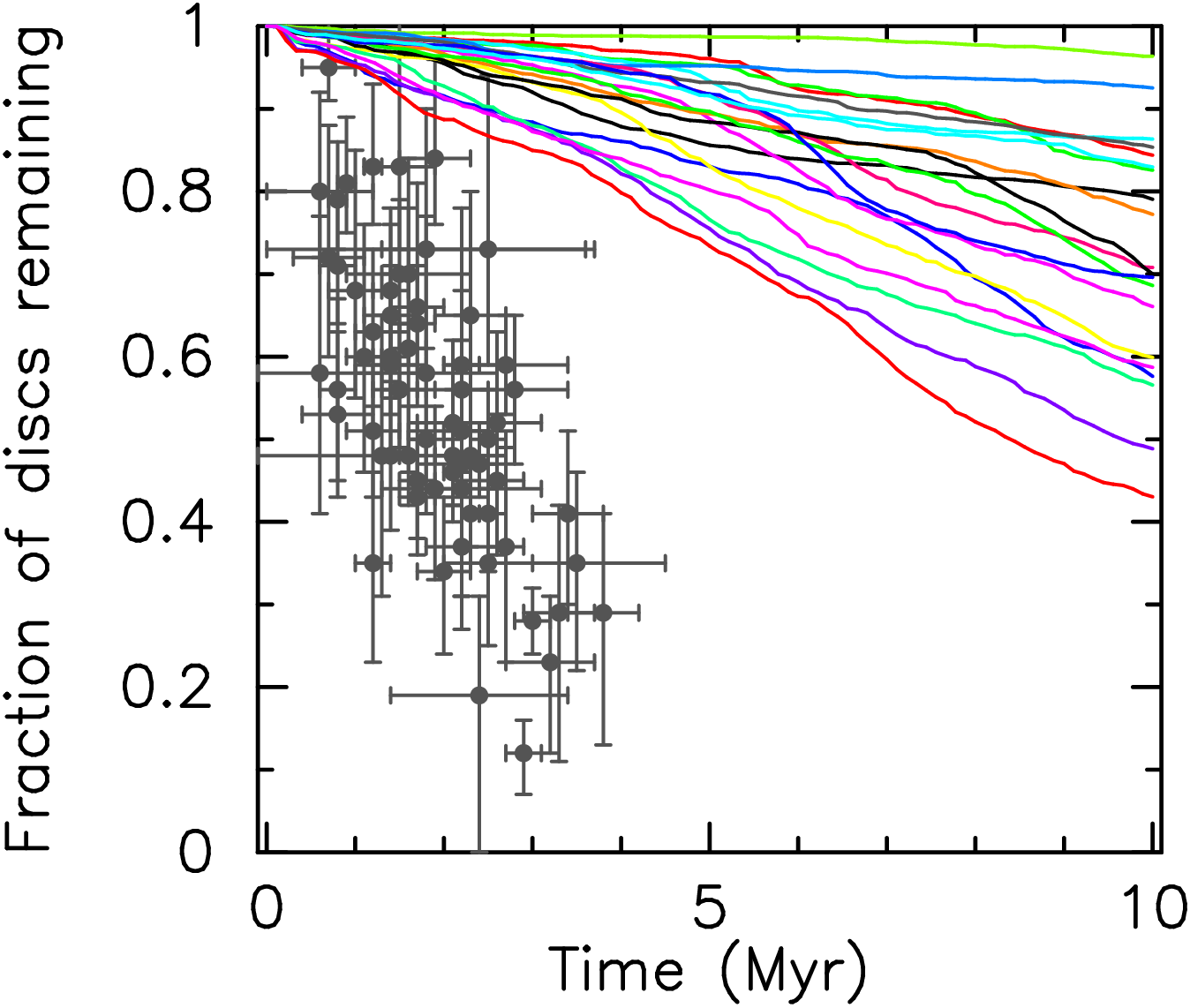}}}
\hspace*{0.2cm}
\subfigure[$\tilde{\rho} \sim 0.2$\,M$_\odot$\,pc$^{-3}$]{\label{G0_densities-p}\rotatebox{270}{\includegraphics[scale=0.19]{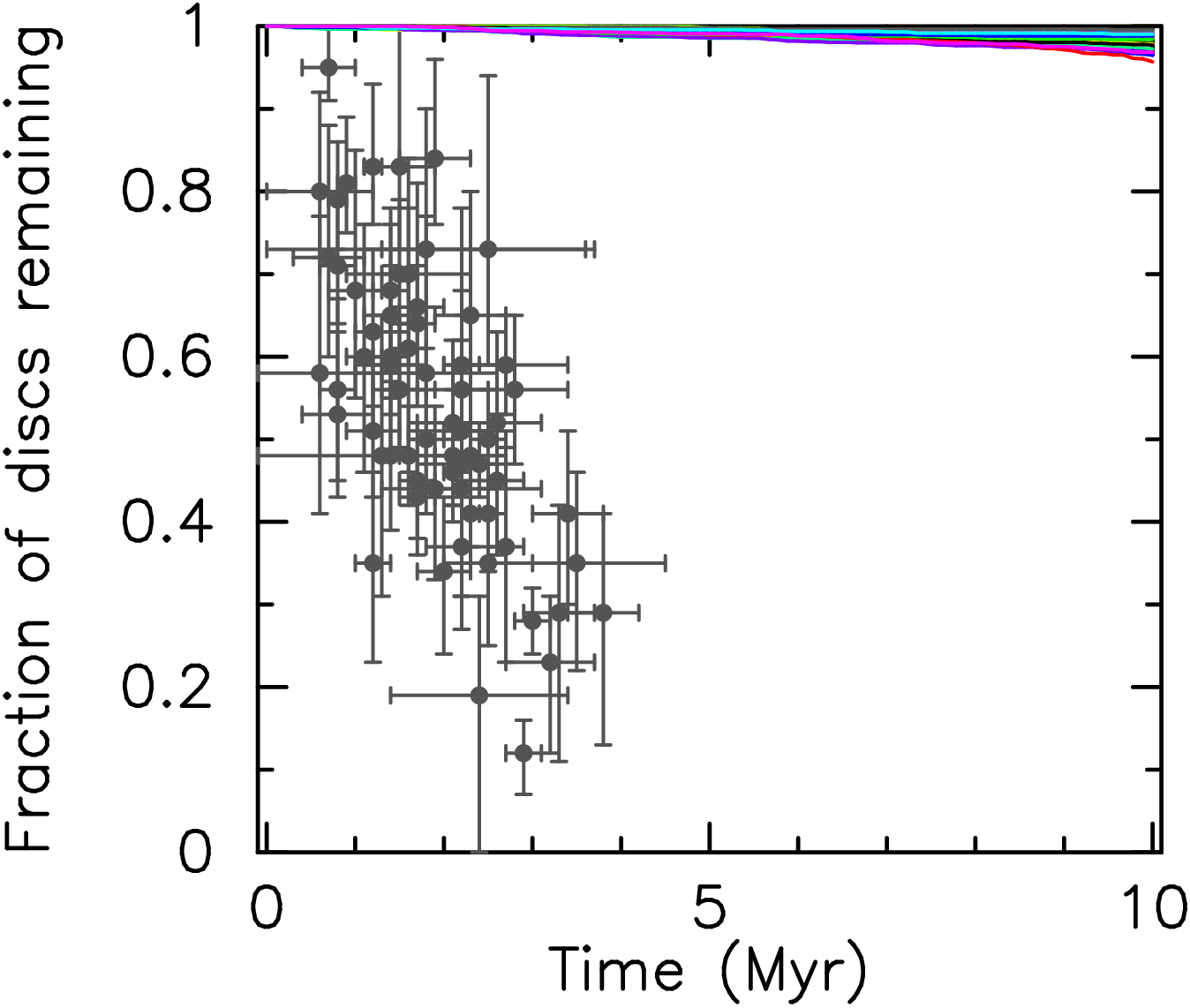}}}

\hspace*{-1.0cm}\subfigure[$\tilde{\rho} \sim 1000$\,M$_\odot$\,pc$^{-3}$]{\label{G0_densities-q}\rotatebox{270}{\includegraphics[scale=0.19]{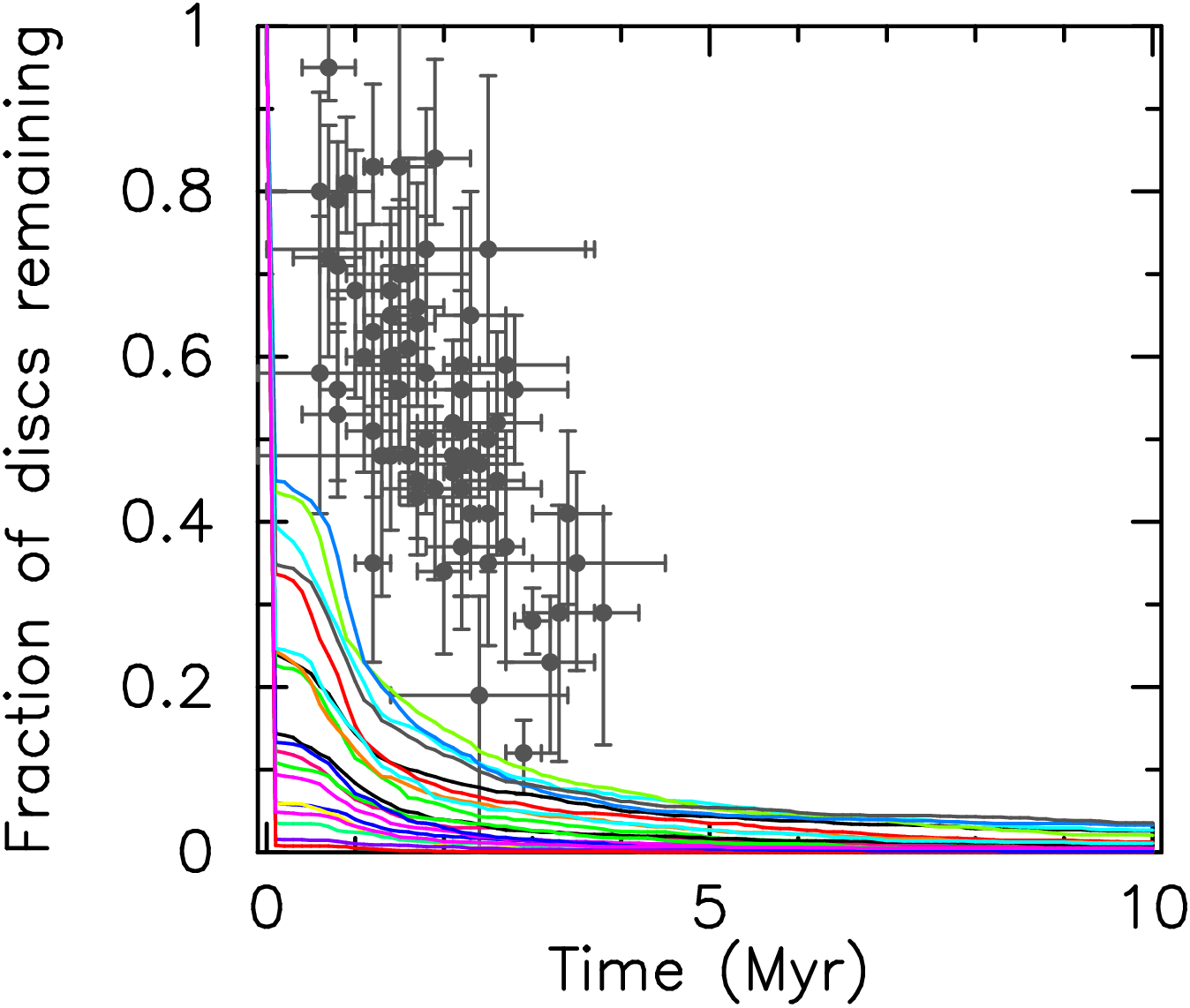}}}
\hspace*{0.2cm}\subfigure[$\tilde{\rho} \sim 100$\,M$_\odot$\,pc$^{-3}$]{\label{G0_densities-r}\rotatebox{270}{\includegraphics[scale=0.19]{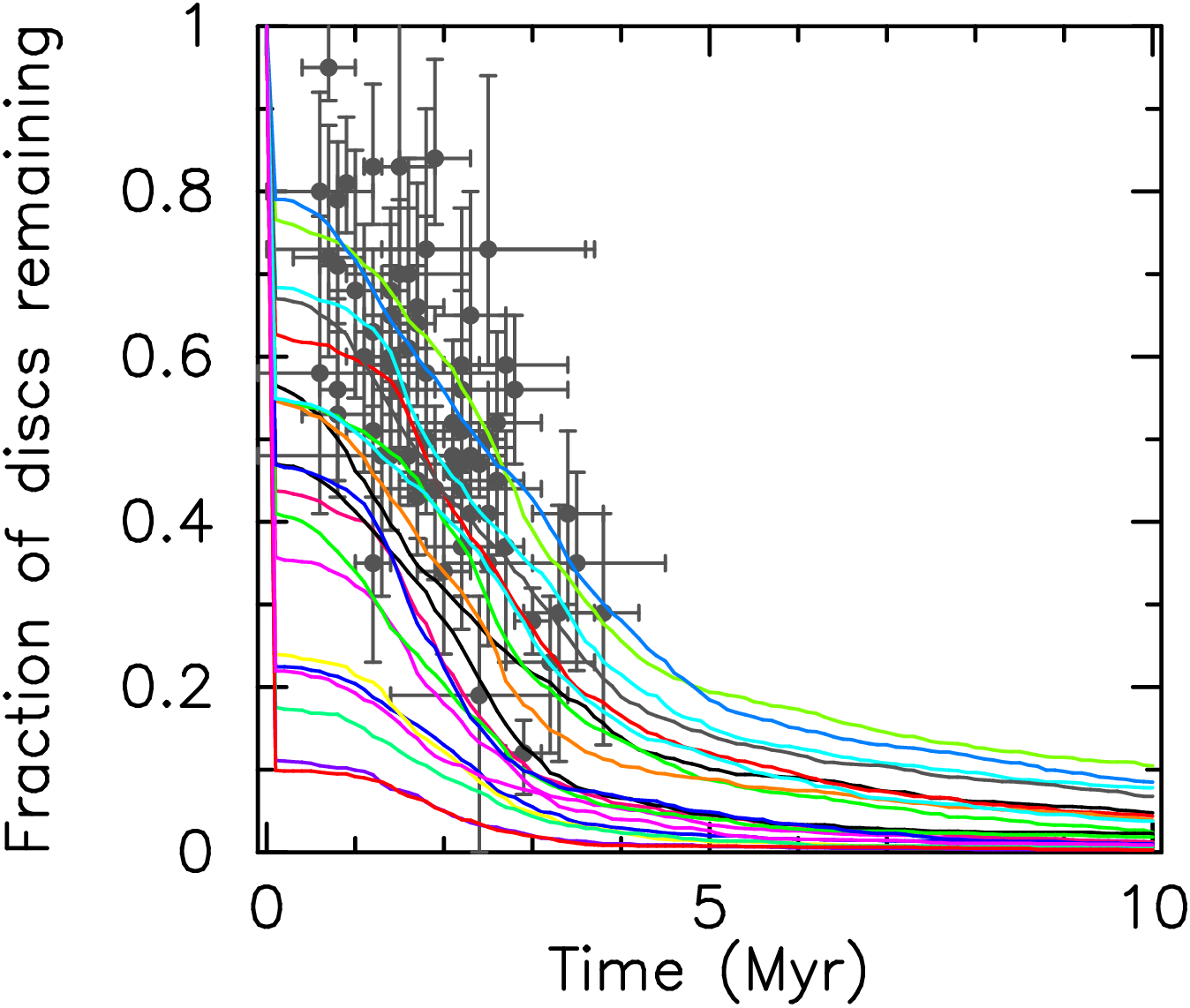}}}
\hspace*{0.2cm}\subfigure[$\tilde{\rho} \sim 10$\,M$_\odot$\,pc$^{-3}$]{\label{G0_densities-s}\rotatebox{270}{\includegraphics[scale=0.19]{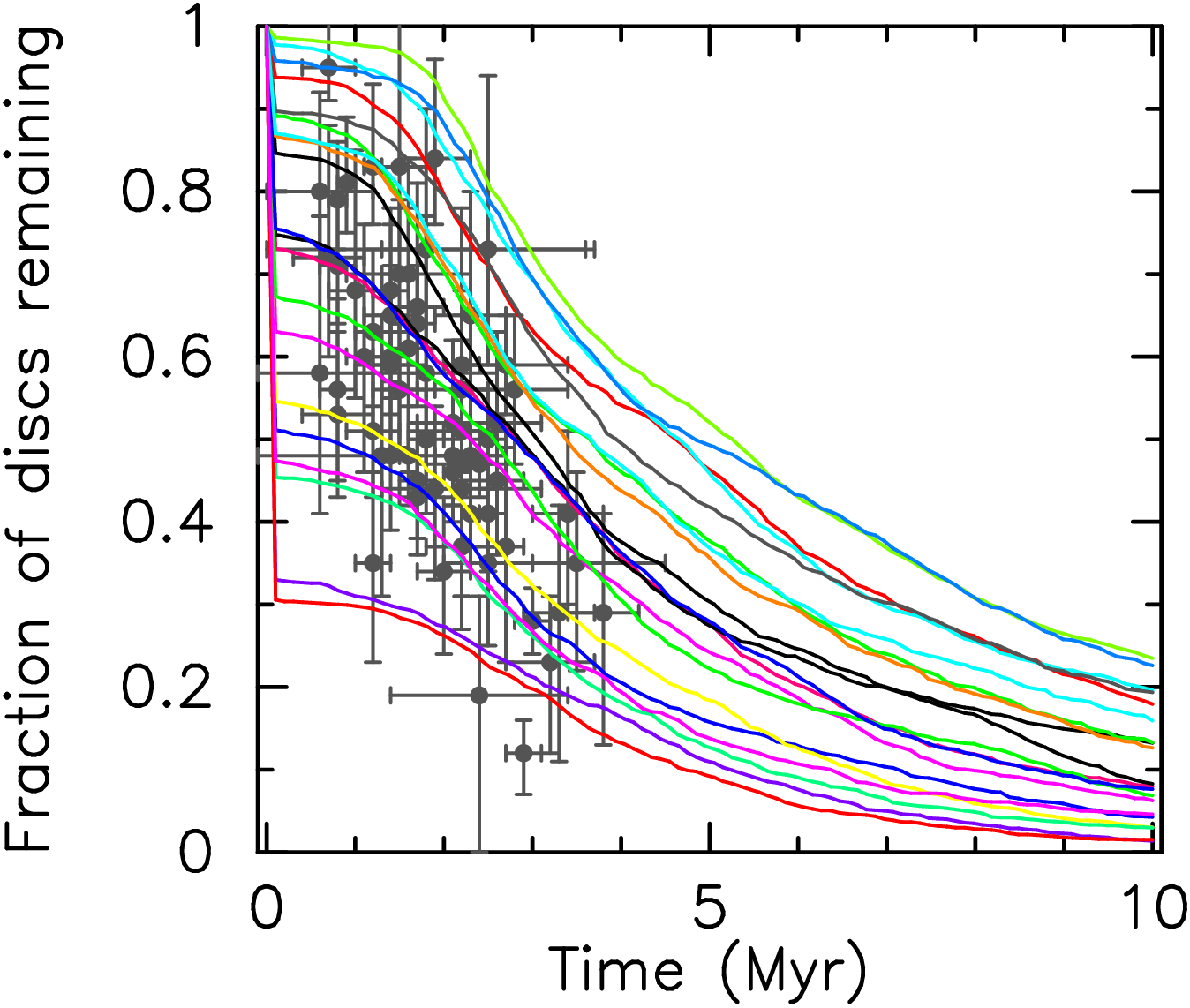}}}
\hspace*{0.2cm}
\subfigure[$\tilde{\rho} \sim 0.2$\,M$_\odot$\,pc$^{-3}$]{\label{G0_densities-t}\rotatebox{270}{\includegraphics[scale=0.19]{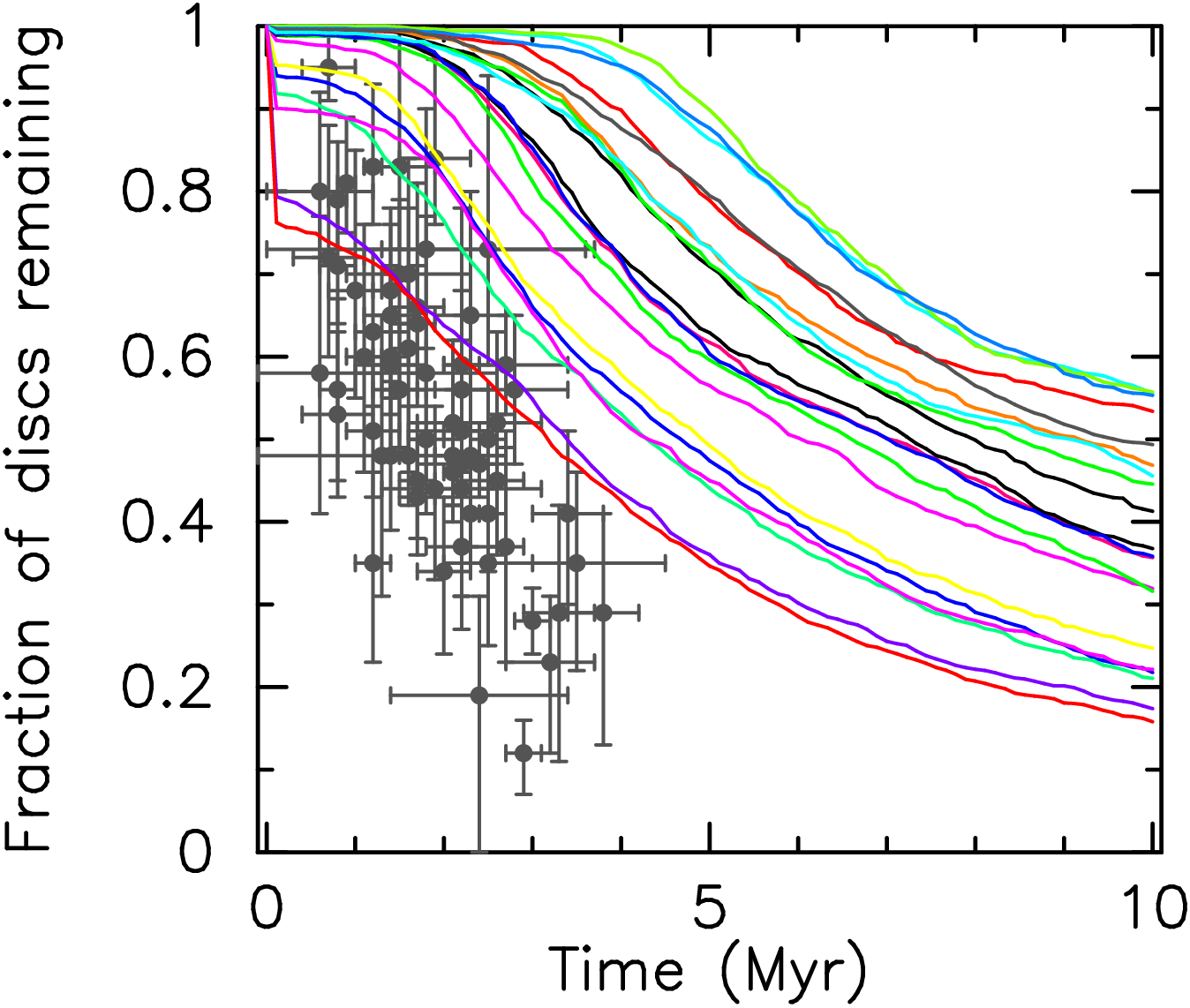}}}

\caption[bf]{The effect of varying the initial local stellar density (the volume density within a sphere that encompasses the ten nearest neighbours to each star). The top row shows the median local density in twenty realisations of the same star-forming region (indicated by the different coloured lines), as well as the mean density within the half-mass radius in all twenty simulations  (the dot-dashed line). The second row shows the median FUV flux, $G_0$, in each simulation and the third row shows the median \\ EUV flux. The fourth row shows the fraction of stars that host gaseous discs in each simulation, where the initial disc radius was 10\,au; the fifth row shows the disc fraction when the initial disc radii were set to 100\,au. These simulations do not include viscous evolution of the discs. The observed disc fractions in star-forming regions from \citet{Richert18} are shown by the dark grey points. }
\label{G0_densities}
  \end{center}
\end{figure*}

In Fig.~\ref{G0_densities} we vary the initial median local density in the star-forming region, whilst keeping the number of stars ($N = 1500$), virial ratio ($\alpha_{\rm vir} = 0.3$) and fractal dimension ($D = 2.0$) constant. The local stellar density is the mass volume density for each star $\rho_{10}$, where the volume is calculated to a fixed number of nearest neighbours (following \citealp{Parker14b} we choose the tenth nearest neighbour, $r_{10}$ but the results tend to be robust for any value in the range $N = 5 - 15$, \citealp{Bressert10,Parker12d}), and the mass is the total mass of the ten nearest neighbours, $M_{10}$:
\begin{equation}
\rho_{10} = \frac{3M_{10}}{4\pi r_{10}^3}.
\end{equation}
We then take the median value of $\rho_{10}$ to determine the median local density for each star-forming region, $\tilde{\rho}$. We show twenty versions of the same initial conditions, identical save for the random number seed used to initialise the mass functions, positions and velocities of the stars. 

In panels (a)--(d) we show the evolution of the median local stellar density by the coloured lines, as well as the mean density within the half-mass radius $\rho_{1/2}$ for all twenty simulations, shown by  dot-dashed black line and defined as
\begin{equation}
\rho_{1/2} = \frac{3M_{F, 1/2}}{4\pi r_{1/2}^3},
\label{half_mass_eq}
\end{equation}
where $M_{F, 1/2}$ half of the total stellar mass of the star-forming region, and $r_{1/2}$ is the radius from the centre that encloses this mass. In all the density regimes, the median local density is higher than the mean central density before dynamical evolution. As the star-forming regions undergo subvirial collapse, the central density can eventually exceed the initial local density. 

This is seen in the evolution of the median FUV flux (second row, panels (e)--(h)) and the EUV flux (third row, panels (i)--(l)), which for the denser regions (the first two columns) have distinct peaks at the point where the region collapses to form a centrally concentrated, spherical star cluster at 1\,Myr for the most dense regions ($\tilde{\rho} = 1000$\,M$_\odot$\,pc$^{-3}$, panel (e)) and 3\,Myr for regions with initial stellar densities of $\tilde{\rho} = 100$\,M$_\odot$\,pc$^{-3}$ (panel (f)). There are also hints at a peak central density, indicative of a collapse, around or just after 10\,Myr in the the simulations that start with median local densities of $\tilde{\rho} = 10$\,M$_\odot$\,pc$^{-3}$ (the third column, panels (g) and (k)). Th{\bf e} reasons for this behaviour are two-fold. First, the disc-hosting low-mass stars are being funnelled into the potential well of the cluster, so they are more likely to be close to the most massive stars. Secondly, the most massive stars undergo dynamical mass segregation \citep{McMillan07,Moeckel09a,Allison10}, which increases the FUV and EUV flux experienced by the majority of the low-mass stars.

There is more variation in the EUV flux values between simulations than the FUV flux (compare e.g. panel (g) with panel (k) in Fig.~\ref{G0_densities}), and this is due to the stochastic sampling of the IMF and the fact that the contribution from EUV comes from more massive (rarer) stars than the equivalent FUV flux. 

In the most dense star-forming regions ($\tilde{\rho} = 1000$\,M$_\odot$\,pc$^{-3}$), the initial FUV flux is $G_0 \sim 10^4$, which increases to  $G_0 \sim 10^5$ during the formation of the cluster and subsequent mass segregation. Strikingly, the peak median $G_0$ value decreases by the same order of magnitude as the decrease in stellar density, so the simulations with stellar densities similar to the Galactic field (0.2\,M$_\odot$\,pc$^{-3}$) still have $G_0$ values between 10 and 100 (panel (h)), i.e. between ten and one hundred times the FUV flux in the interstellar medium. 

The high FUV and EUV fluxes, even at lower densities, have severely detrimental effects on the survival of the gas content within protoplanetary discs. When we implement the \texttt{FRIED} grid of photoevaporation models, we see a significant depletion in discs with initial radii of more than 10\,au.  Panels (m)--(p) show the fraction of stars that host protoplanetary discs with initial radii $r_{\rm disc} = 10$\,au in each of the twenty simulations as a function of time. This fraction rapidly drops to 60\,per cent in the densest simulations ($\tilde{\rho} = 1000$\,M$_\odot$\,pc$^{-3}$), with final fractions between 10 to  50\,per cent. As the stellar density (and $G_0$ fields) decrease, the fraction of surviving discs increases, as can be seen in panels (n)--(p), with almost no mass-loss due to photoevaporation in 10\,au discs when the stellar density is field-like ($\tilde{\rho} = 0.2$\,M$_\odot$\,pc$^{-3}$) and the $G_0$ fields are only a factor of 10 -- 100 higher than the ISM (panel p).   

However, when the disc radii are initially 100\,au (with initial disc masses 0.1\,M$_\star$), the mass-loss due to photoevaporation is very drastic, with disc fractions dropping to between zero and 50\,per cent in the first 0.1\,Myr, depending on the initial stellar density.  If we take an example calculation from the \texttt{FRIED} grid, an average mass star (0.5\,M$_\odot$) with a 100\,au radius disc  of mass 0.05\,M$_\odot$ in an FUV field of $G_0 = 100$ will experience a mass-loss rate of $\dot{M} = 1.43 \times 10^{-6}$\,M$_\odot$\,yr$^{-1}$, i.e.\,\,will lose 0.143\,M$_\odot$ in 0.1\,Myr. It is therefore unsurprising that we see such significant disc depletion in our simulated star-forming regions when using these models.  In Appendix~\ref{appendix:resolution} we show further examples of the evolution of discs subject to mass-loss from the \texttt{FRIED} grid.

\subsubsection{Initial spatial structure}

\begin{figure*}
  \begin{center}
  \vspace*{-0.5cm}
\setlength{\subfigcapskip}{10pt}
\hspace*{0.3cm}\subfigure[$D = 1.6$]{\label{fractal_dimensions-a}\rotatebox{270}{\includegraphics[scale=0.23]{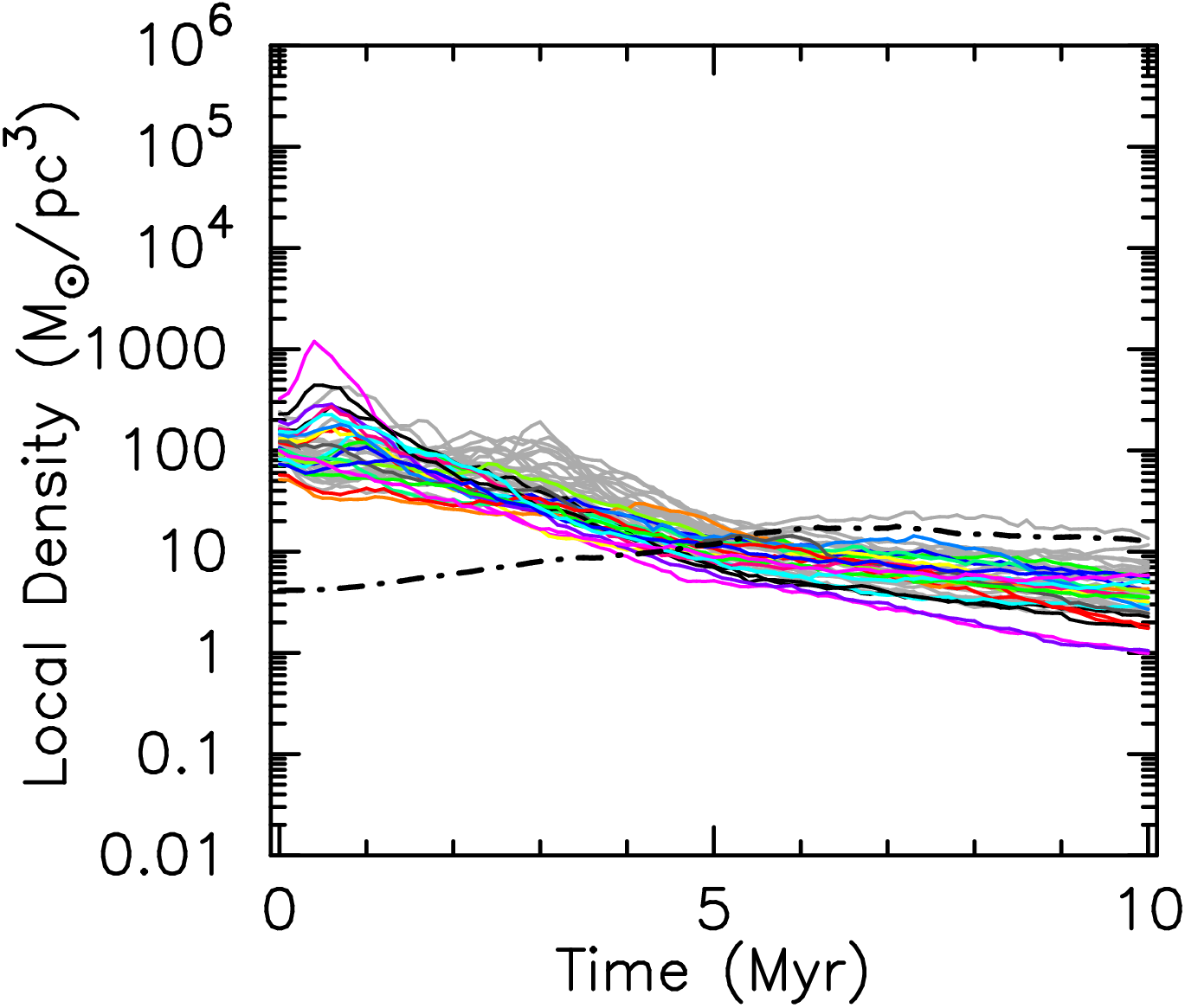}}}
\hspace*{0.5cm}\subfigure[$D = 3.0$]{\label{fractal_dimensions-c}\rotatebox{270}{\includegraphics[scale=0.23]{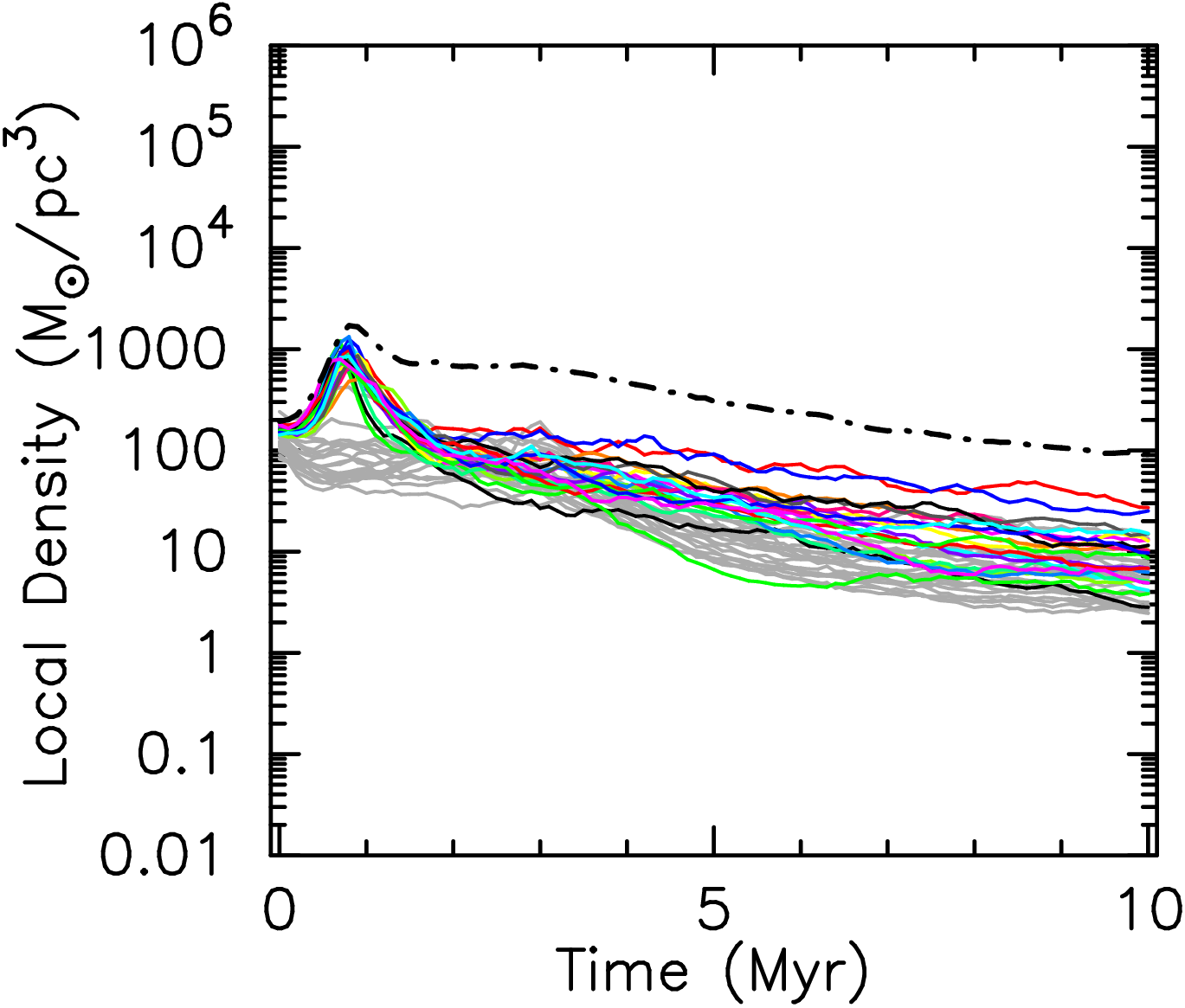}}} \newline
\hspace*{0.3cm}\subfigure[$D = 1.6$]{\label{fractal_dimensions-d}\rotatebox{270}{\includegraphics[scale=0.23]{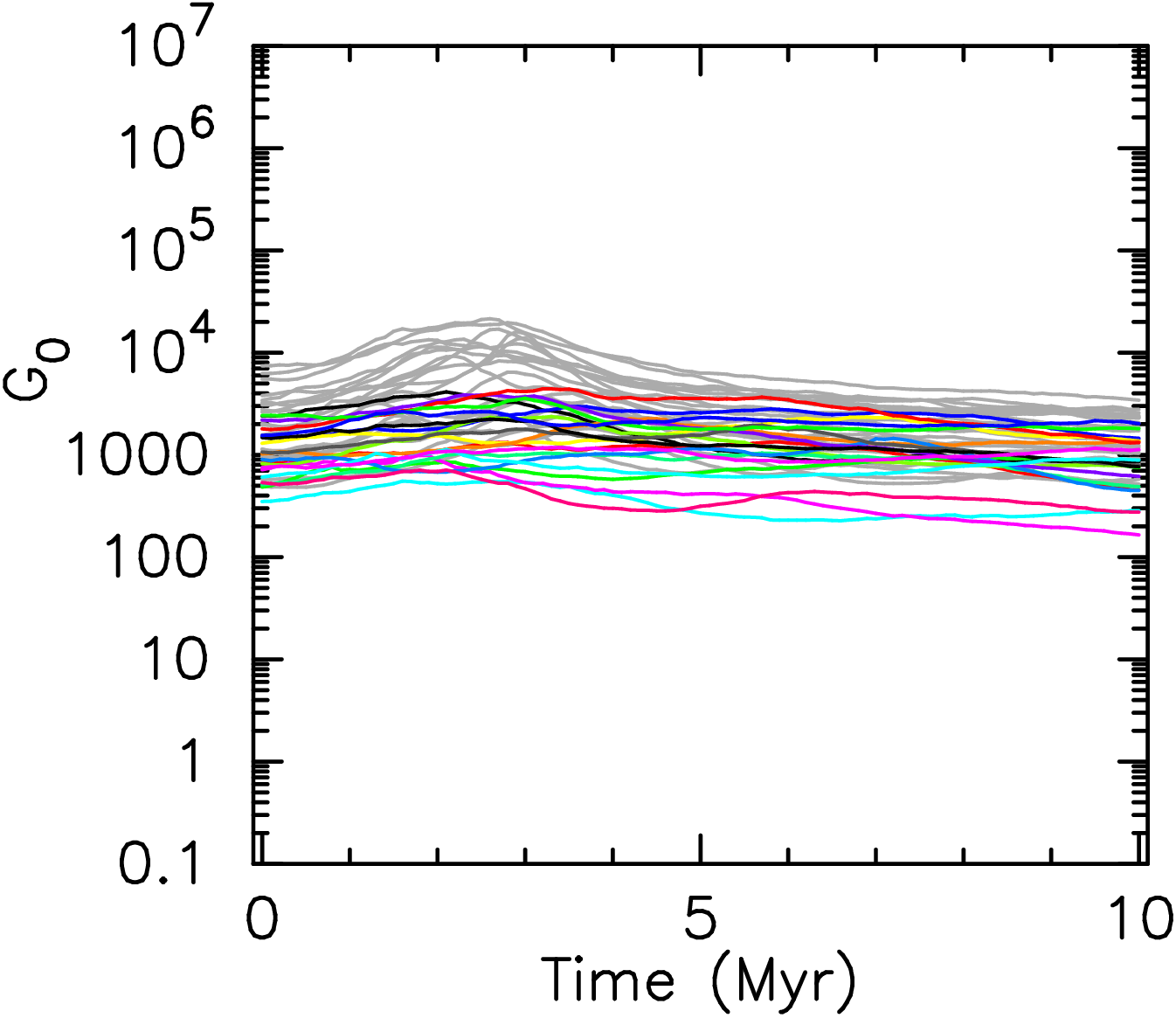}}}
\hspace*{0.5cm}\subfigure[$D = 3.0$]{\label{fractal_dimensions-f}\rotatebox{270}{\includegraphics[scale=0.23]{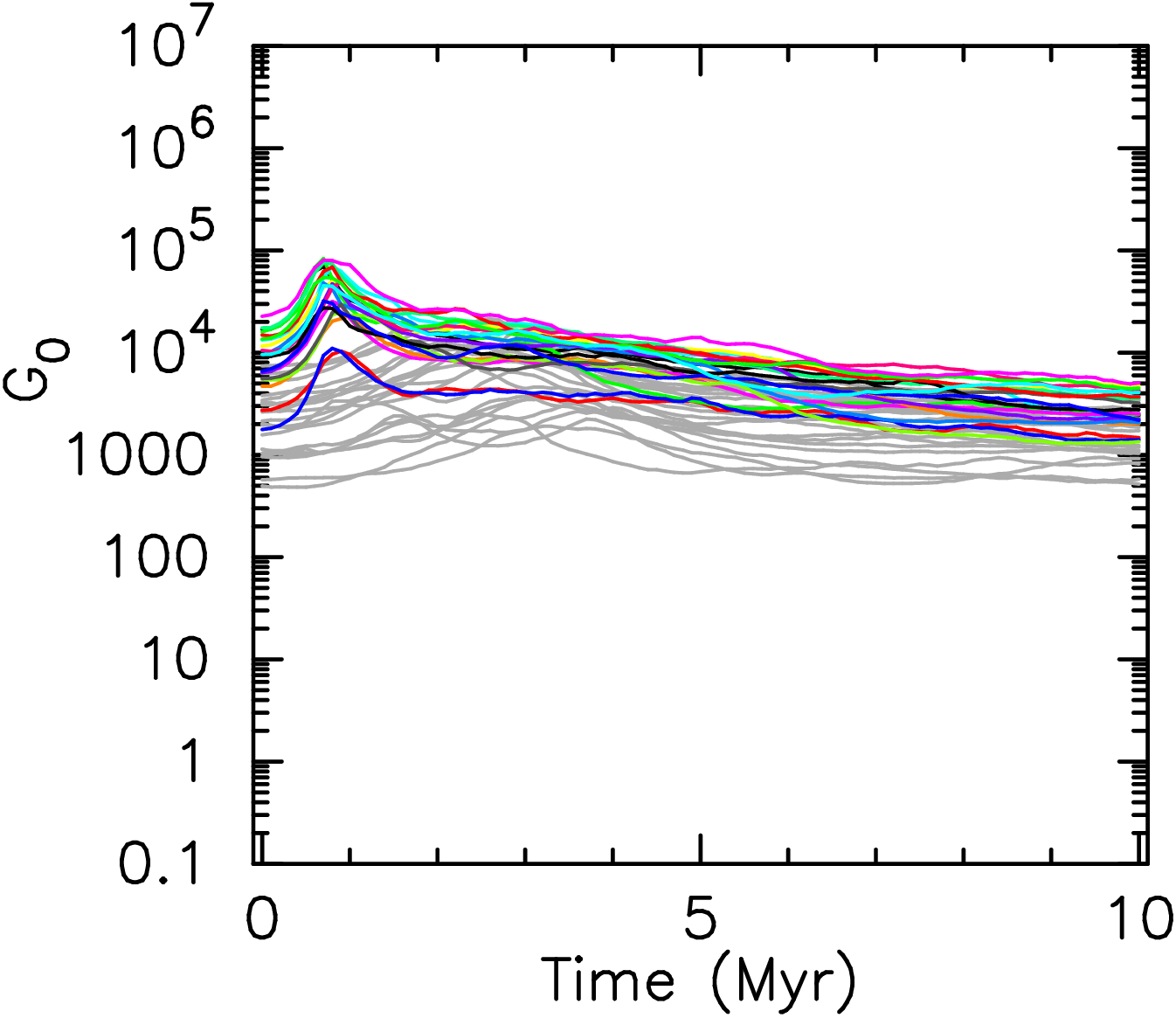}}} \newline
\hspace*{0.3cm}\subfigure[$D = 1.6$]{\label{fractal_dimensions-g}\rotatebox{270}{\includegraphics[scale=0.23]{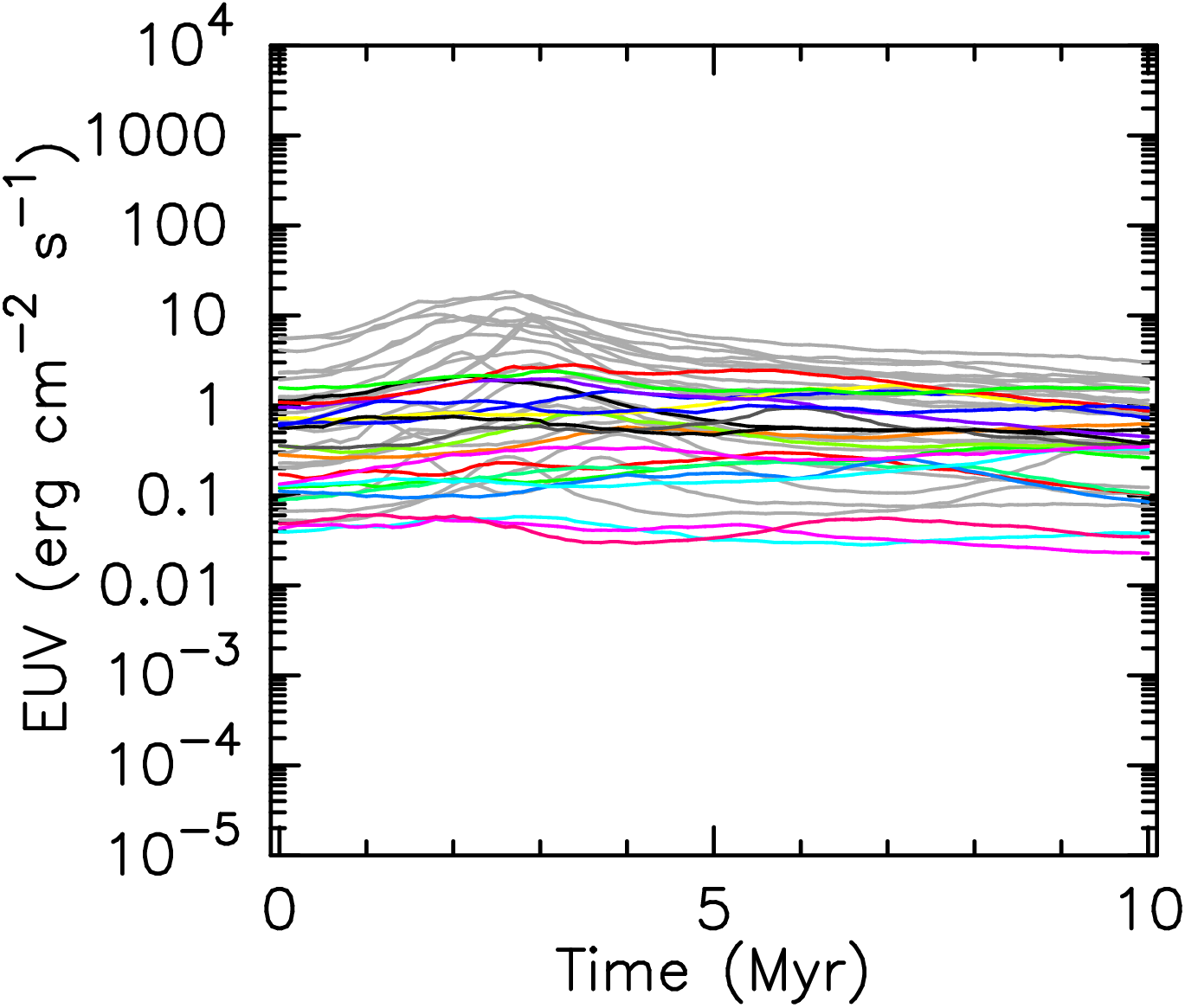}}}
\hspace*{0.5cm}\subfigure[$D = 3.0$]{\label{fractal_dimensions-i}\rotatebox{270}{\includegraphics[scale=0.23]{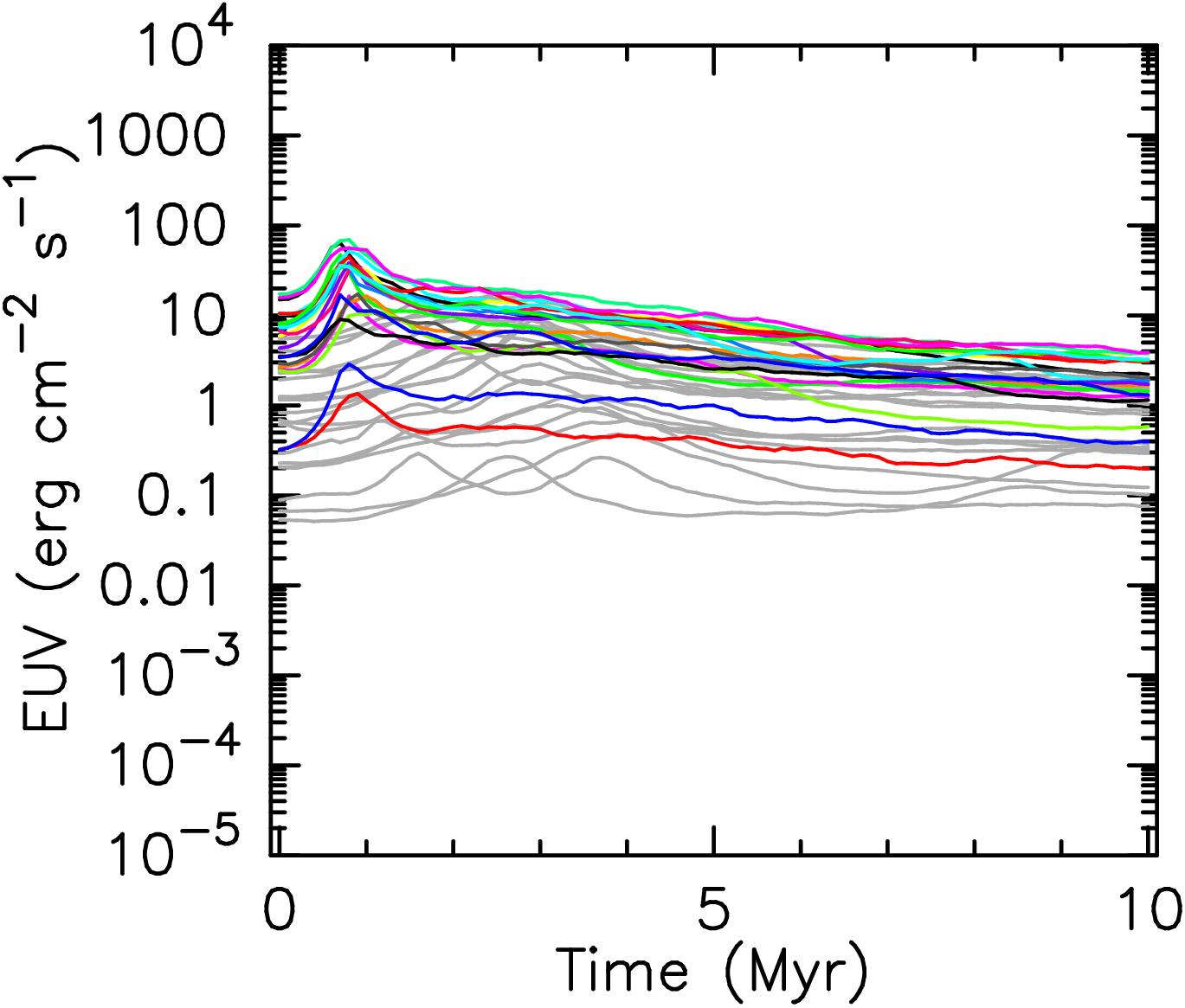}}} \newline
\hspace*{-1.7cm}\subfigure[$D = 1.6$]{\label{fractal_dimensions-j}\rotatebox{270}{\includegraphics[scale=0.23]{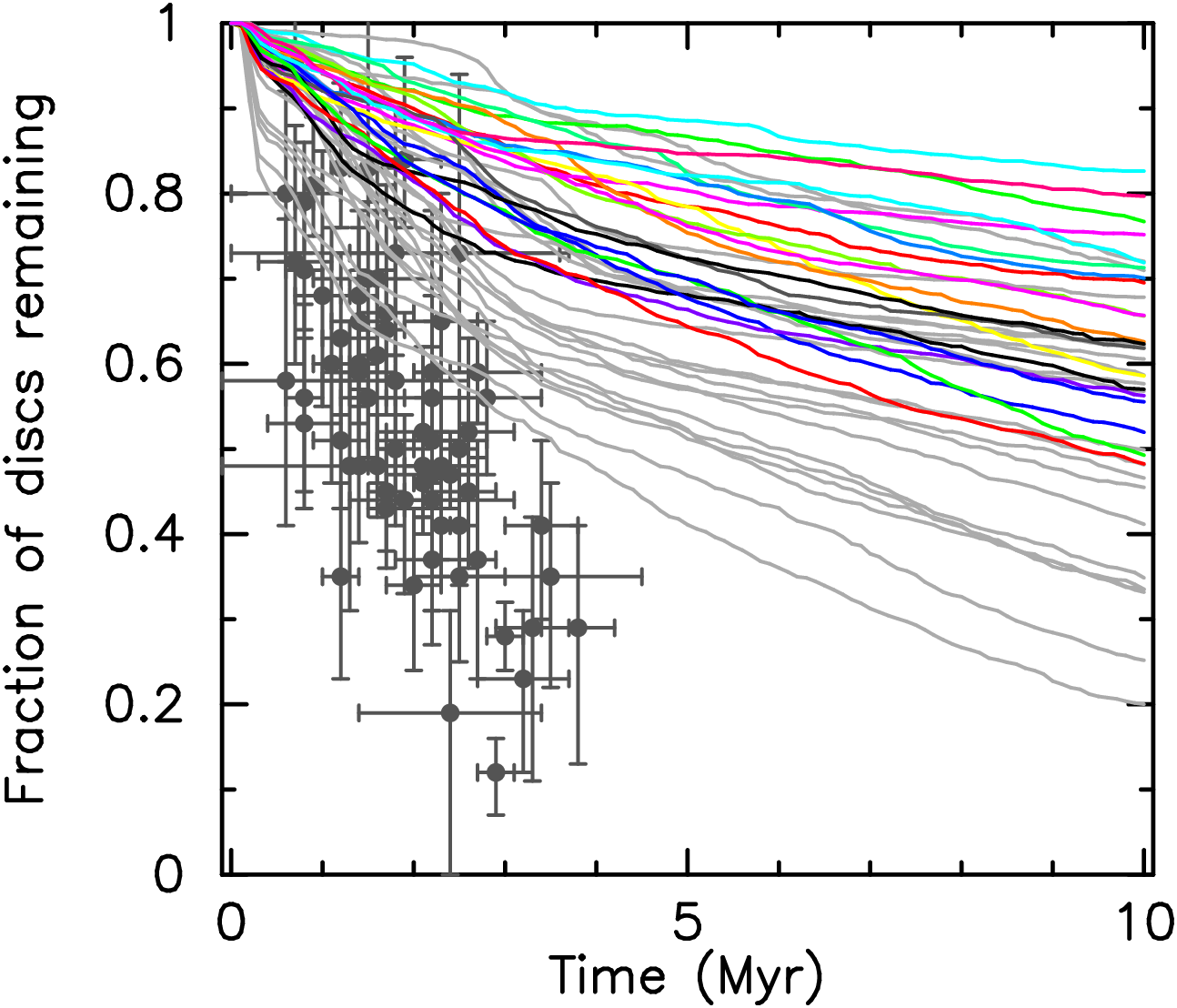}}}
\hspace*{0.9cm}\subfigure[$D = 3.0$]{\label{fractal_dimensions-l}\rotatebox{270}{\includegraphics[scale=0.23]{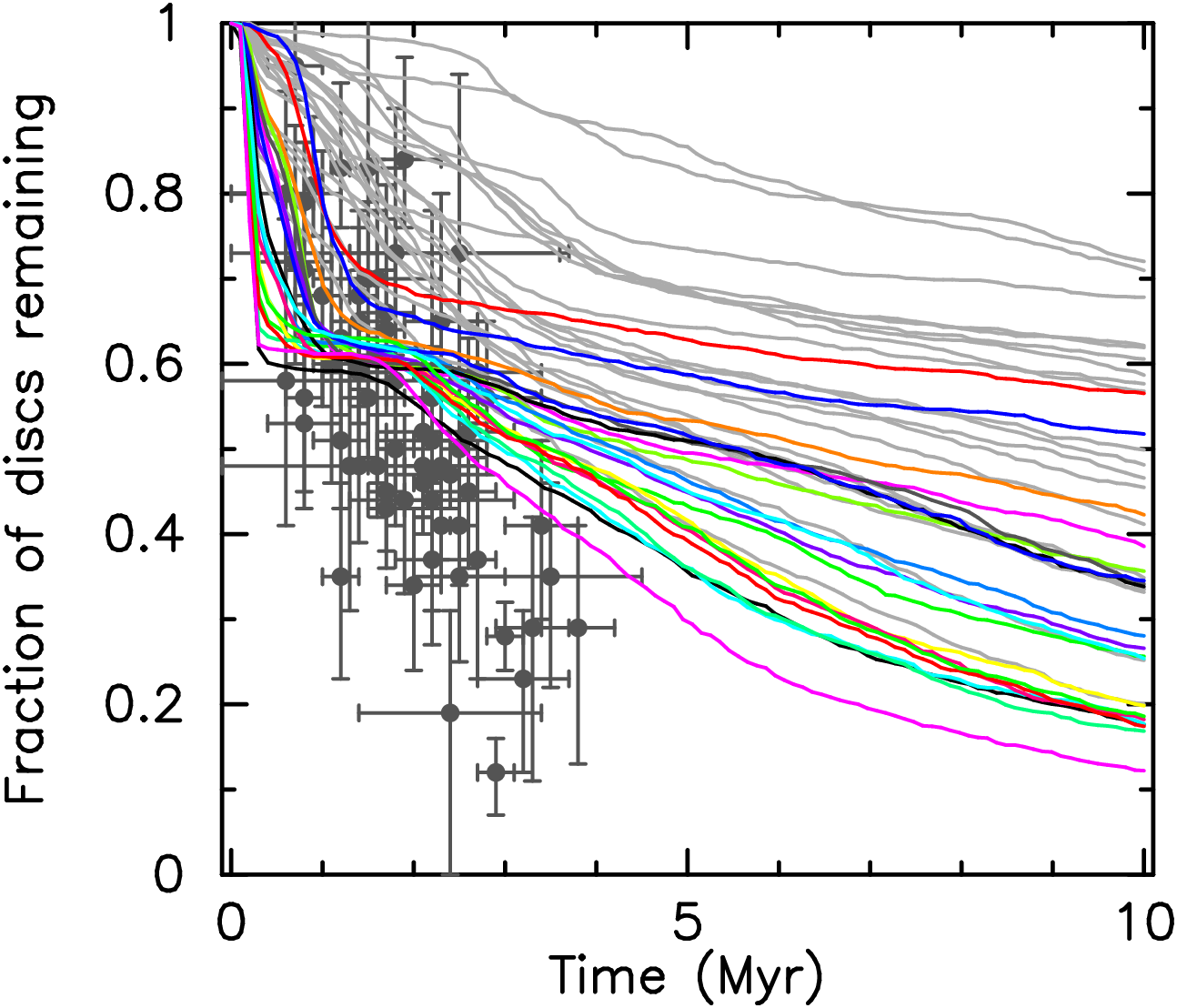}}} 
\caption[bf]{Comparison of the evolution of star-forming regions with different fractal dimensions but with constant initial stellar density ($\tilde{\rho} \sim 100$\,M$_\odot$\,pc$^{-3}$). The top row shows the median local density in twenty realisations of the same star-forming region (indicated by the different coloured lines), as well as the mean density within the half-mass radius in all twenty simulations  (the dot-dashed line). The second row shows the median FUV flux, $G_0$, in each simulation and the third row shows the median EUV flux. The fourth row shows the fraction of stars that host gaseous discs in each simulation, where the initial disc radius was 10\,au. The simulations with a high fractal dimension (less spatial and kinematic substructure) have higher $G_0$ values as the simulation progresses, because the substructure does not dynamically evolve, but the region collapses to form a cluster faster than the substructured regions (because the overall density is higher to begin with). The grey lines are the default simulations (moderate substructure, $D = 2.0$). The observed disc fractions in star-forming regions from \citet{Richert18} are shown by the dark grey points. }
\label{fractal_dimensions}
  \end{center}
\end{figure*}

In Fig.~\ref{fractal_dimensions} we show the effects of varying the initial degree of substructure by changing the fractal dimension of the star-forming regions. We fix the median local density to be $\tilde{\rho} = 100$\,M$_\odot$\,pc$^{-3}$ in each simulation, which means the initial radii are quite different; $r_F = 5$\,pc for the highly substructured simulations ($D = 1.6$), $r_F = 2.5$\,pc for the moderately substructured simulations ($D = 2.0$) and $r_F = 1.1$\,pc for the non-substructured simulations ($D = 3.0$).  In this figure the lefthand column shows the results for simulations with a high degree of substructure ($D = 1.6$) and the righthand column shows the results for simulations with no substructure ($D = 3.0$). In each panel, the results from simulations with an intermediate amount of substructure ($D = 2.0$), but otherwise identical initial conditions, are shown by the background grey lines.

These very different initial radii affect the global evolution of the star-forming regions. The gravitational potential is $\psi \propto M_F/r_F$, where  $M_F$ and $r_F$ are the mass and radius of the star-forming region, respectively. Therefore, a region with a smaller radius will have a deeper gravitational potential, which facilitates a deeper collapse of a subvirial star-forming region. 

This is seen in the evolution of the central density of our star-forming regions, shown in panels (a) and (b) of Fig.~\ref{fractal_dimensions}. These three regions all have the same local stellar density (which reflects the density in the substructure), but the high degree of clumpiness for the regions with a low fractal dimension ($D = 1.6$, panel (a)) results in a large amount of empty space.  Despite the common initial stellar density, the regions with no substructure (panel (b)) initially are able to collapse into a deeper potential well, and attain both local and central densities of $\tilde{\rho} = 1000$\,M$_\odot$\,pc$^{-3}$, i.e.\,\,a factor of ten higher than the initial density. 

This behaviour significantly affects the radiation fields. First, the higher the substructure, the lower the initial $G_0$ and EUV fields. This is because the photoionising stars are on average further away from the majority of stars than is the case for a uniform sphere (no substructure). Secondly, because the regions with no substructure can collapse to higher central densities (where the massive stars are located), the $G_0$ field reaches a maximum  of nearly $G_0 = 10^5$ after the region collapses to form a cluster (panel (d)). Contrast this with the region with a high degree of initial substructure ($D = 1.6$), where the $G_0$ field remains constant at $G_0 = 1000$ for the entirety of the simulation (panel(c)). This behaviour is also the same for the EUV radiation fields.

The impact of the higher $G_0$ and EUV fields in the less substructured simulations can be seen in the evolution of the disc fractions for discs with initial radii $r_{\rm disc} = 10$\,au (the trends are also similar for discs with larger radii, which we do not show here). The disc fractions in the simulations with $D = 1.6$ (substructured) drop to between 50 and 80\,per cent, whereas in the simulations with no initial substructure the disc fractions drop to between 10 and 60\,per cent.

Note that if these regions had similar initial \emph{volume averaged densities} (Eq.~\ref{half_mass_eq}), then the radii of the highly substructured ($D = 1.6$) simulations would be smaller, and in that case the more substructured regions would likely lead to more disc destruction than in the smoother regions. However, it is the \emph{local} density the more accurately traces the dynamical evolution of these star-forming regions \citep{Parker20}, despite this not being commonly adopted by observers or simulators to characterise the density of star-forming regions.

Interestingly, the simulations with no initial substructure, shown in panel (h), display a flattening of the disc fractions after $\sim$0.5\,Myr, before decreasing again. Here, the initial radiation field rapidly destroys discs in the first 0.5\,Myr, but then as the central densities of the star-forming regions increase further disc destruction occurs. This behaviour is not present in the substructured simulations because the density of (and hence FUV fields within)  the star-forming regions are highest at $t = 0$\,Myr. 

\subsubsection{Initial virial ratio}

\begin{figure*}
  \begin{center}
\setlength{\subfigcapskip}{10pt}
\vspace*{-0.5cm}
\hspace*{0.3cm}\subfigure[$\alpha_{\rm vir} = 0.5$]{\label{virial_ratios-b}\rotatebox{270}{\includegraphics[scale=0.23]{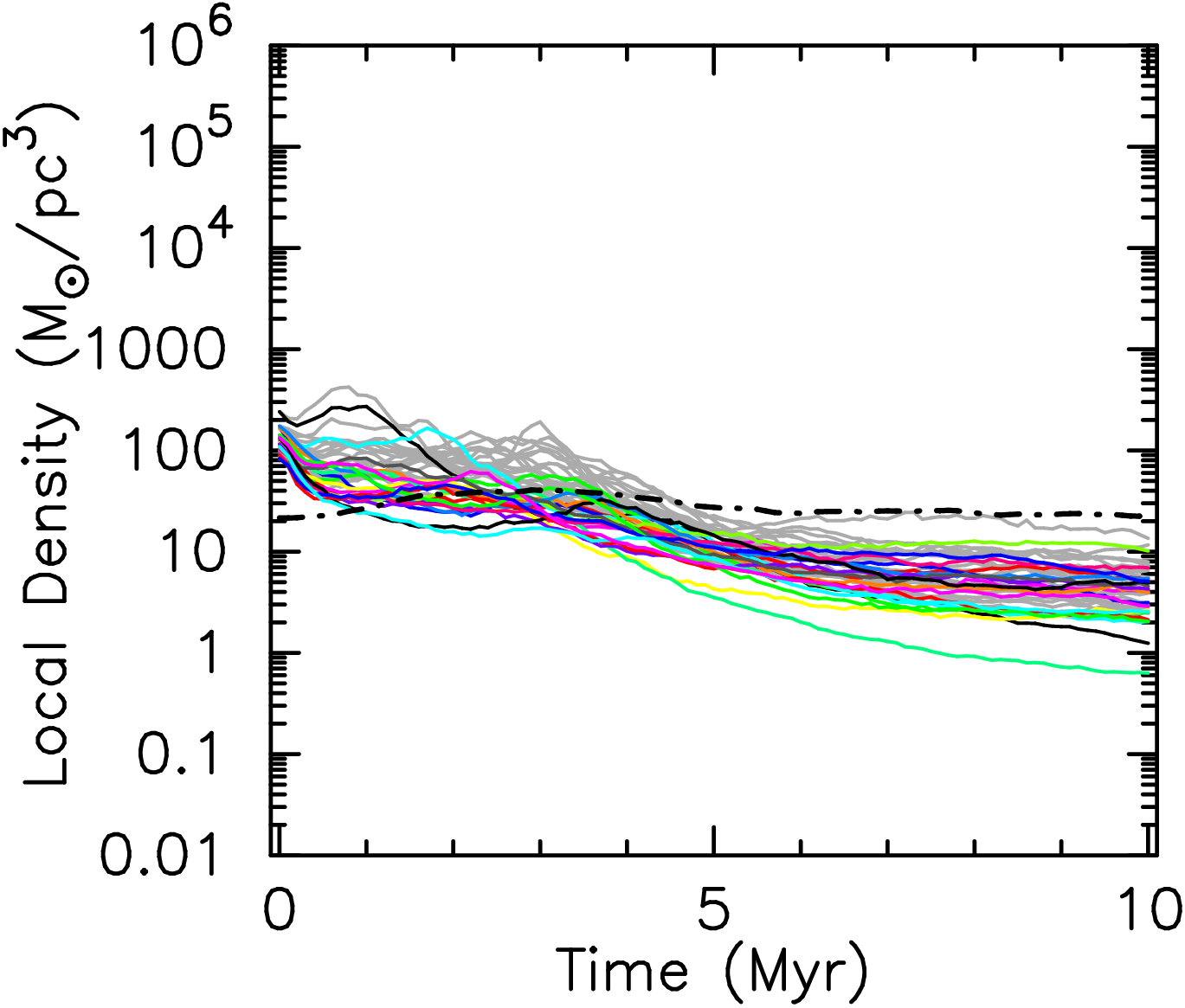}}}
\hspace*{0.5cm}\subfigure[$\alpha_{\rm vir} = 1.5$]{\label{virial_ratios-c}\rotatebox{270}{\includegraphics[scale=0.23]{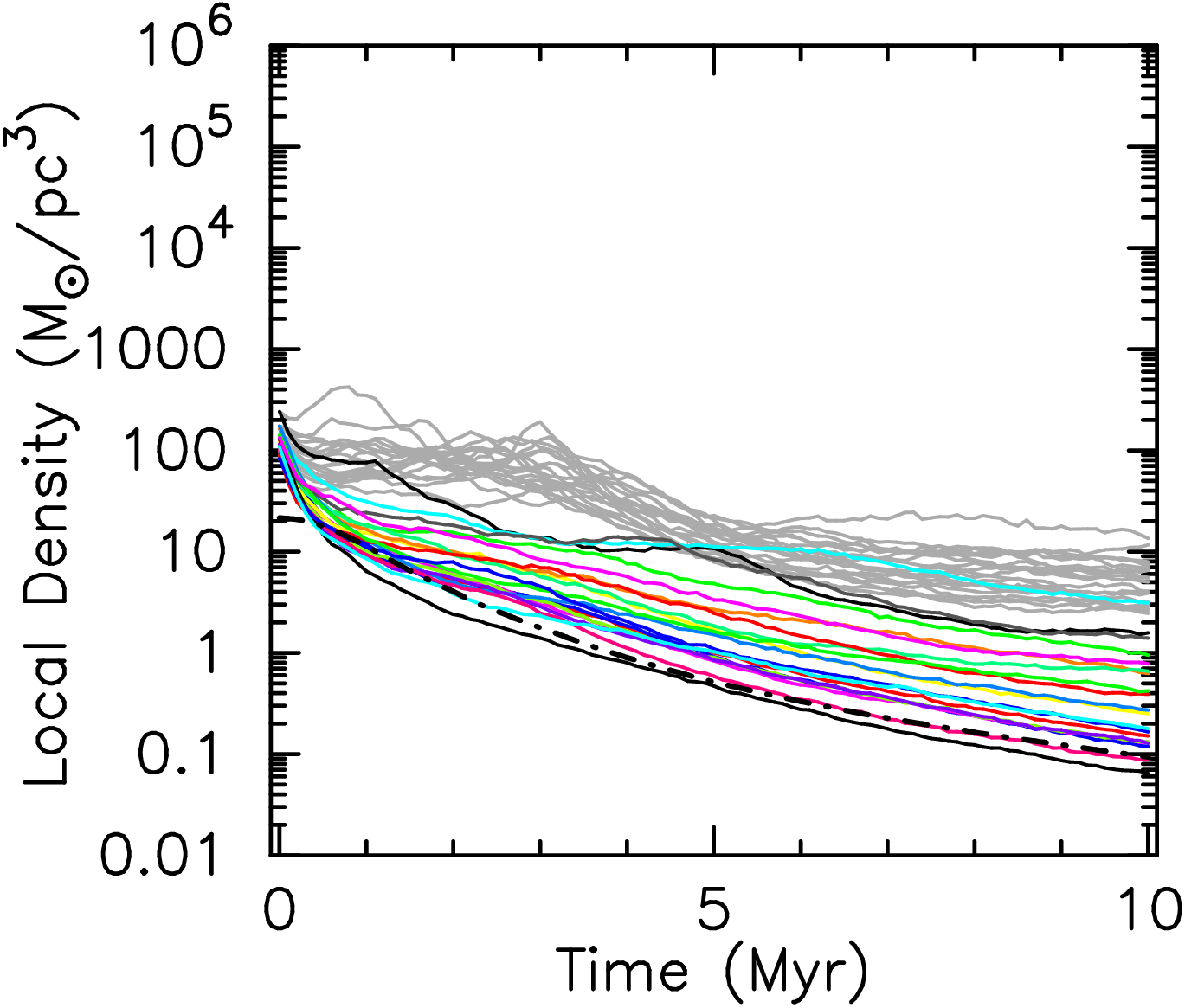}}} \newline
\hspace*{0.3cm}\subfigure[$\alpha_{\rm vir} = 0.5$]{\label{virial_ratios-e}\rotatebox{270}{\includegraphics[scale=0.23]{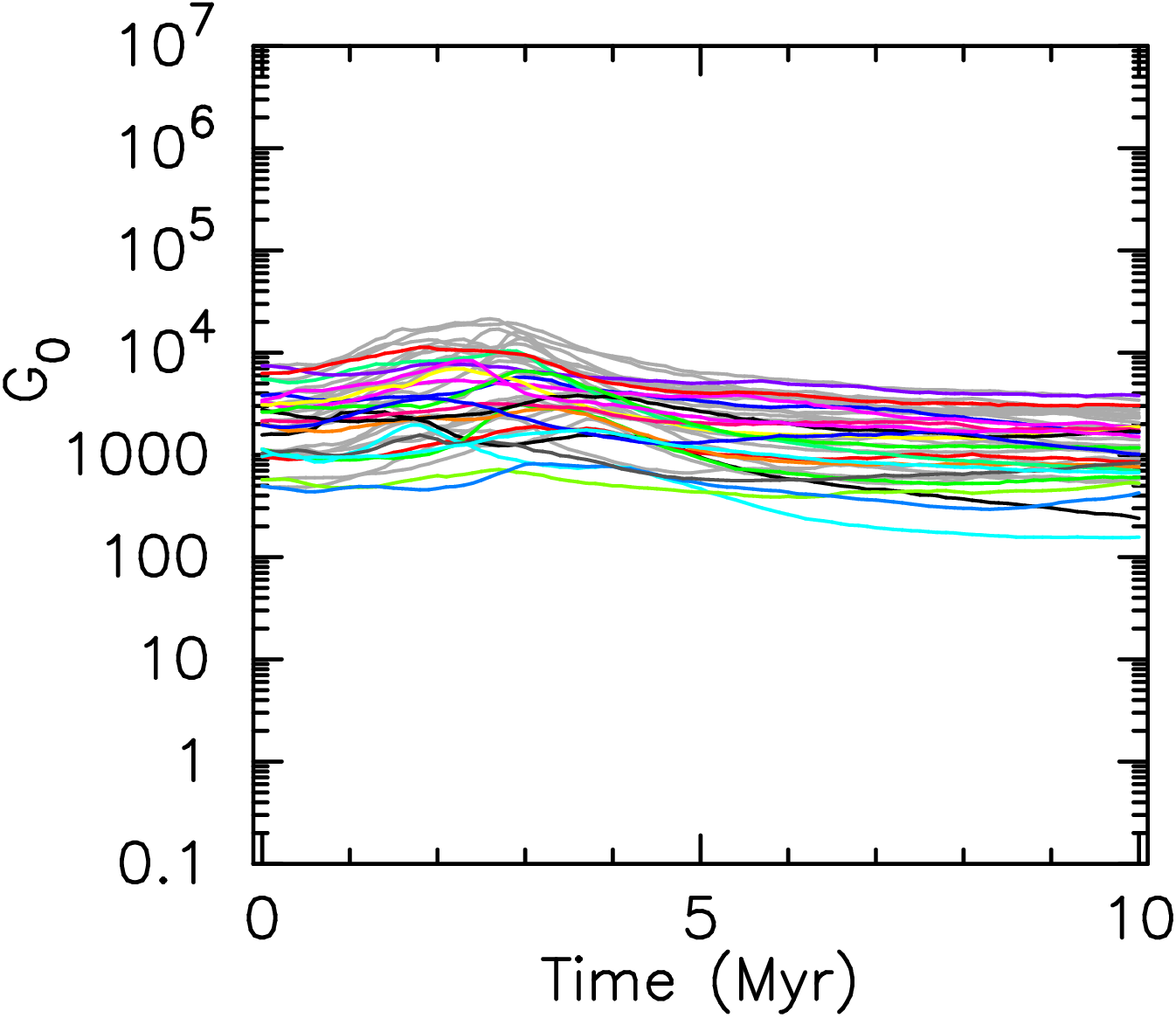}}} 
\hspace*{0.5cm}\subfigure[$\alpha_{\rm vir} = 1.5$]{\label{virial_ratios-f}\rotatebox{270}{\includegraphics[scale=0.23]{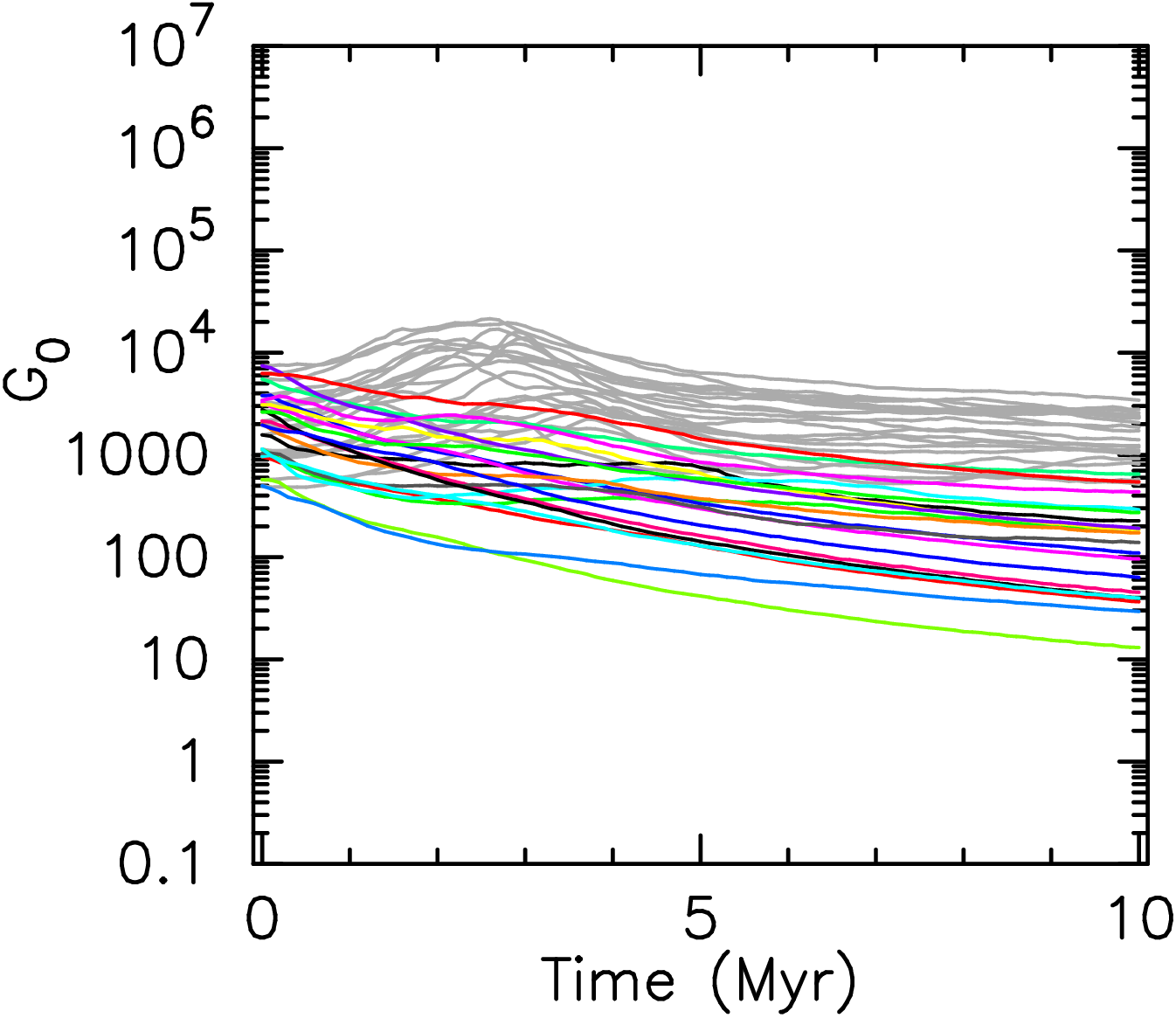}}} \newline
\hspace*{0.3cm}\subfigure[$\alpha_{\rm vir} = 0.5$]{\label{virial_ratios-h}\rotatebox{270}{\includegraphics[scale=0.23]{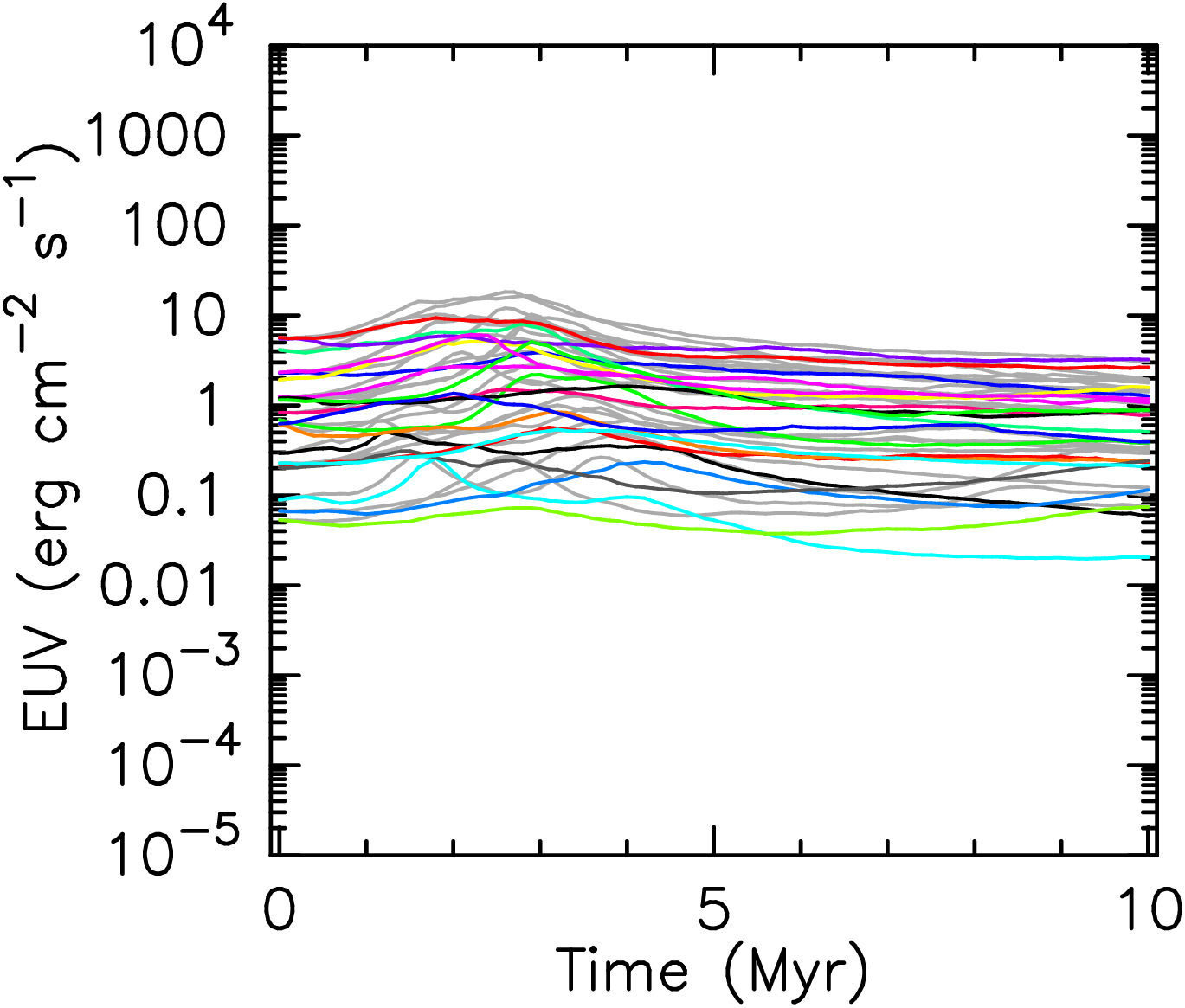}}} 
\hspace*{0.5cm}\subfigure[$\alpha_{\rm vir} = 1.5$]{\label{virial_ratios-i}\rotatebox{270}{\includegraphics[scale=0.23]{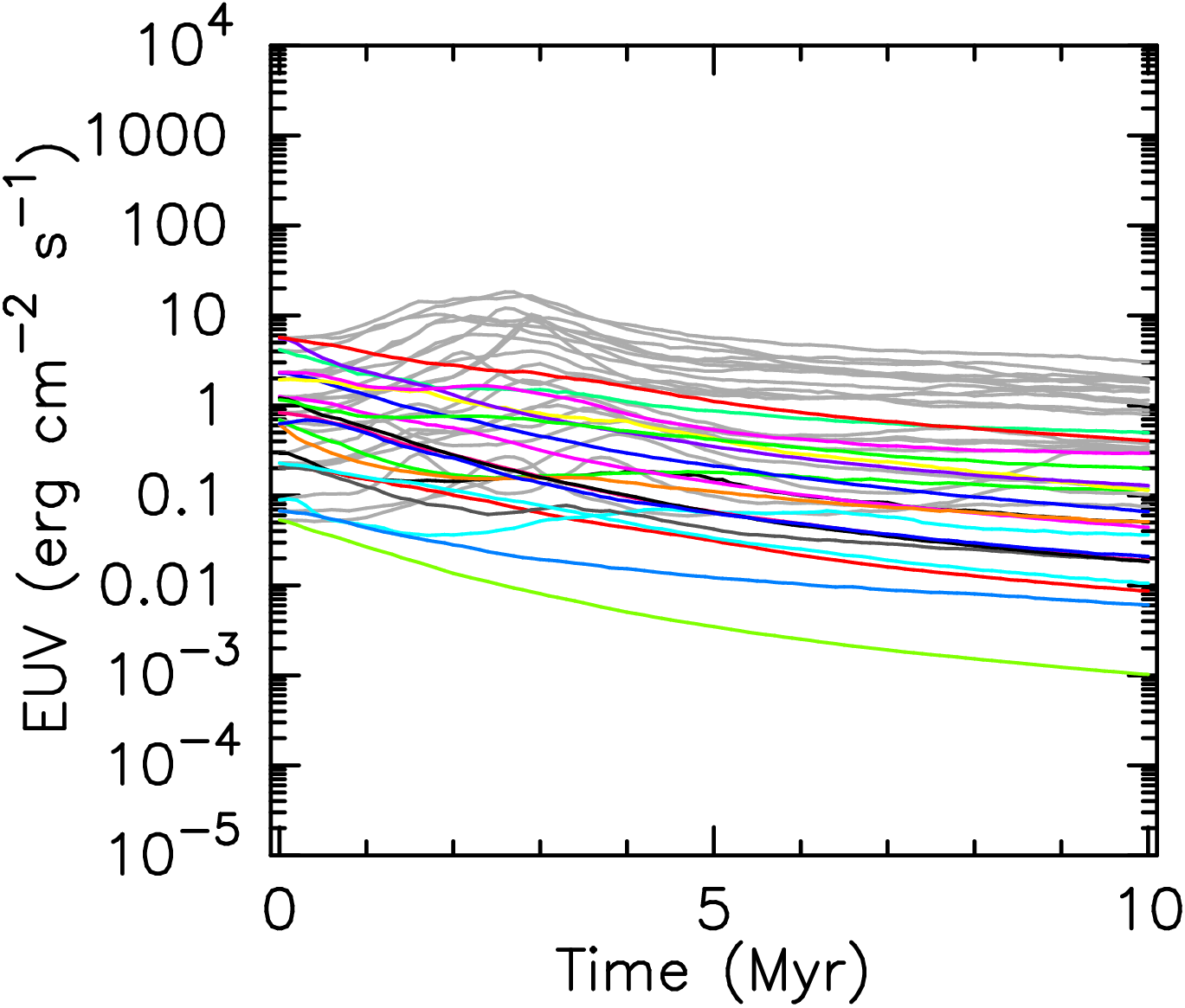}}} \newline
\hspace*{0.5cm}\subfigure[$\alpha_{\rm vir} = 0.5$]{\label{virial_ratios-k}\rotatebox{270}{\includegraphics[scale=0.23]{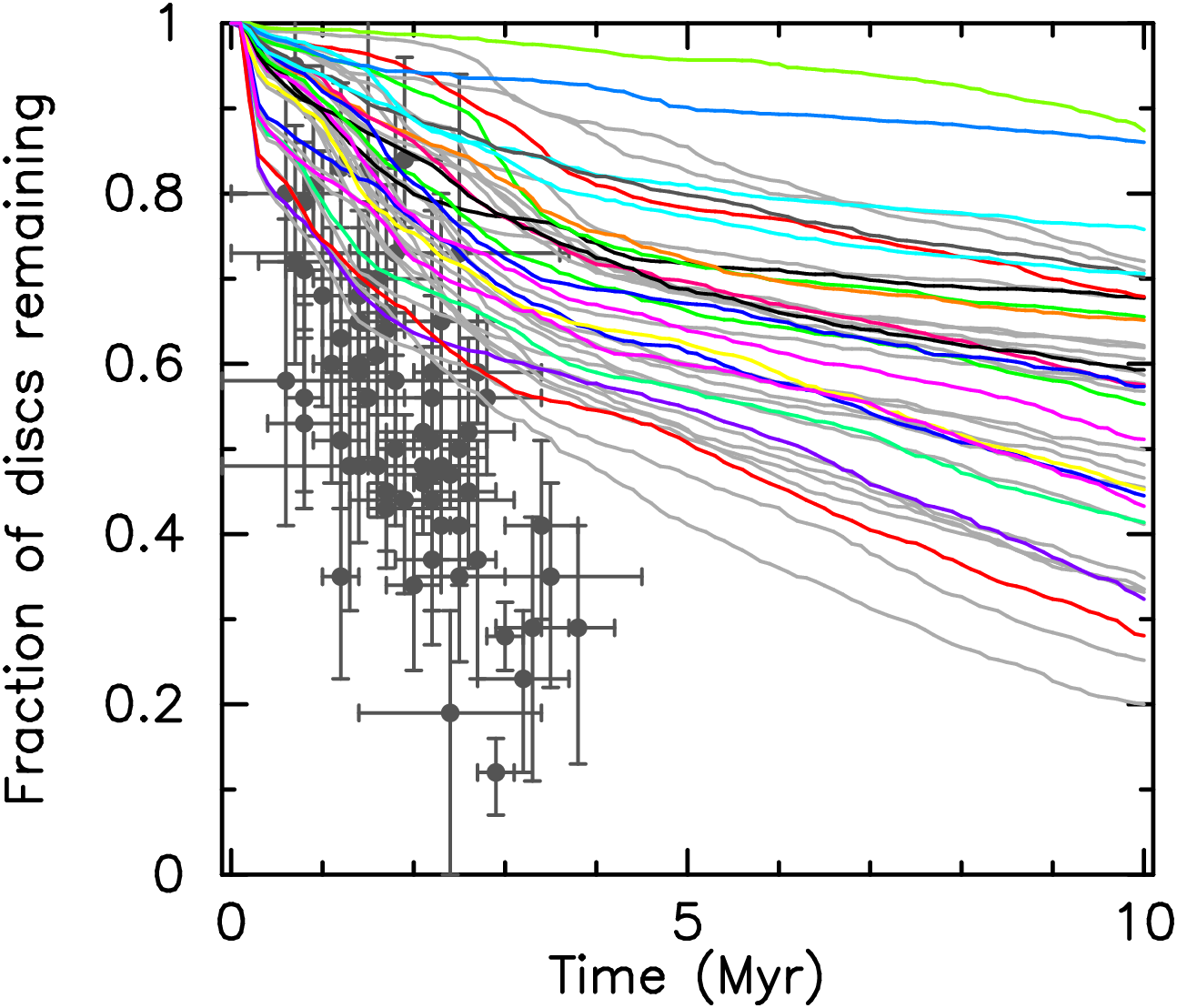}}}
\hspace*{0.9cm}\subfigure[$\alpha_{\rm vir} = 1.5$]{\label{virial_ratios-l}\rotatebox{270}{\includegraphics[scale=0.23]{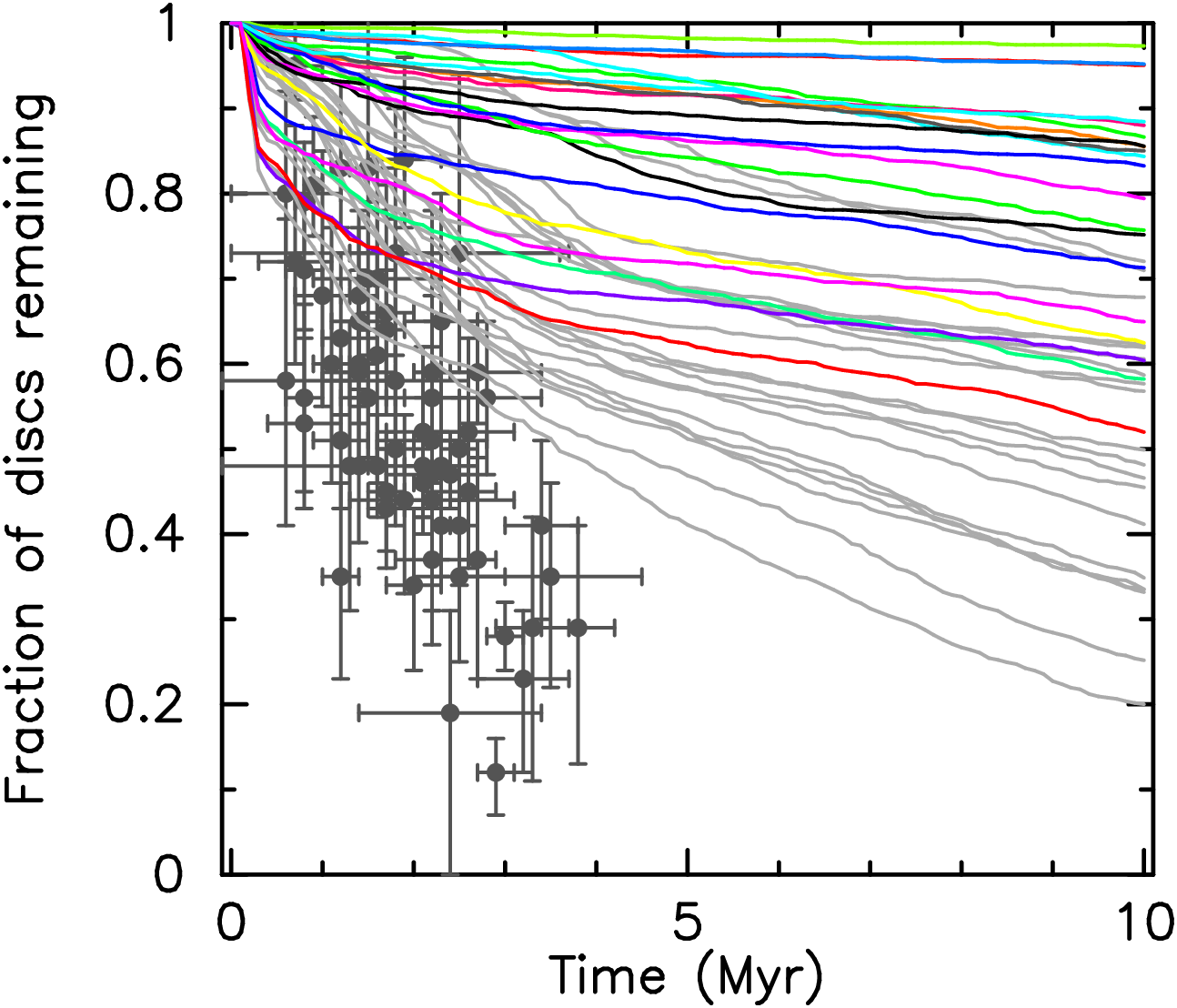}}} \newline

\caption[bf]{The effect of varying the initial virial ratios when the initial stellar density is kept constant ($\tilde{\rho} \sim 100$\,M$_\odot$\,pc$^{-3}$). We also use the same initial fractal dimension ($D = 2.0$).  The top row shows the median local density in twenty realisations of the same star-forming region (indicated by the different coloured lines), as well as the mean density within the half-mass radius in all twenty simulations  (the dot-dashed line). The second row shows the median FUV flux, $G_0$, in each simulation and the third row shows the median EUV flux. The fourth row shows the fraction of stars that host gaseous discs in each simulation, where the initial disc radius was 10\,au. The lefthand column shows simulations where the initial virial ratio is $\alpha_{\rm vir} = 0.5$ (virial equilibrium). The righthand column shows simulations where the initial virial ratio is $\alpha_{\rm vir} = 1.5$ (supervirial). The default simulations, which are subvirial  ($\alpha_{\rm vir} = 0.3$), are shown by the background grey lines. The observed disc fractions in star-forming regions from \citet{Richert18} are shown by the \newline dark grey points. }
\label{virial_ratios}
  \end{center}
\end{figure*}

We vary the initial virial ratio of the star-forming regions to determine the effect of the bulk motion on the survival of protoplanetary discs. Many star-forming regions are observed to be subvirial, which means they may collapse to form a bound cluster, but observations indicate that most star-forming regions have dispersed after 10\,Myr. It is unclear what the main mechanism for dispersal is, but many authors have investigated the hypothesis that regions disperse with supervirial velocities, following the rapid removal of the gas potential leftover from the star formation process \citep{Tutukov78,Lada84,Goodwin97,Goodwin06,Baumgardt07,Pfalzner13c,Shukirgaliyev18}. 

We mimic this process by setting our initial velocities to be supervirial initially, as well as running a set of simulations where the star-forming region is in global virial equilibrium. In Fig.~\ref{virial_ratios}, panel (a) we see that the density evolution of virialised  star-forming regions (the coloured lines) is very similar to the subvirial regions (our default simulations, shown by the grey lines), apart from the subvirial regions attain higher central densities due to the more violent nature of the collapse. As we would expect, the supervirial regions (panel b) expand rapidly to low densities.

Interestingly, the different virial ratios lead to little variation in the radiation fields. The subvirial and virial regions tend to have slightly higher $G_0$ and EUV fields after 1\,Myr (panel (c) and panel (e)), with quite similar fractions of surviving discs (panel (g) of Fig.~\ref{virial_ratios}). The supervirial regions have high $G_0$ values to begin with, and this largely governs the disc fractions over time, as most of the mass-loss due to photoevaporation occurs in the first 0.5\,Myr. However, the $G_0$ and EUV fields are only a factor of ten lower than in the (sub)virial regions after 10\,Myr, despite the local density in the supervirial star-forming regions being a factor of 100 lower than at birth. The reason for this is that supervirial star-forming regions dynamically evolve so that the most massive stars sweep up retinues of low-mass stars \citep{Parker14b,Rate20}, meaning that the most massive stars will almost exclusively reside in the denser locations of the star-forming regions, where there are lots of low-mass stars that will experience strong radiation fields.  However, there are also many low-mass stars that do not reside near to massive stars and so the fraction of discs that survive in supervirial regions can be 25\,per cent higher than in the (sub)virial star-forming region (compare the coloured lines with the grey lines in panel (h)). 

\subsubsection{IMF sampling and low $N$}

\begin{figure*}
  \begin{center}
\setlength{\subfigcapskip}{10pt}
\hspace*{0.3cm}\subfigure[Constant mass function, $N = 1500$]{\label{mass_distributions-a}\rotatebox{270}{\includegraphics[scale=0.23]{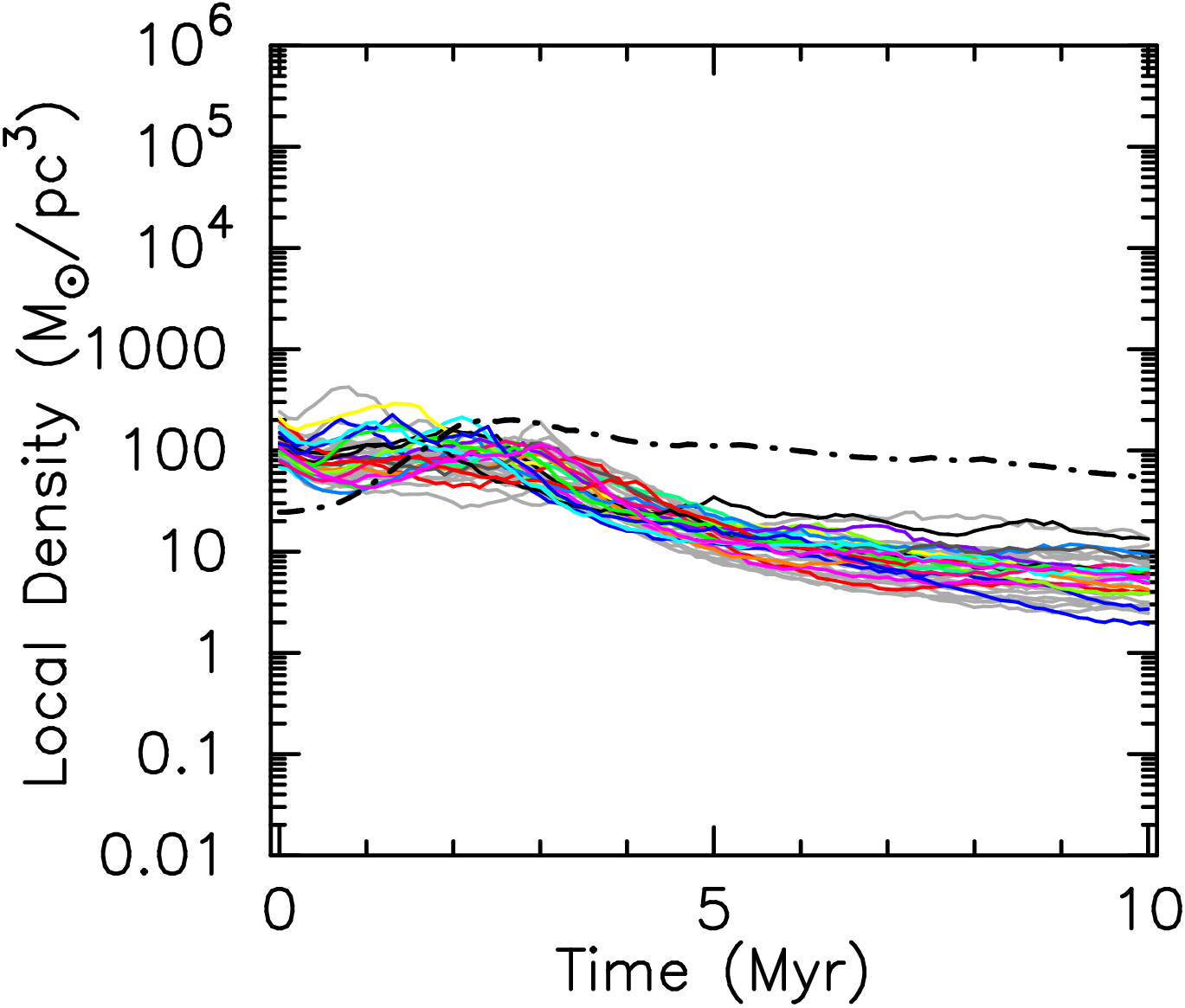}}}
\hspace*{0.5cm}\subfigure[Stochastic IMF sampling, $N = 150$]{\label{mass_distributions-b}\rotatebox{270}{\includegraphics[scale=0.23]{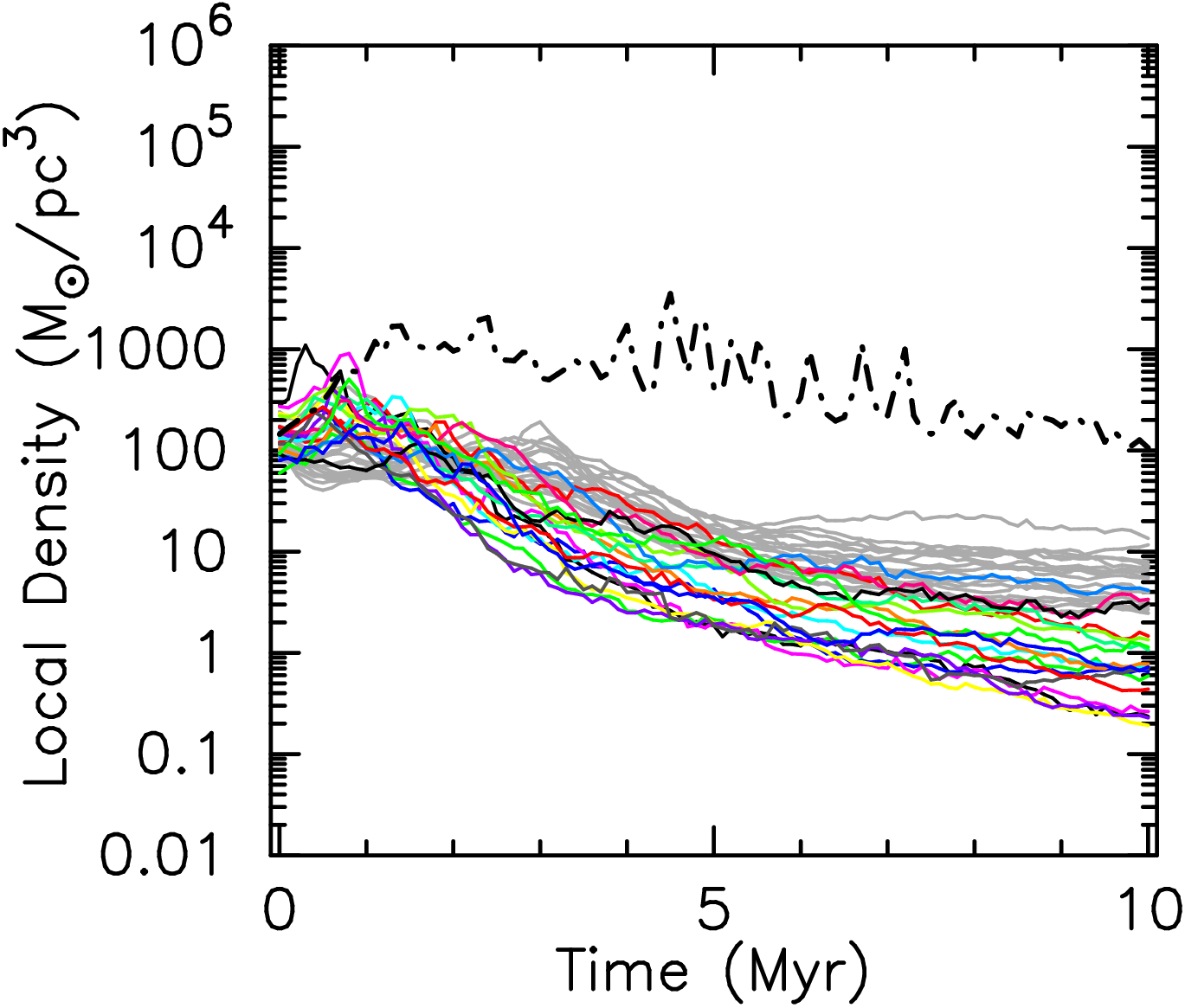}}} \newline
\hspace*{0.3cm}\subfigure[Constant mass function, $N = 1500$]{\label{mass_distributions-c}\rotatebox{270}{\includegraphics[scale=0.23]{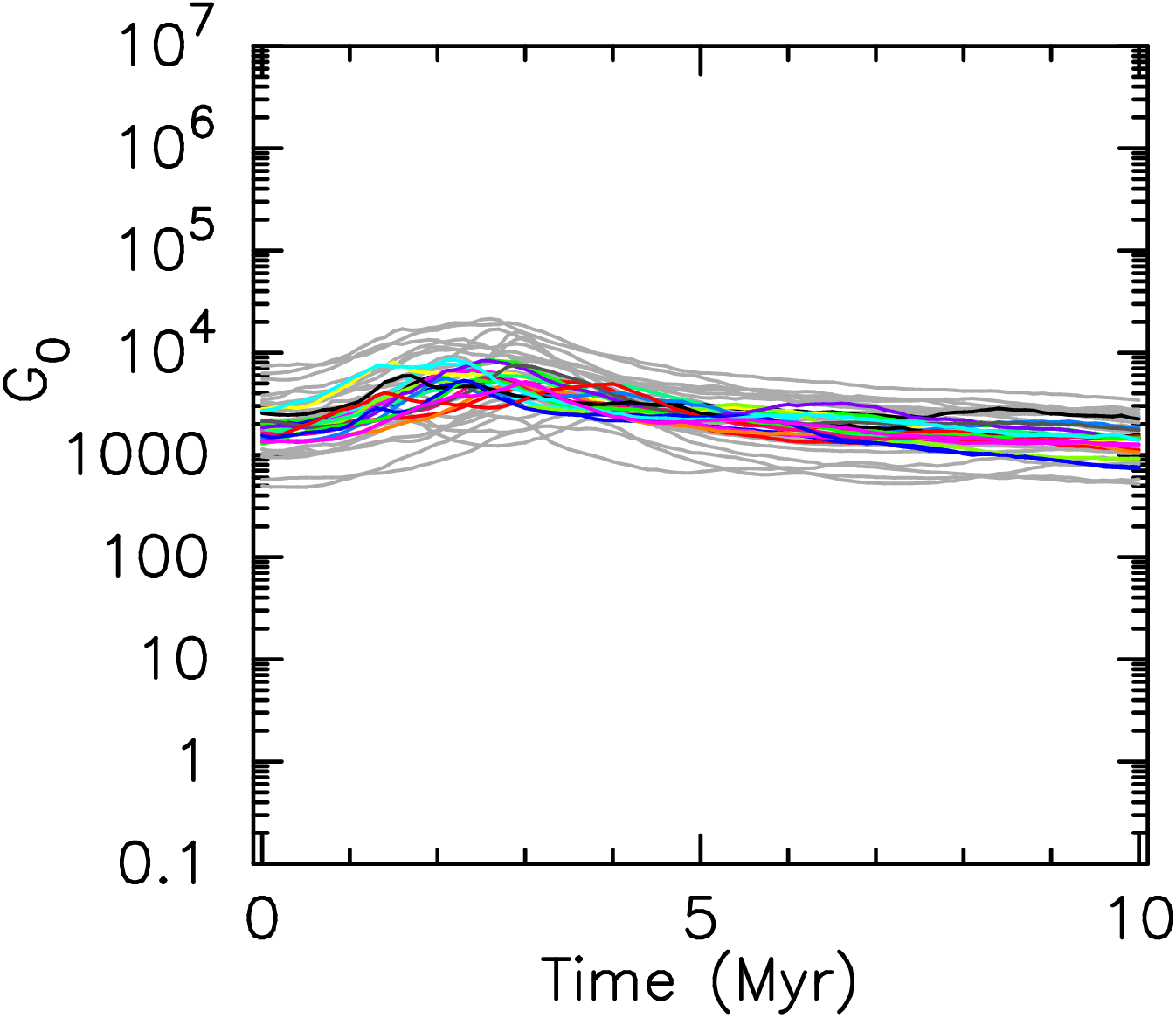}}}
\hspace*{0.5cm}\subfigure[Stochastic IMF sampling, $N = 150$]{\label{mass_distributions-d}\rotatebox{270}{\includegraphics[scale=0.23]{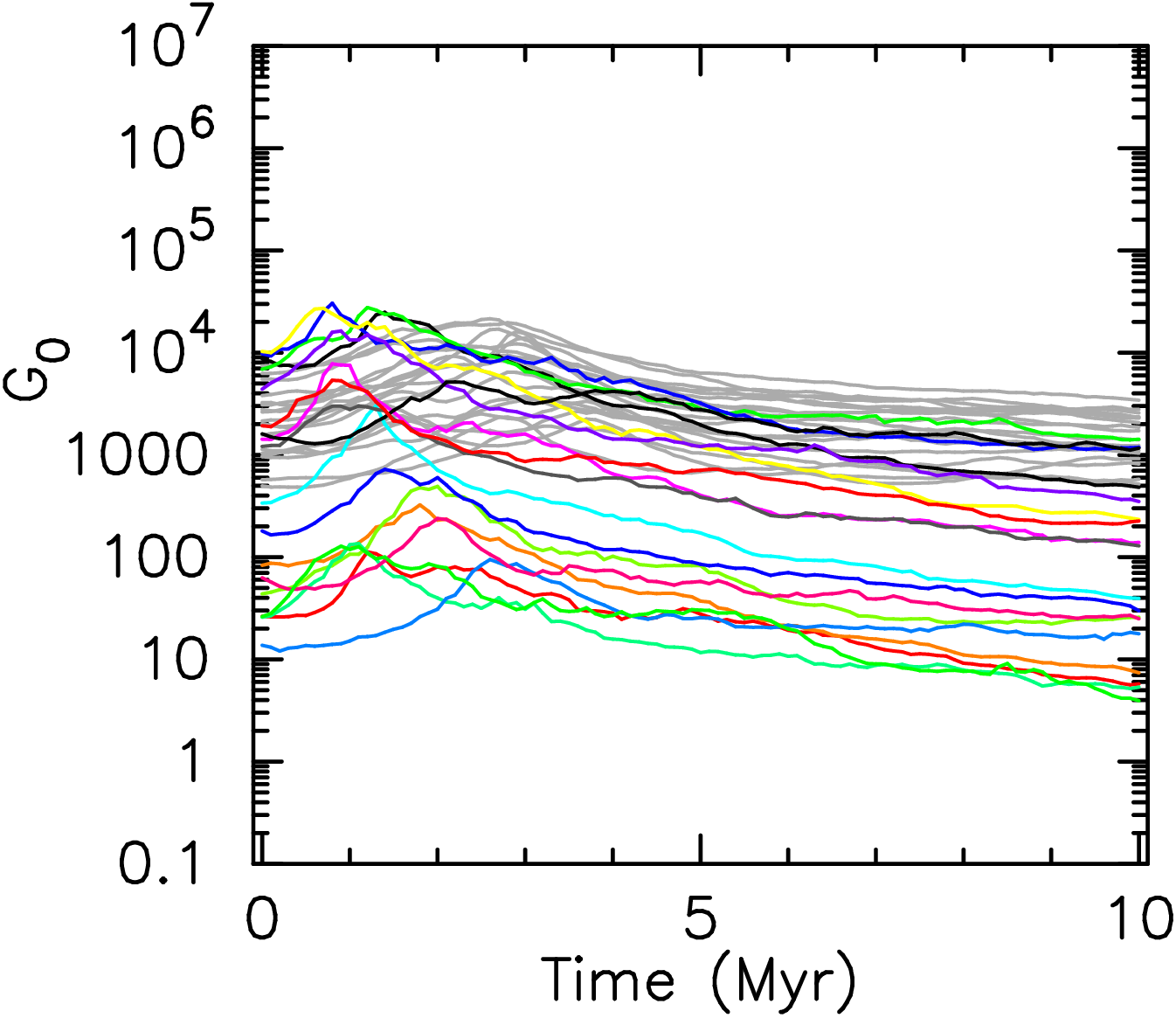}}} \newline
\hspace*{0.3cm}\subfigure[Constant mass function, $N = 1500$]{\label{mass_distributions-e}\rotatebox{270}{\includegraphics[scale=0.23]{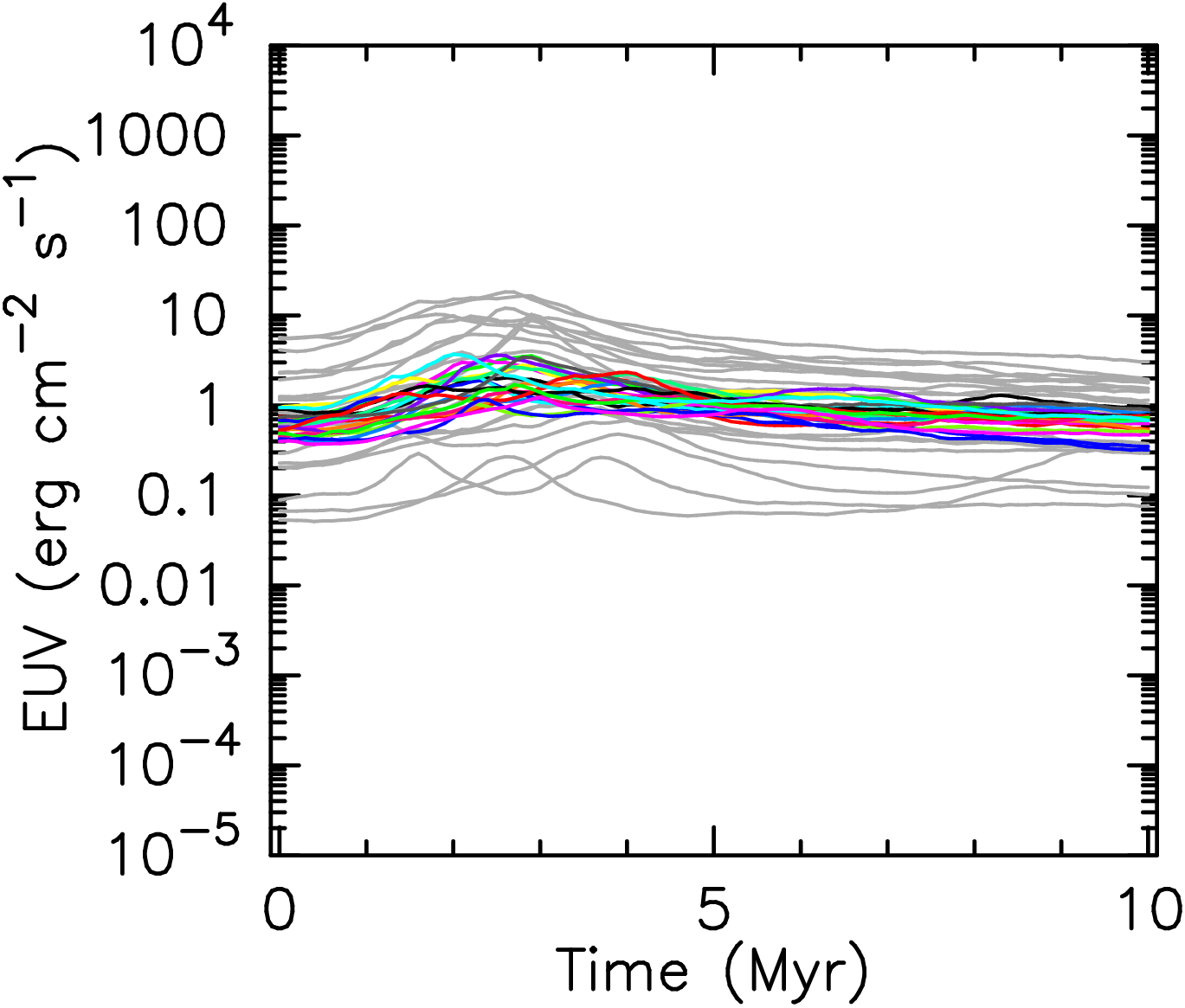}}}
\hspace*{0.5cm}\subfigure[Stochastic IMF sampling, $N = 150$]{\label{mass_distributions-f}\rotatebox{270}{\includegraphics[scale=0.23]{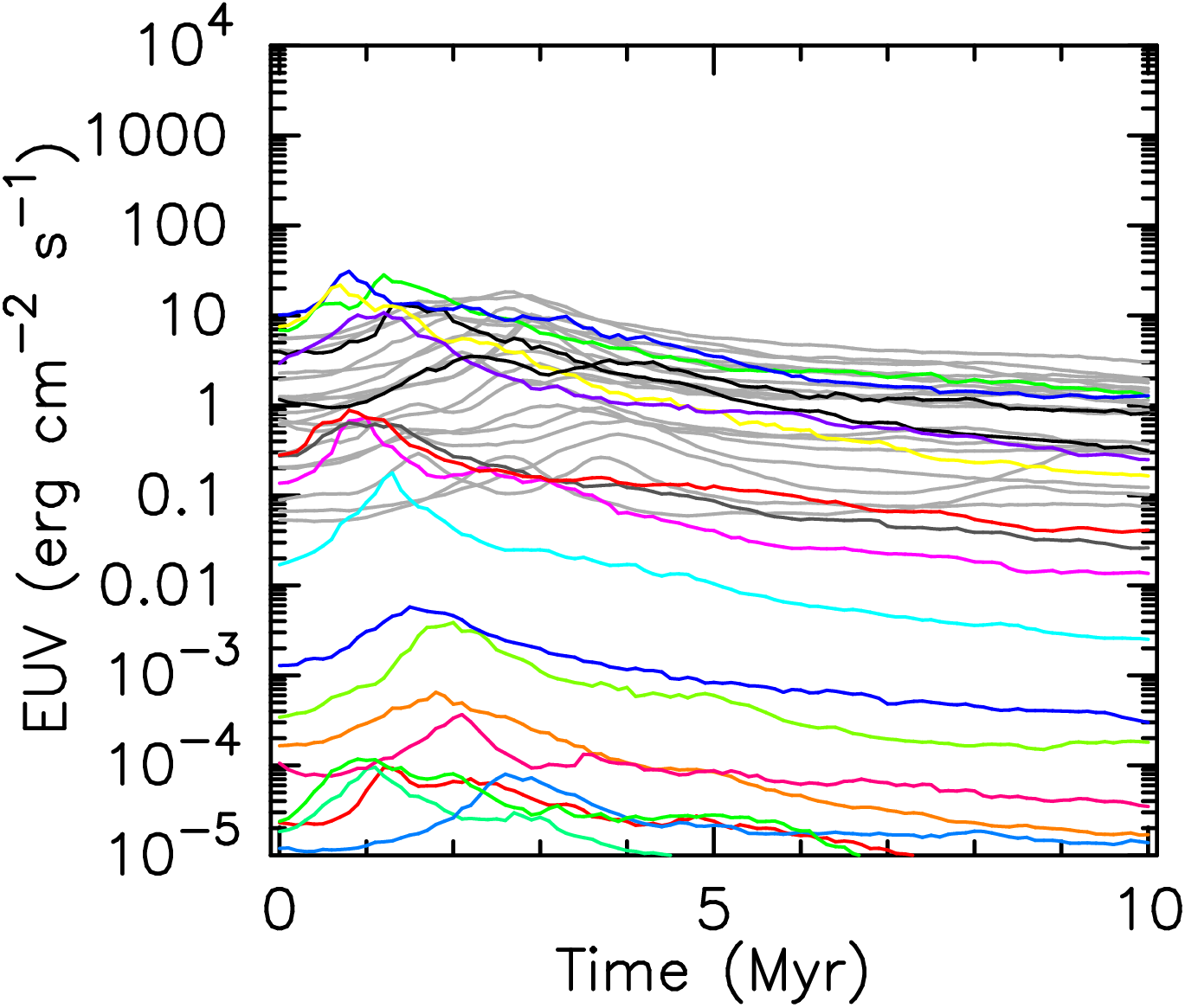}}} \newline
\hspace*{-1.9cm}\subfigure[Constant mass function, $N = 1500$]{\label{mass_distributions-g}\rotatebox{270}{\includegraphics[scale=0.23]{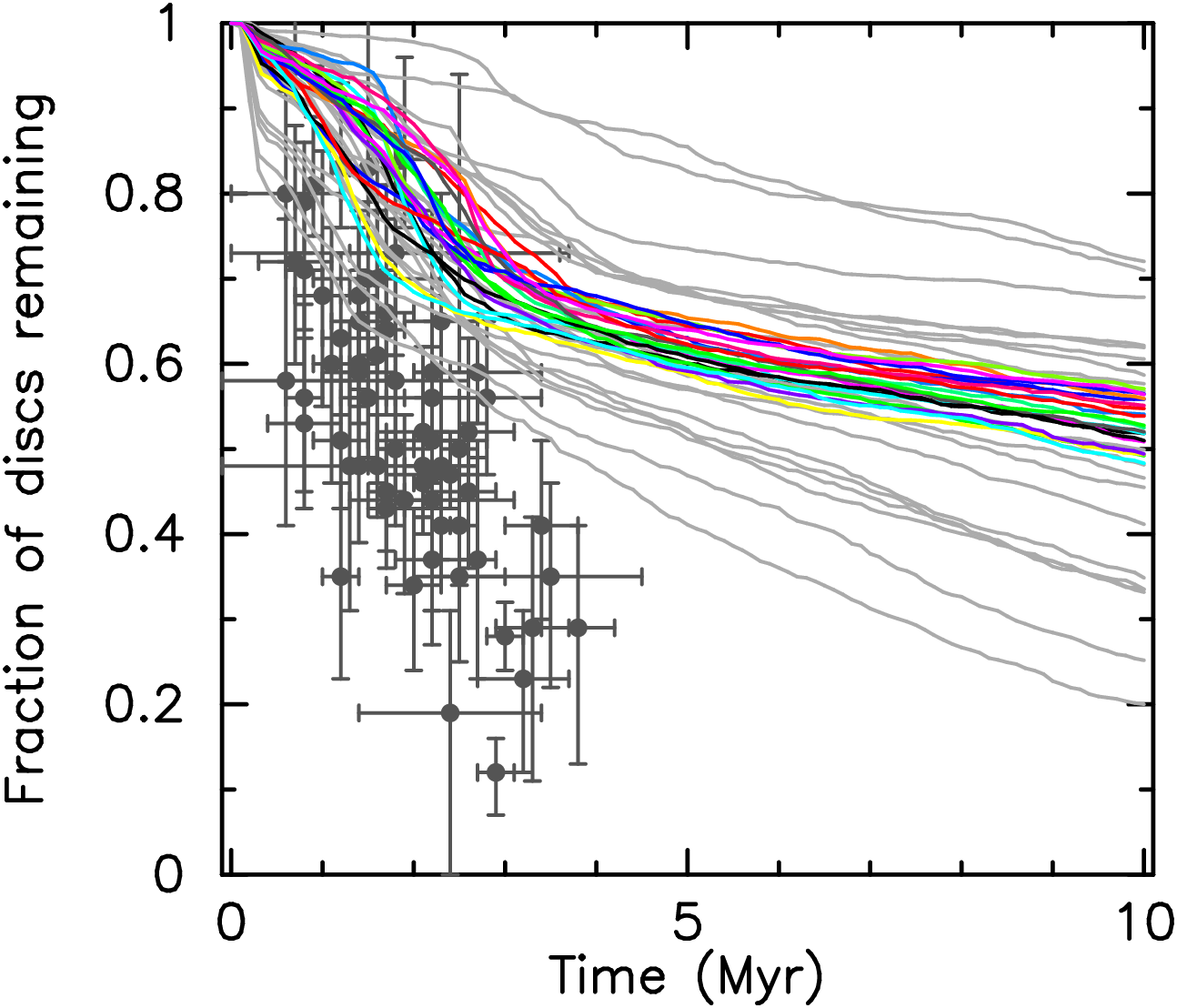}}}
\hspace*{0.9cm}\subfigure[Stochastic IMF sampling, $N = 150$]{\label{mass_distributions-h}\rotatebox{270}{\includegraphics[scale=0.23]{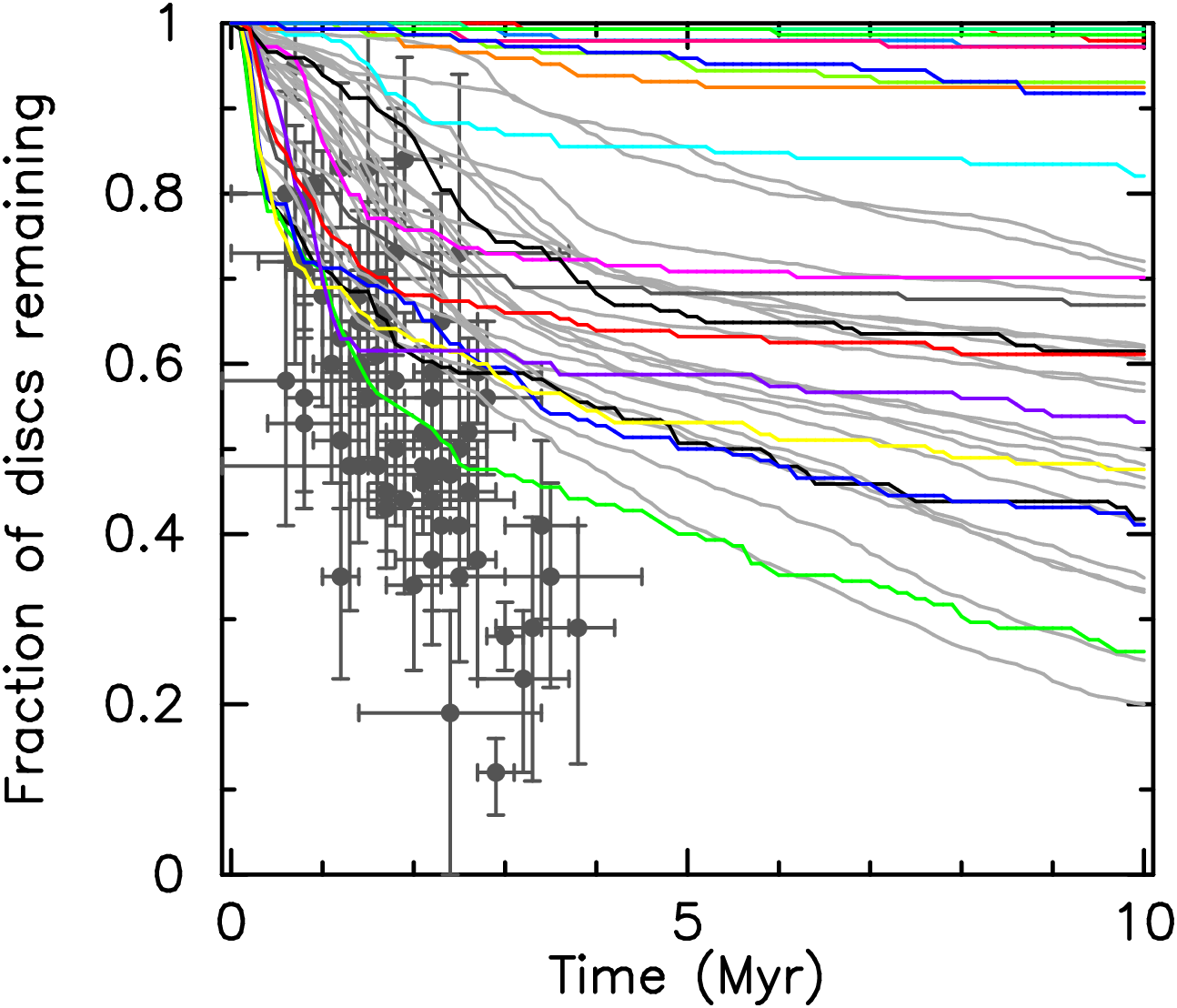}}}

\caption[bf]{Comparison of different mass distributions with constant initial stellar density ($\tilde{\rho} \sim 100$\,M$_\odot$\,pc$^{-3}$). The top row shows the median local density in twenty realisations of the same star-forming region (indicated by the different coloured lines), as well as the mean density within the half-mass radius in all twenty simulations  (the dot-dashed line). The second row shows the median FUV flux, $G_0$, in each simulation and the third row shows the median EUV flux. The fourth row shows the fraction of stars that host gaseous discs in each simulation, where the initial disc radius was 10\,au. The lefthand column shows the results for simulations that are identical to our default simulation ($\alpha_{\rm vir} = 0.3$, $D = 2.0$, $N = 1500$, $r_F = 2.5$\,pc), but where the mass distribution of stars is identical in each simulation (whereas \hspace*{4cm} the positions and velocities of the stars are randomly different). The righthand panel shows simulations with our default density, virial ratio and degree of substructure, but now the total number of stars is only $N = 150$. The grey lines indicate the results for our default simulations. The observed disc fractions in star-forming regions from \citet{Richert18} are shown by the dark grey points. }
\label{mass_distributions}
  \end{center}
\end{figure*}

We now take our default simulation ($\tilde{\rho} \sim 100$\,M$_\odot$\,pc$^{-3}$, $D = 2.0$, $\alpha_{\rm vir}$ = 0.3) and remove the variation of the stellar IMF between different realisations of the same simulation. We adopt a single IMF, and the only parameters that vary randomly are the positions and velocities of the individual stars. We show the results of this in the lefthand column of Fig.~\ref{mass_distributions}. The grey lines show the values from the default simulation, where the numbers of massive stars are allowed to vary between simulations with the statistically same initial conditions. 

Whilst the evolution of the local density varies between the different simulations, the spread in $G_0$ and EUV fields is noticeably narrower than in the simulations where the IMF varies between realisations. This also leads to a narrower range of disc fractions,  suggesting it is not the individual dynamics of statistically similar star-forming regions that dominates disc photoevaporation, but rather the mass distribution of stars.

We further demonstrate this point by examining the photoevaporation in star-forming regions with only $N = 150$ stars (but again, with initial densities of $\tilde{\rho} \sim 100$\,M$_\odot$\,pc$^{-3}$). These regions expand faster than the higher-mass regions, but it is striking that there is a significant spread in the $G_0$ and EUV fields. This is because the IMF is not fully sampled; in some instances the region contains several stars above 5\,M$_\odot$ that produce radiation fields, but in two of our simulations no such stars are produced. Of the regions that do contain photoionising stars, the median $G_0$ fields range from $G_0 = 10$ to $G_0 = 10^4$. This translates into a huge range in the fractions of surviving discs, from 100\,per cent in the simulations with no intermediate or massive stars, to 83\,per cent in a simulation that only contains a 6\,M$_\odot$ star, to as low as 25\,per cent for a region containing stars with masses 14, 19 and 44\,M$_\odot$. 


\subsection{Viscous evolution in discs}

\begin{figure*}
  \begin{center}
\setlength{\subfigcapskip}{10pt}
\hspace*{-1.0cm}\subfigure[Radius fixed to initial value]{\label{disc_evolution-a}\rotatebox{270}{\includegraphics[scale=0.26]{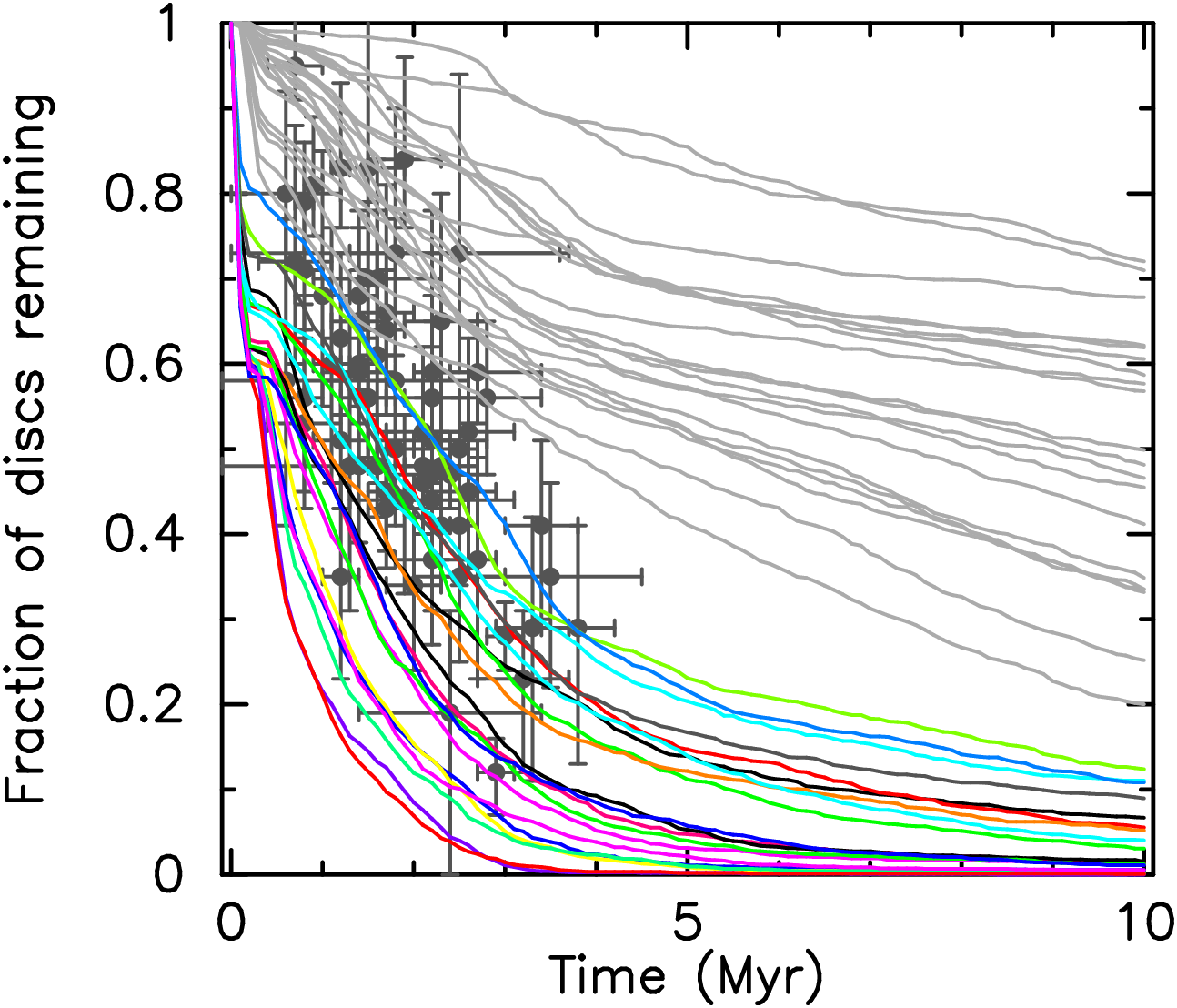}}}
\hspace*{0.3cm}\subfigure[Evolving disc radius, viscous evolution]{\label{disc_evolution-b}\rotatebox{270}{\includegraphics[scale=0.26]{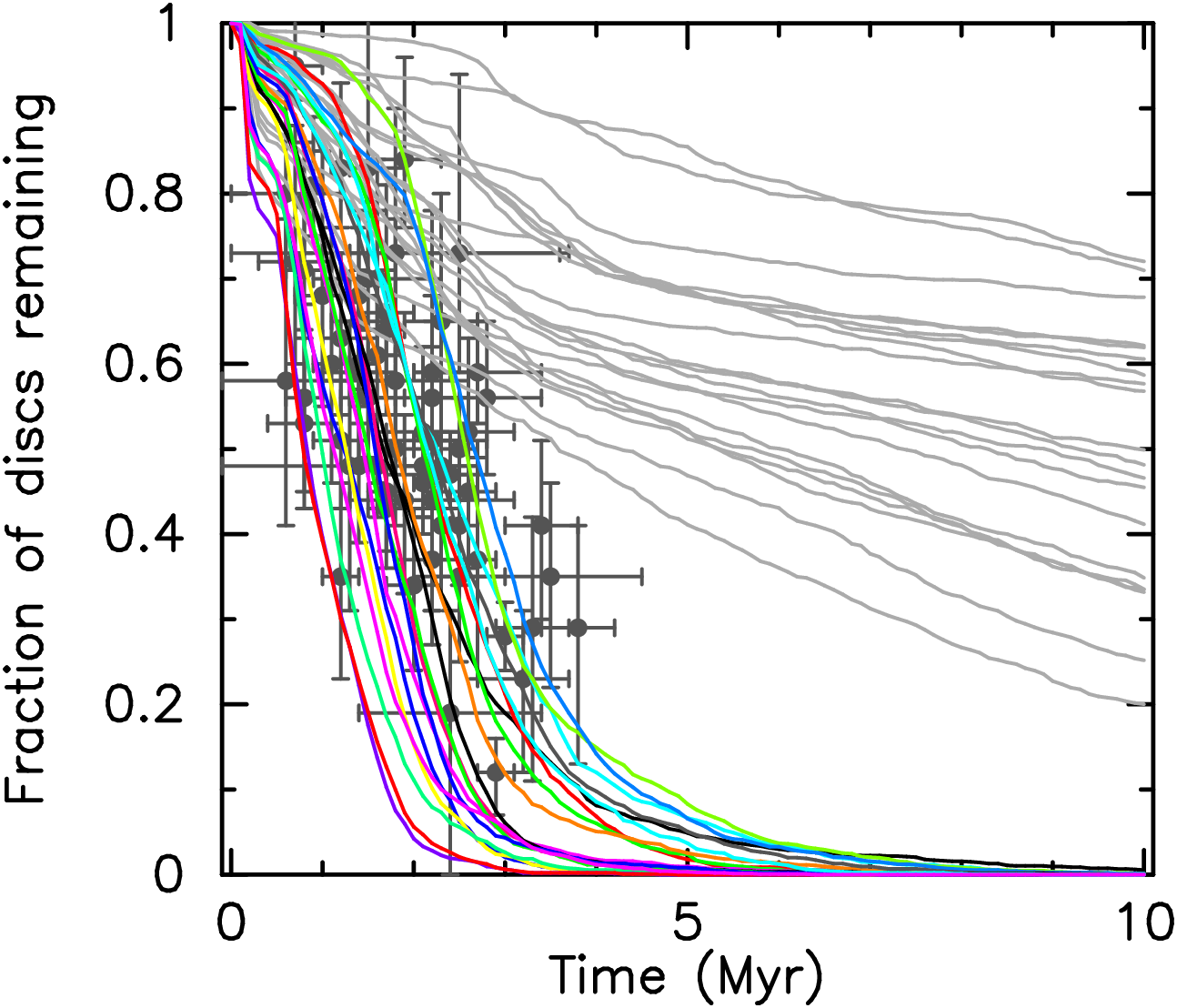}}}
\hspace*{0.3cm}\subfigure[Evolving disc radius, viscous evolution, accretion onto central star]{\label{disc_evolution-c}\rotatebox{270}{\includegraphics[scale=0.26]{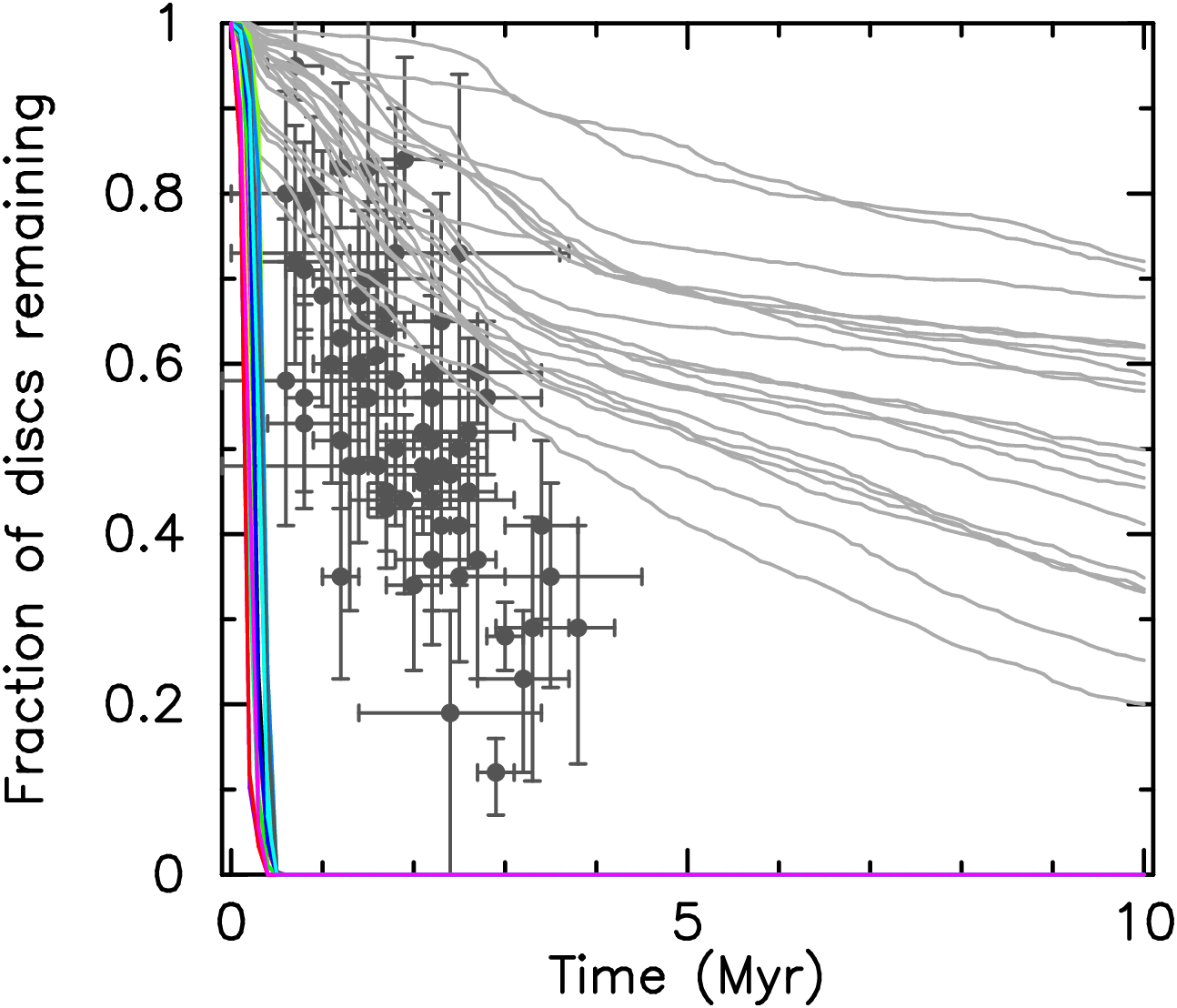}}}
\caption[bf]{The evolution of discs in our default simulation ($\tilde{\rho} \sim 100$\,M$_\odot$\,pc$^{-3}$, $D = 2.0$, $\alpha_{\rm vir}$ = 0.3) where the initial disc radii are all 10\,au. In panel (a) we show the fraction of discs over time following mass-loss due to photoevaporation, but the radius is kept constant. The grey lines indicate the disc fractions when the disc radius is allowed to decrease. In panel (b) the disc radii are allowed to decrease during photoevaporation (by keeping the surface density of the disc at 1\,au constant), and then increase again due to viscous spreading. In panel (c) we also allow mass-loss in the inner edge of the disc due to accretion on to the central star. The grey lines indicate the results for our default simulations. The observed disc fractions in star-forming regions from \citet{Richert18} are shown by the dark grey points. }
\label{disc_evolution}
  \end{center}
\end{figure*}

Until now the only internal disc evolution we have included is the disc's response to losing mass through photoevaporation. We have fixed the surface density of the disc to be constant at 1\,au from the star, so that when the disc loses mass the radius must decrease. In Fig.~\ref{disc_evolution-a} we show the evolution of the disc fraction in our default simulation ($\tilde{\rho} \sim 100$\,M$_\odot$\,pc$^{-3}$, $D = 2.0$, $\alpha_{\rm vir}$ = 0.3) but we do not allow the outer radius to evolve inwards, Instead, the outer radius is set to be constant, and the result is a much more rapid destruction of the discs (compare the coloured lines with the grey lines, which are discs with the same initial radius, but whose radii evolve inwards in the same simulations (Fig.~\ref{G0_densities-n})).

We then allow the radius to evolve inwards due to the disc preferentially losing material from its edges during photoevaporation (as in our default calculations), but this time we implement viscous spreading in the disc, so that the outer radius can increase. This viscous spreading causes the fraction of discs to drop significantly, because the discs have lower mass (due to the photoevaporative mass-loss), and an increased radius and reduced mass makes the disc even more susceptible to further mass loss. This is shown in Fig.~\ref{disc_evolution-b}, where the disc fractions decrease to zero after $\sim$6\,Myr. At first glance, the combination of photoevaporation and viscous evolution appears to reproduce the observed disc fractions extremely well. However, if we also allow mass-loss due to accretion onto the central star from the inner edge of the disc, then the discs are depleted on much faster timescales ($\sim$0.5\,Myr, see Fig.~\ref{disc_evolution-c}) because their further mass loss makes them more susceptible to subsequent future photoevaporation.

We emphasise that our analytic method to model the viscous evolution of the discs may not capture all of the physics in a full simulation \citep{Krumholz15}, but in this regard it is no worse than our prescription for the mass-loss due to photoevaporation, which is essentially an interpolation of more complex simulation results \citep{Haworth18b}. The point is that viscous spreading would increase the radius of the disc, thereby lowering the surface density and making the disc more susceptible to mass-loss due to photoevaporation.

\section{Discussion}

As one would expect, in our simulations the radiation field is lower for star-forming regions with similar stellar populations but with lower densities. Our simulation parameter space encompasses very dense  ($\tilde{\rho} = 1000$\,M$_\odot$\,pc$^{-3}$) regions, all the way down to regions with densities similar to the Galactic field \citep[$\sim 0.2$\,M$_\odot$\,pc$^{-3}$][]{Korchagin03}. Due to the presence of massive stars, the FUV flux in our simulations exceeds the FUV flux in the interstellar medium, $G_0$ \citep{Habing68}, by a factor of 10 -- 100. This means that in any star-forming region with stars $>$10\,M$_\odot$, the environment experienced by a protoplanetary disc is much more destructive than in a star-forming region without massive stars \citep[e.g. Taurus;][]{Luhman03,Luhman04,Guedel07}. 

\citet{Fatuzzo08} calculated the FUV flux for star-forming regions in the Solar neighbourhood, and due to the dearth of massive stars in these nearby regions, obtained fluxes that are considerably lower than those we find in our simulations.  Our calculated FUV and EUV fluxes are probably more appropriately compared to the Orion Nebula Cluster, which contains several stars above 20\,M$_\odot$. This then poses the interesting question of what star-forming region did the Sun (and other extrasolar planet host stars) form in -- if massive stars were present in these regions during planet formation, then the EUV and FUV fluxes will have been much higher than in the nearby star-forming regions we observe today. 

Interestingly, even a low-mass star-forming region (i.e.\,\,$N = 150$ stars) will still have a large EUV and FUV flux if it contains any massive stars. This is shown in the righthand panel of Fig.~\ref{mass_distributions}, where the EUV and FUV fluxes are comparable to those in much more populous star forming regions. In fact, some low-mass regions have higher FUV and EUV fluxes than more populous regions, simply due to their having more massive stars. If the mass functions of the regions are set to be constant, there is very little variation in the FUV and EUV fluxes between simulations with the same initial stellar densities. 

The most notable result in our parameter space study is how effective FUV radiation is at destroying discs. The \texttt{FRIED} grid simulations from \citet{Haworth18b} predict the almost total destruction of $r_{\rm disc} = 100$\,au discs within less than 1\,Myr. When the initial disc radius is set to just 10\,au, photoevaporation still leads to the destruction of at least 45\,per cent of discs (and sometimes much more) in the most dense star-forming regions ($\tilde{\rho} = 1000$\,M$_\odot$\,pc$^{-3}$). 

The \texttt{FRIED} models tend to be more destructive than earlier models of disc mass loss due to photoevaporation \citep{Scally01,Nicholson19a}, and we compare the disc fractions between the \texttt{FRIED} grid and the earlier models in Appendix~\ref{appendix:scally_comp}. These models have the same prescription for mass-loss due to EUV radiation, so the only difference is in the mass-loss due to FUV radiation. 

An implication of this rapid destruction of protoplanetary disc is that gas giant planets like Jupiter and Saturn must have to form close to their host star because discs with radii $>$10\,au do not survive in our simulations, and within 1 -- 2\,Myr \citep[cf.][]{Nicholson19a}, which appears to be corroborated by recent observational studies that find evidence for extremely rapid planet formation \citep{Alves20,SeguraCox20}. Alternatively, perhaps gas giant planets exclusively form in star-forming regions where there are no massive stars. This latter hypothesis is however in tension with the idea that the Sun formed in the vicinity of massive stars that enriched the Sun's protoplanetary disc (or protosolar nebula) in the short-lived radioisotopes $^{26}$Al and $^{60}$Fe \citep{Adams10,Gounelle12,Parker14a,Adams14,Lichtenberg16b,Nicholson17,Lichtenberg19}.

In our simulations, the initial stellar density is the most important factor in determining if a disc will be destroyed. As Fig.~\ref{G0_densities} shows, usually more than 50\,per cent of 10\,au discs are destroyed in the most dense regions, whereas in very low-density regions almost all of these discs survive. 

There is considerable debate in the literature as to the initial density of star-forming regions, as several authors have pointed out that the present-day density cannot be reliably used as a proxy for the initial density \citep{Marks12,Parker14e}. However, a combination of different structural and kinematic analyses seems to suggest that most star-forming regions probably have initial densities of at least $\tilde{\rho} = 100$\,M$_\odot$\,pc$^{-3}$ \citep{Wright14,Parker14e,Parker17a,Schoettler20}. If we adopt an initial density of $\tilde{\rho} = 100$\,M$_\odot$\,pc$^{-3}$ as our `default' density, then we would expect between 20 to 70\,percent of 10\,au discs to survive after a few Myr, but 100\,au discs would almost all be destroyed. The implication of this is that if gas giant planets are forming in environments that contain massive stars, they need to form in the inner (sub-10\,au) regions of discs, but presumably further out than the snow/ice line(s). 

Varying the other initial conditions besides stellar density generally results in only a minimal difference to the fraction of surviving discs. In particular, despite the expansion in supervirial simulations, most of the discs are destroyed early on, when the star-forming region is still compact. The degree of initial spatial and kinematic substructure can affect the fraction of surviving discs. Simulations with less substructure have a more uniform density profile than substructured regions (even though the median \emph{initial} stellar density is the same), and therefore the stars are on average closer to more of the ionising stars. This results in more photoevaporation of the discs, and additionally, the non-substructured regions have a smaller radius and therefore deeper gravitational potential during the subvirial collapse, leading to higher central densities and further destruction of the discs. 

In most of our calculations we have not included the effects of viscous evolution, and instead have just adjusted the radius of the disc so that evolves inwards following mass-loss due to photoevaporation \citep{Haworth19}. When we include viscous evolution, the disc radius is first adjusted inwards and we then calculate the subsequent expansion of the disc due to viscous evolution. Although the outer radius tends to increase by only a small amount during this viscous spreading, the disc is more susceptible to photoevaporation because its mass and therefore surface density have decreased. 

Our implementation of viscous evolution is purely analytical and may be too simplistic,  but appears to be in reasonable agreement with simulations conducted with the \texttt{VADER} code \citep{Krumholz15,ConchaRamirez19}, as well as other analytic estimates \citep{Hartmann98,Lichtenberg16b,ConchaRamirez19a}.

Whilst it is not the intention of this paper to attempt to match the disc fractions for individual star-forming regions, we show the observed disc fractions in nearby star-forming regions from \citet{Richert18} in our plots. A general comparison should not be made between the observations and our simulations because the observed regions in the \citet{Richert18} data may have different initial stellar densities from one another, as well as different disc radii (both in terms of the present-day and initial radii).

  However, we may draw some tentative conclusions. First, if the disc radius decreases due to photoevaporation, and viscous evolution does not subsequently increase the radius, too many discs survive if their initial radii are 10\,au. However, if viscous evolution is effective, then the disc fractions in our simulations match those in the observational data in regions with initial stellar densities of  $\tilde{\rho} = 100$\,M$_\odot$\,pc$^{-3}$ (compare the grey lines to the coloured lines, and the observational data, in Fig.~\ref{disc_evolution-b}), with the caveat that accretion of mass onto the central star would also increase the rate of photoevaporation (Fig.~\ref{disc_evolution-c}).

  Discs with larger initial radii (100\,au) lose mass much more rapidly, and the disc fractions in these simulations match the observed fractions in regions with initial stellar densities of $\tilde{\rho} = 10$\,M$_\odot$\,pc$^{-3}$. Future numerical work could in principle be tailored to specific star-forming regions to test the initial disc radius distributions, and the role of different internal physics (e.g.\,\,viscous evolution) on the evolution of the discs.

\section{Conclusions}

We have performed $N$-body simulations of the dynamical evolution of star-forming regions with a wide range of initial conditions. We have varied the initial stellar density, degree of spatial and kinematic substructure, virial ratio and initial mass distribution. We then calculated the the far ultraviolet (FUV) and extreme ultraviolet (EUV) fluxes within these star-forming regions and used these to determine the mass-loss due to photoevaporation from protoplanetary discs within these regions. Our conclusions are the following. 

(i) In all of our star-forming regions, the  FUV flux is significantly higher than the value measured in the interstellar medium ($G_0 = 1.8 \times 10^{-3}$ erg\,s$^{-1}$\,cm$^{-2}$). Even when our simulations start with stellar densities similar to the Galactic field (0.1\,M$_\odot$\,pc$^{-3}$), the FUV flux can be as high as 100\,$G_0$. This is caused by the presence of massive stars, and even regions with intermediate-mass (B-type stars, 5 -- 15\,M$_\odot$) experience very high radiation fields. 

(ii) We determine the mass-loss in protoplanetary discs due to external radiation fields by using the new \texttt{FRIED} grid of photoevaporative mass-loss models \citep{Haworth18b}. In the radiation fields present in our star-forming regions, mass-loss due to photoevaporation would destroy the gas component of discs if those discs have radii of more than 10\,au. 

(iii) Whilst the initial stellar density is the biggest factor in determining the fate of the discs, subtle changes to the initial conditions of star-forming regions with identical initial densities can also affect the survival chances of discs. Star-forming regions with low levels of spatial and kinematic substructure lead to a more uniform, and longer lasting exposure, to radiation fields from massive stars. This leads to more discs being destroyed. The overall bulk motion of the region, set by the initial virial ratio, has a much more minimal affect of the fraction of surviving discs. 

(iv) If we implement a simple prescription of the viscous evolution of the discs, the outward spreading of the disc radius, combined with mass accretion onto the central star, severely exacerbates the destruction of discs. 

(v) Taken together, this suggests that gas giant planets such as Jupiter and Saturn must either form extremely rapidly ($<$1\,Myr -- for recent observational evidence see  \citealp{Alves20,SeguraCox20}, and for further theoretical evidence see also \citealp{ConchaRamirez19,Nicholson19a}),  and relatively close to the parent star (i.e.\,\,beyond the snow line, but within 10\,au of the star), or these planets exclusively form in star-forming regions like Taurus where there are no photoionising sources. 

(vi) The latter scenario is in significant tension with the evidence from short-lived radioisotopes (SLRs) in meteorites that suggest that the Sun formed in a star-forming region that contained one or more massive stars that enriched the Sun's protoplanetary disc \citep{Ouellette07,Parker14a,Lichtenberg16b,Nicholson17}, or the prestellar core from which the Sun formed \citep{Gounelle12}. Further investigation is required to determine whether a Solar System analogue can survive photoevaporative mass-loss \emph{and} be enriched in SLRs. 

\section*{Acknowledgements}

We are grateful to the anonymous referee for their prompt, constructive and helpful reports. RJP acknowledges support from the Royal Society in the form of a Dorothy Hodgkin Fellowship. HLA was supported by the 2018 Sheffield Undergraduate Research Experience (SURE) scheme.

\section*{Data availability statement}

The data used to produce the plots in this paper will be shared on reasonable request to the corresponding author.

\bibliographystyle{mnras}  
\bibliography{general_ref}

\appendix

\section{Simulation resolution}
\label{appendix:resolution}

Given the fast rate of destruction in our simulations, we deemed it prudent to check whether the time resolution of our simulations was adequate. For example, using a snapshot output of 0.1\,Myr may be too coarse to determine the true effects of photoevaporation.

\begin{figure*}
  \begin{center}
\setlength{\subfigcapskip}{10pt}
\hspace*{-1.0cm}\subfigure[Disc mass]{\label{disc_comparison_resolution_10000G0_100_p1-a}\rotatebox{270}{\includegraphics[scale=0.26]{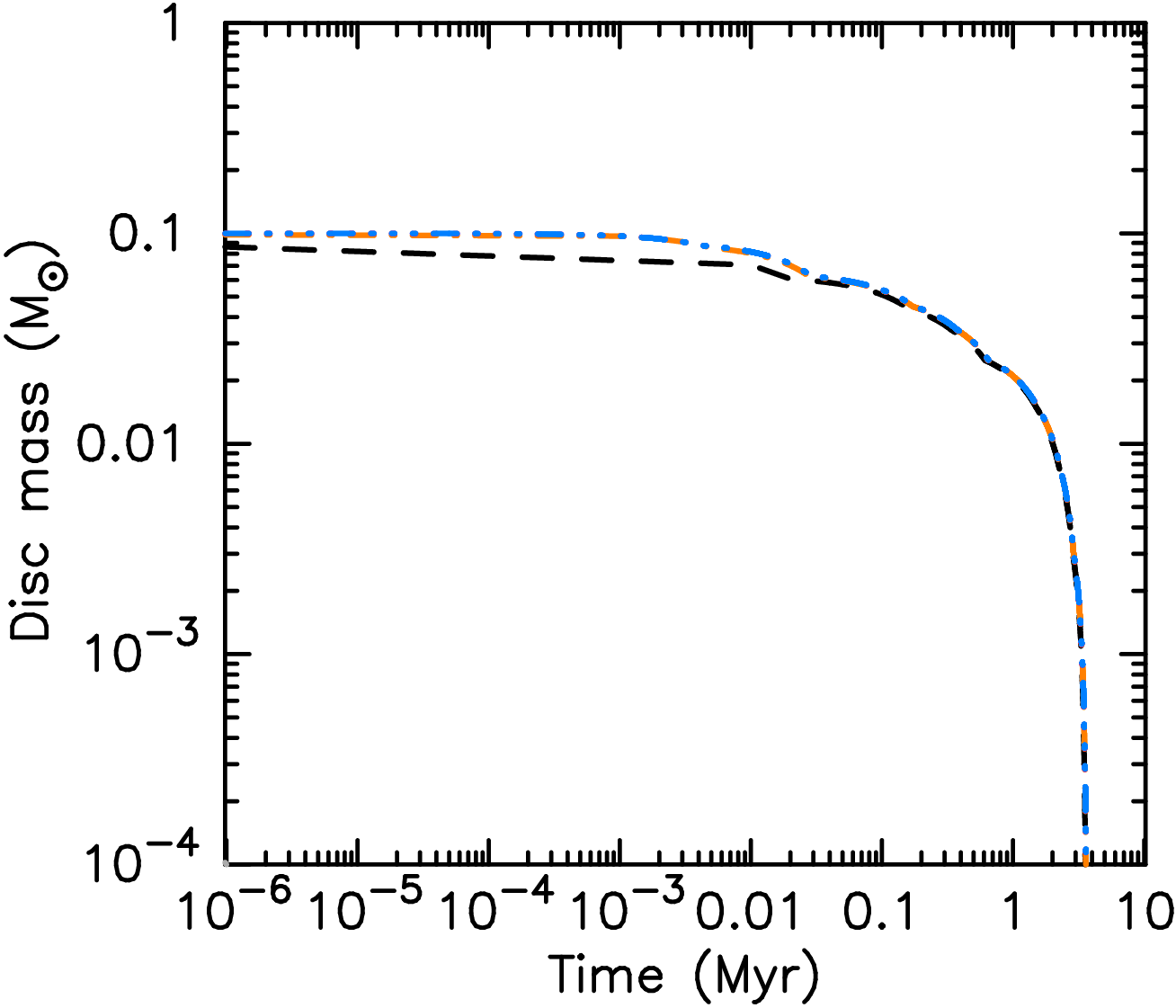}}}
\hspace*{0.3cm}\subfigure[Disc radius]{\label{disc_comparison_resolution_10000G0_100_p1-b}\rotatebox{270}{\includegraphics[scale=0.26]{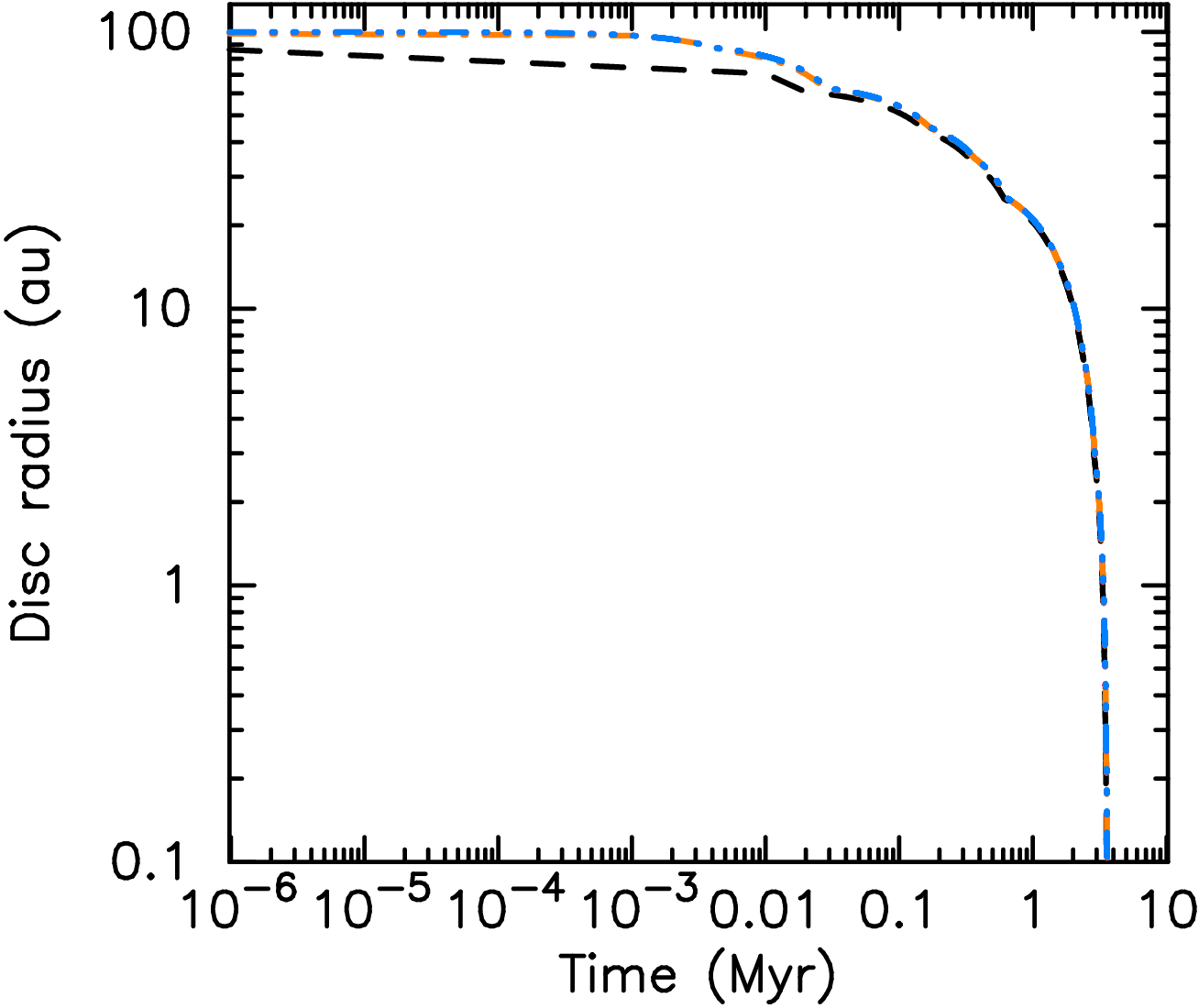}}}
\hspace*{0.3cm}\subfigure[Mass loss rate]{\label{disc_comparison_resolution_10000G0_100_p1-c}\rotatebox{270}{\includegraphics[scale=0.26]{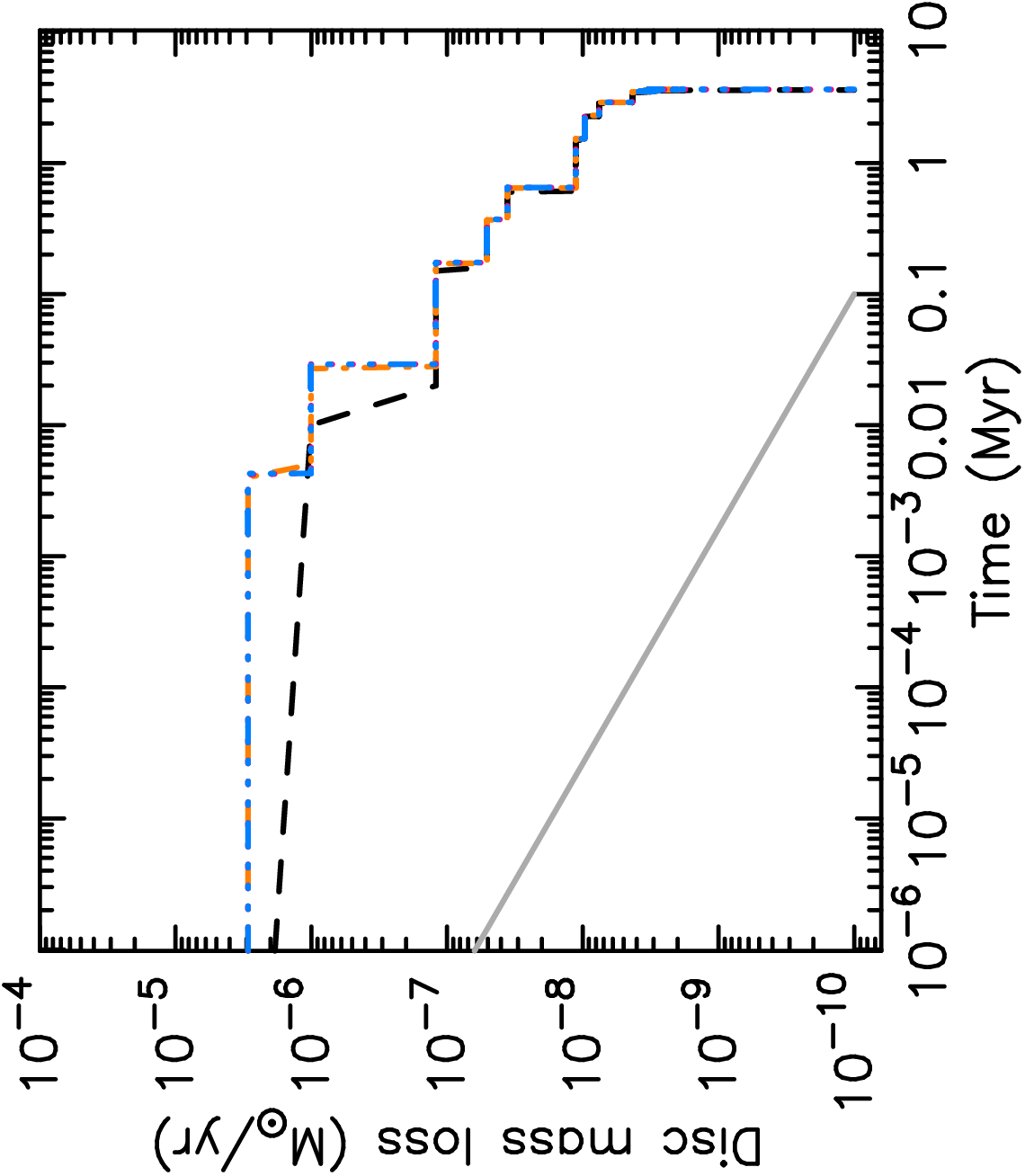}}}
\caption[bf]{Evolution of a single disc in a radiation field where the disc radius is allowed to evolve according to Eqn.~\ref{rescale_disc}. In this simulation the radiation field is $10^4 G_0$, the stellar mass is 1\,M$_\odot$, the initial disc mass is 0.1\,M$_\odot$ and the initial disc radius is 100\,au. We show the evolution of the disc mass in panel (a), the evolution of the disc radius in panel (b) and the mass-loss rate responsible for this evolution in panel (c). The solid grey lines are the results for a timestep of 0.1\,Myr, the dashed black lines are 0.01\,Myr, the dot-dashed orange lines are 0.001\,Myr, the dotted pink lines are $10^{-4}$\,Myr and the dot-dot-dot-dashed blue lines are $10^{-5}$\,Myr.}
\label{disc_comparison_resolution_10000G0_100_p1}
  \end{center}
\end{figure*}

\begin{figure*}
  \begin{center}
\setlength{\subfigcapskip}{10pt}
\hspace*{-1.0cm}\subfigure[Disc mass]{\label{disc_comparison_resolution_10000G0_100_p01-a}\rotatebox{270}{\includegraphics[scale=0.26]{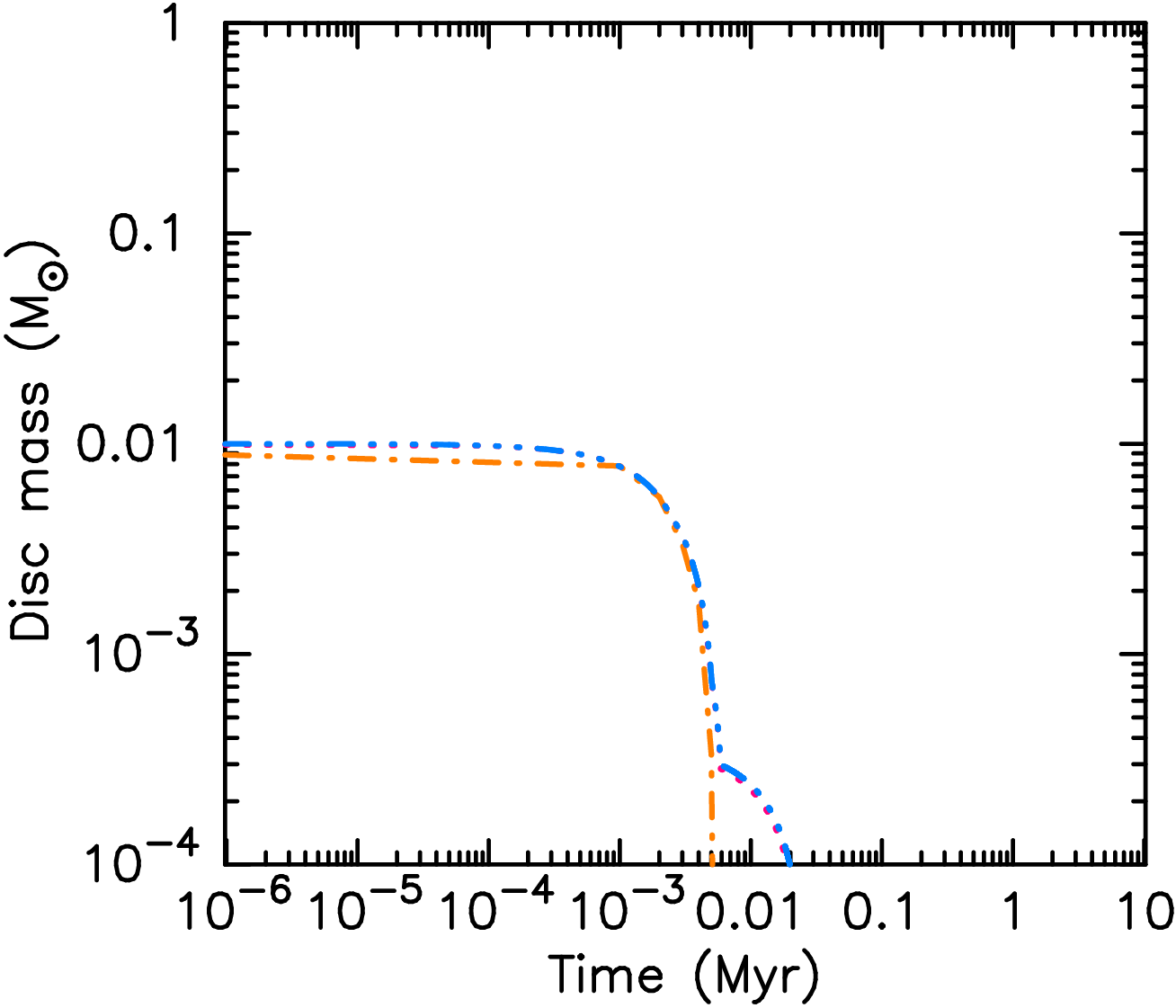}}}
\hspace*{0.3cm}\subfigure[Disc radius]{\label{disc_comparison_resolution_10000G0_100_p01-b}\rotatebox{270}{\includegraphics[scale=0.26]{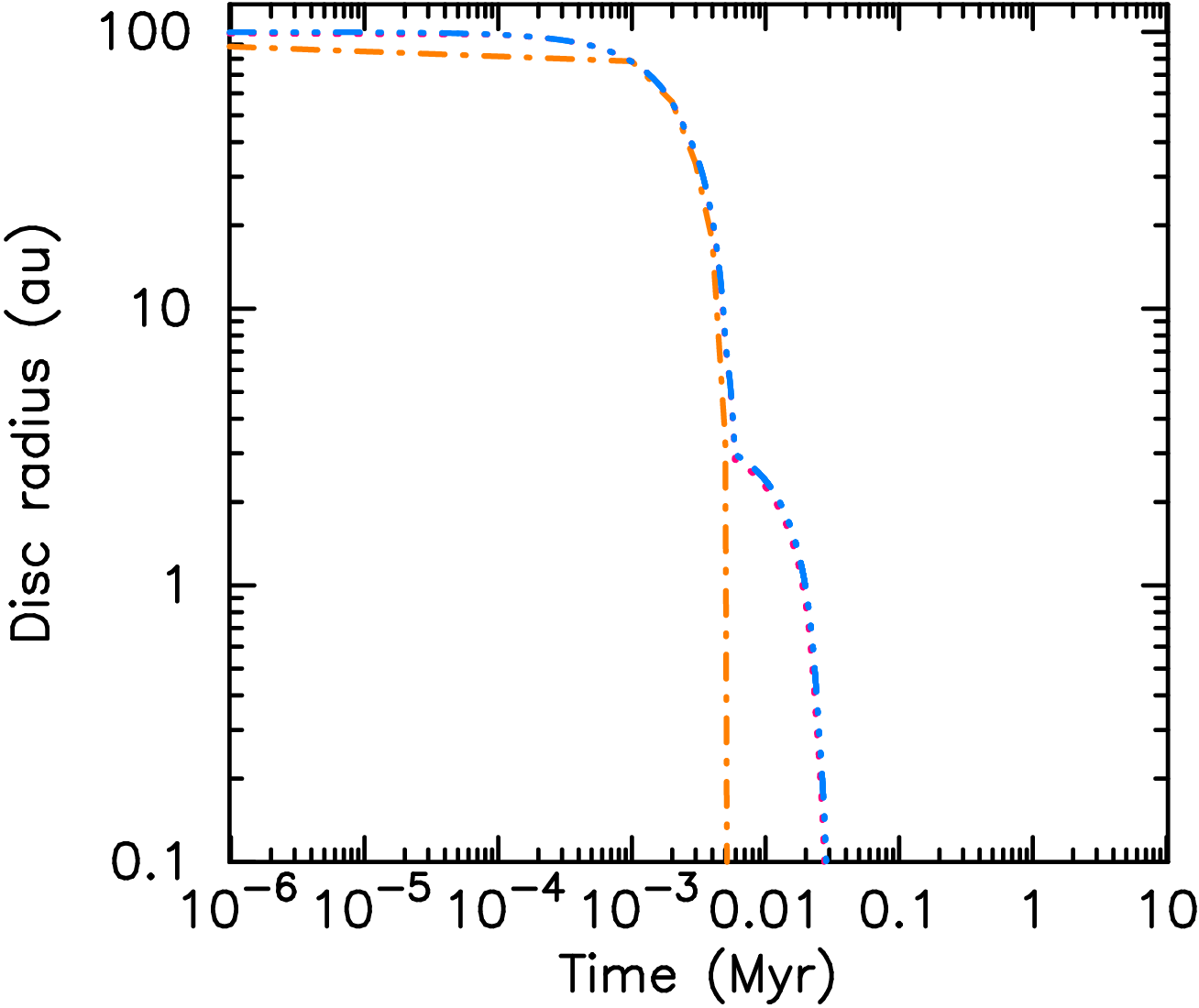}}}
\hspace*{0.3cm}\subfigure[Mass loss rate]{\label{disc_comparison_resolution_10000G0_100_p01-c}\rotatebox{270}{\includegraphics[scale=0.26]{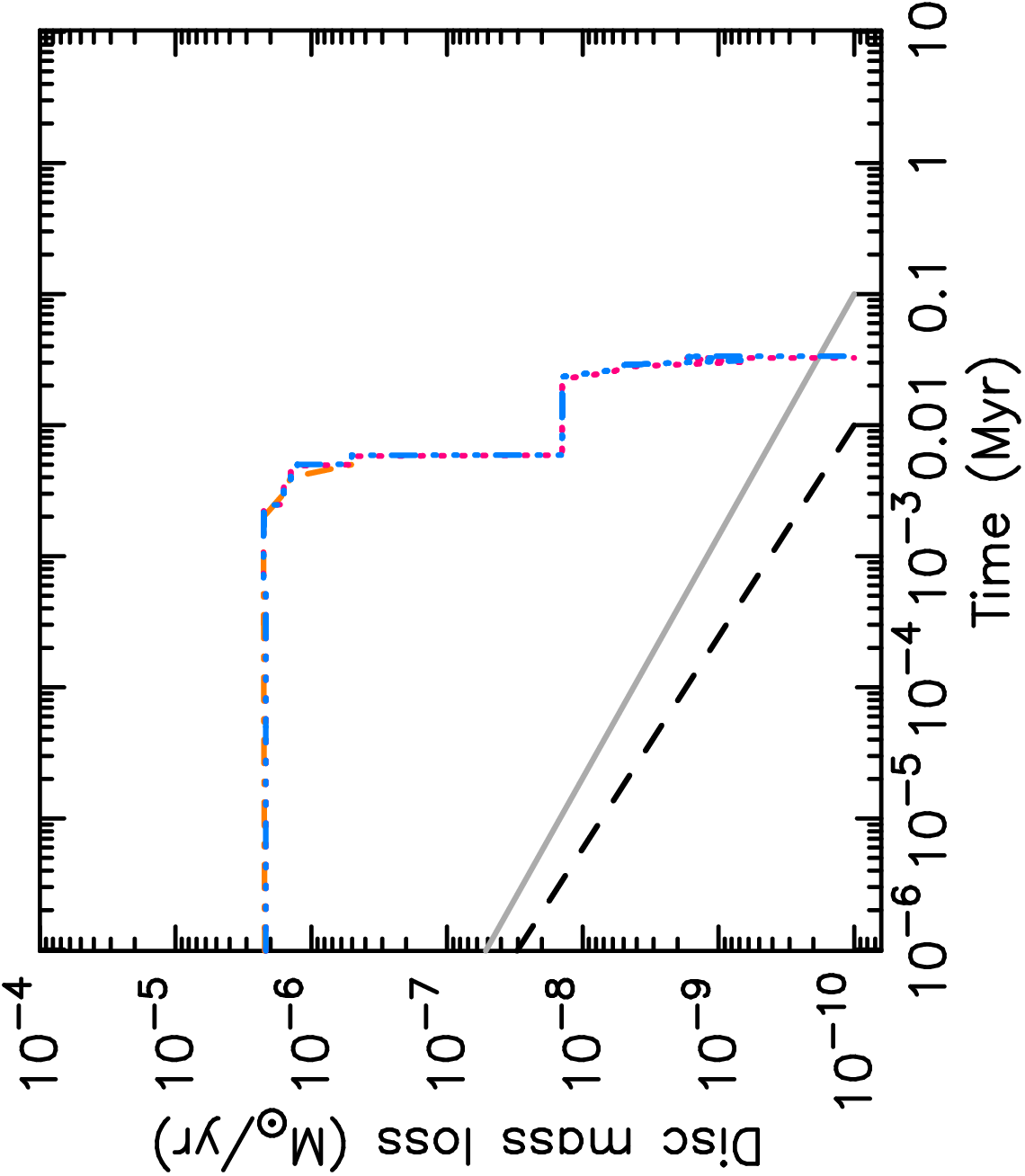}}}
\caption[bf]{Evolution of a single disc in a radiation field where the disc radius is allowed to evolve according to Eqn.~\ref{rescale_disc}. In this simulation the radiation field is $10^4 G_0$, the stellar mass is 0.1\,M$_\odot$, the initial disc mass is 0.01\,M$_\odot$ and the initial disc radius is 100\,au. We show the evolution of the disc mass in panel (a), the evolution of the disc radius in panel (b) and the mass-loss rate responsible for this evolution in panel (c). The solid grey lines are the results for a timestep of 0.1\,Myr, the dashed black lines are 0.01\,Myr, the dot-dashed orange lines are 0.001\,Myr, the dotted pink lines are $10^{-4}$\,Myr and the dot-dot-dot-dashed blue lines are $10^{-5}$\,Myr.}
\label{disc_comparison_resolution_10000G0_100_p01}
  \end{center}
\end{figure*}

\begin{figure*}
  \begin{center}
\setlength{\subfigcapskip}{10pt}
\hspace*{-1.0cm}\subfigure[Disc mass]{\label{disc_comparison_resolution_1000G0_100_p01-a}\rotatebox{270}{\includegraphics[scale=0.26]{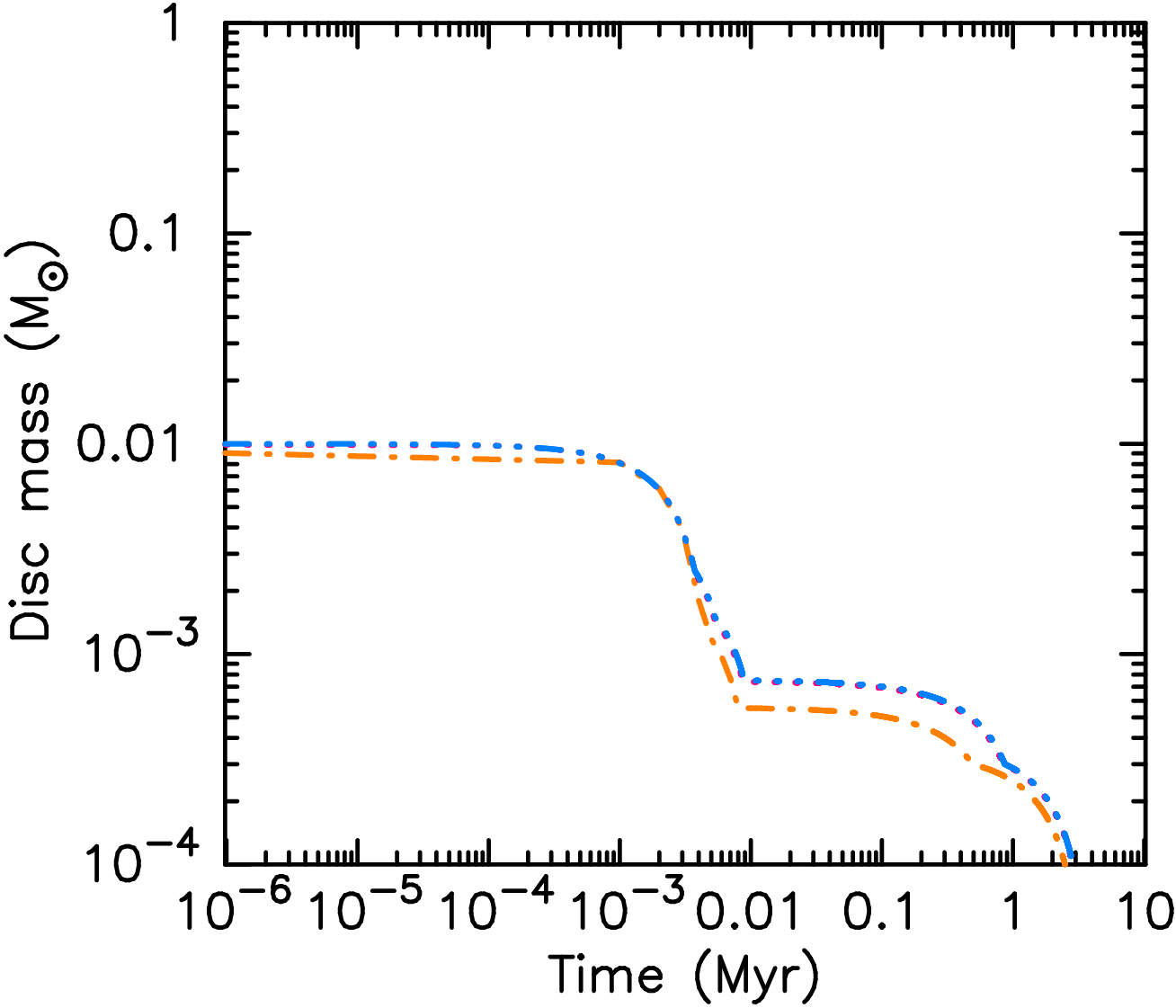}}}
\hspace*{0.3cm}\subfigure[Disc radius]{\label{disc_comparison_resolution_1000G0_100_p01-b}\rotatebox{270}{\includegraphics[scale=0.26]{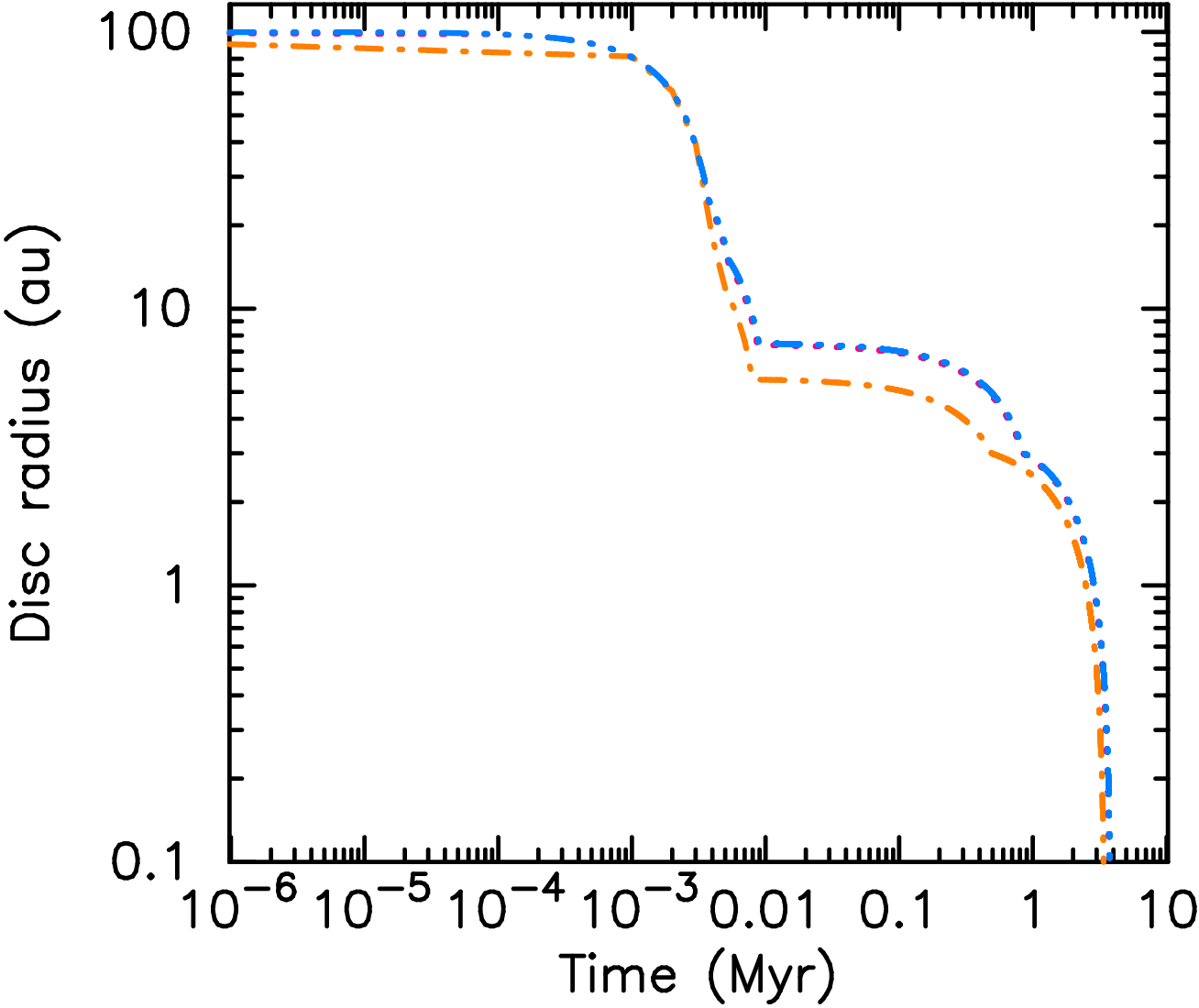}}}
\hspace*{0.3cm}\subfigure[Mass loss rate]{\label{disc_comparison_resolution_1000G0_100_p01-c}\rotatebox{270}{\includegraphics[scale=0.26]{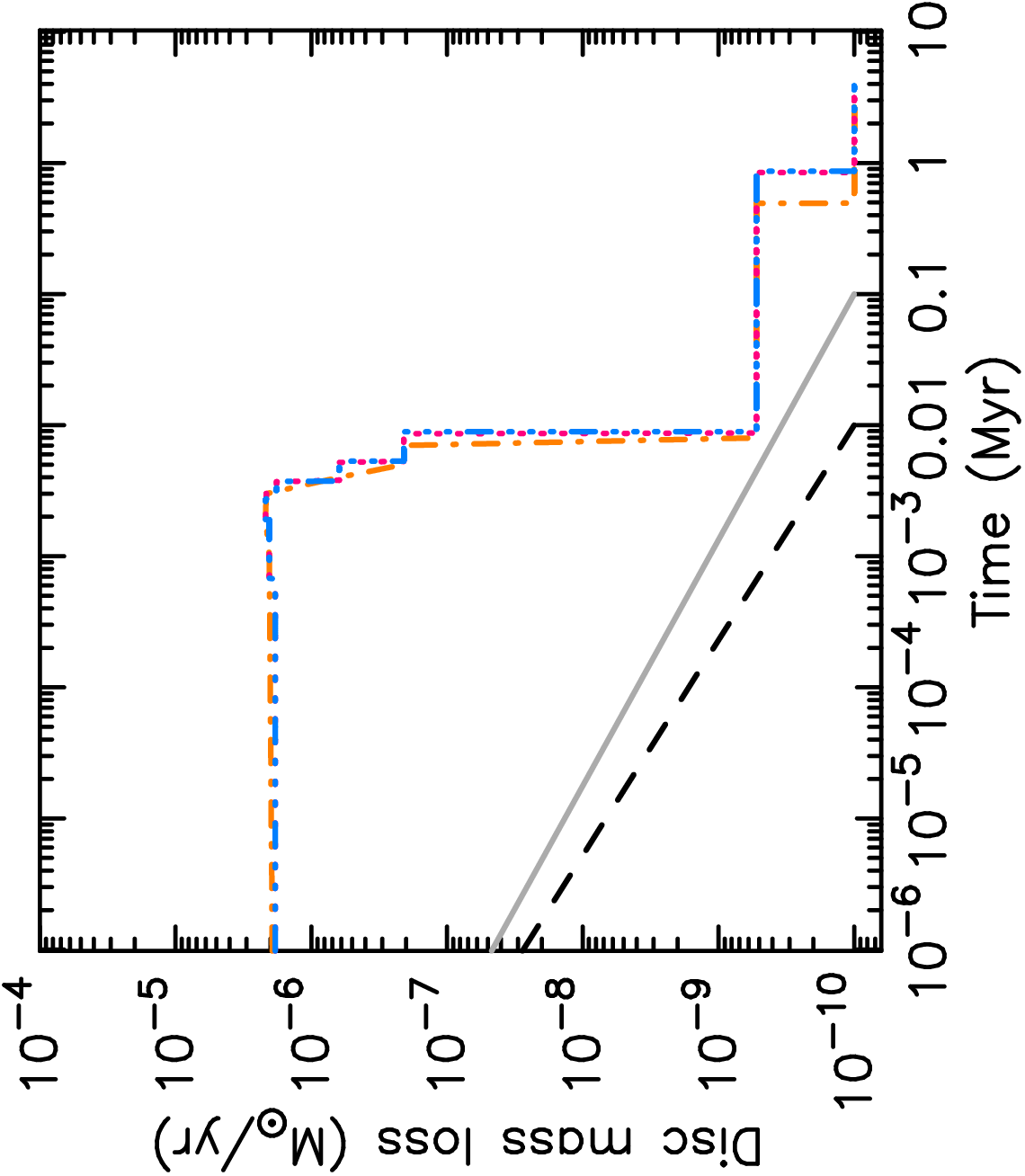}}}
\caption[bf]{Evolution of a single disc in a radiation field where the disc radius is allowed to evolve according to Eqn.~\ref{rescale_disc}. In this simulation the radiation field is $10^3 G_0$, the stellar mass is 0.1\,M$_\odot$, the initial disc mass is 0.01\,M$_\odot$ and the initial disc radius is 100\,au. We show the evolution of the disc mass in panel (a), the evolution of the disc radius in panel (b) and the mass-loss rate responsible for this evolution in panel (c). The solid grey lines are the results for a timestep of 0.1\,Myr, the dashed black lines are 0.01\,Myr, the dot-dashed orange lines are 0.001\,Myr, the dotted pink lines are $10^{-4}$\,Myr and the dot-dot-dot-dashed blue lines are $10^{-5}$\,Myr.}
\label{disc_comparison_resolution_1000G0_100_p01}
  \end{center}
\end{figure*}

\begin{figure*}
  \begin{center}
\setlength{\subfigcapskip}{10pt}
\hspace*{-1.0cm}\subfigure[Disc mass]{\label{disc_comparison_resolution_10000G0_10_p1-a}\rotatebox{270}{\includegraphics[scale=0.26]{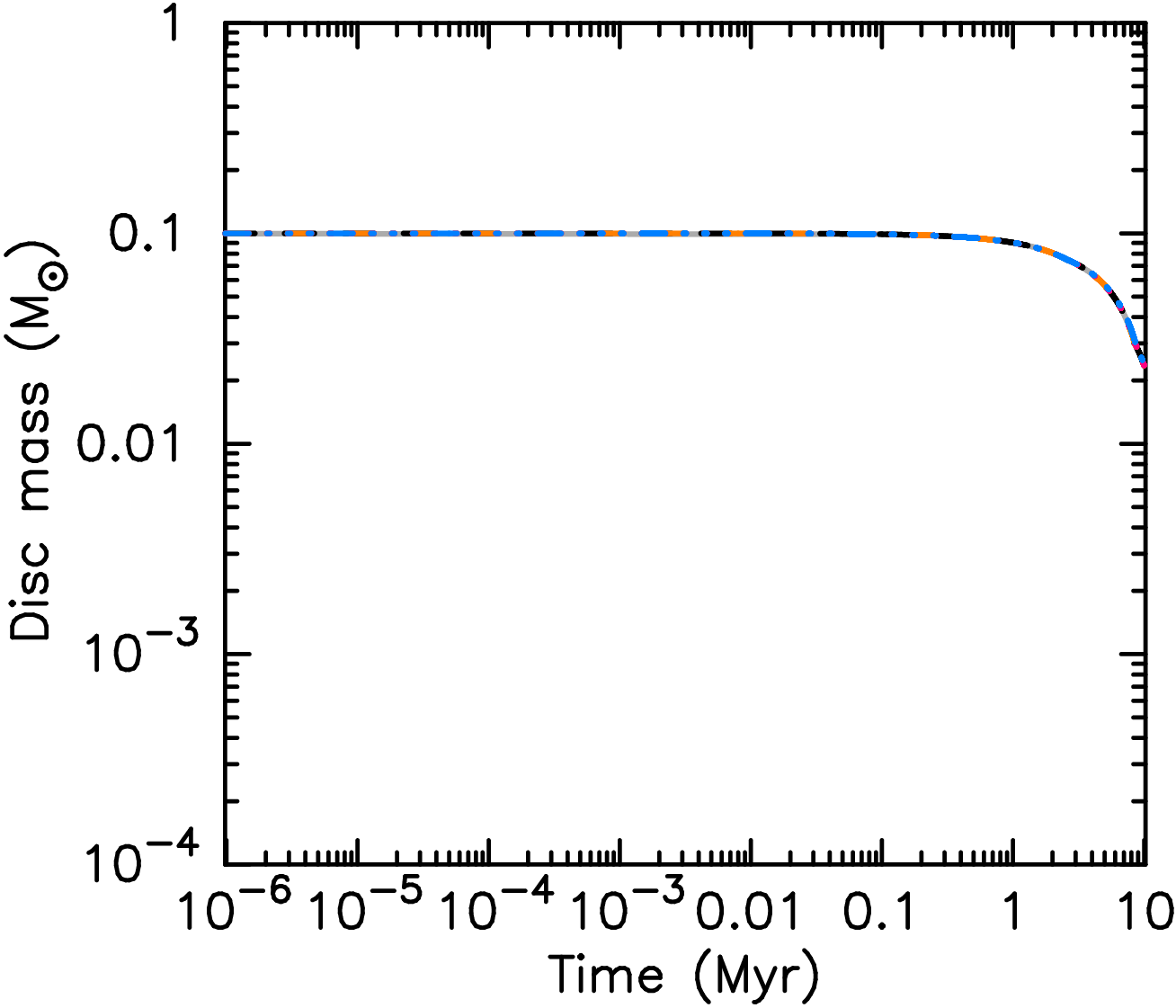}}}
\hspace*{0.3cm}\subfigure[Disc radius]{\label{disc_comparison_resolution_10000G0_10_p1-b}\rotatebox{270}{\includegraphics[scale=0.26]{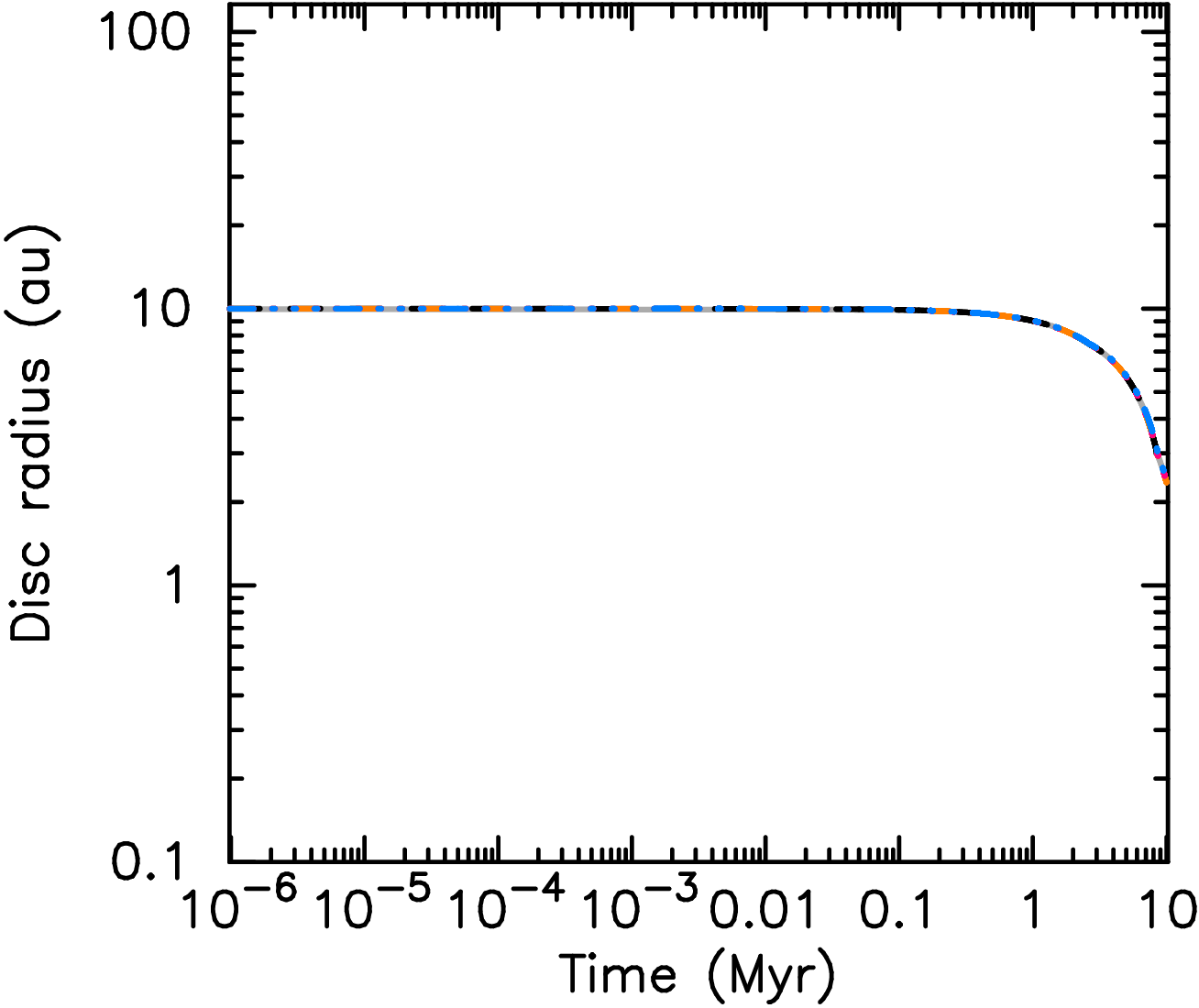}}}
\hspace*{0.3cm}\subfigure[Mass loss rate]{\label{disc_comparison_resolution_10000G0_10_p1-c}\rotatebox{270}{\includegraphics[scale=0.26]{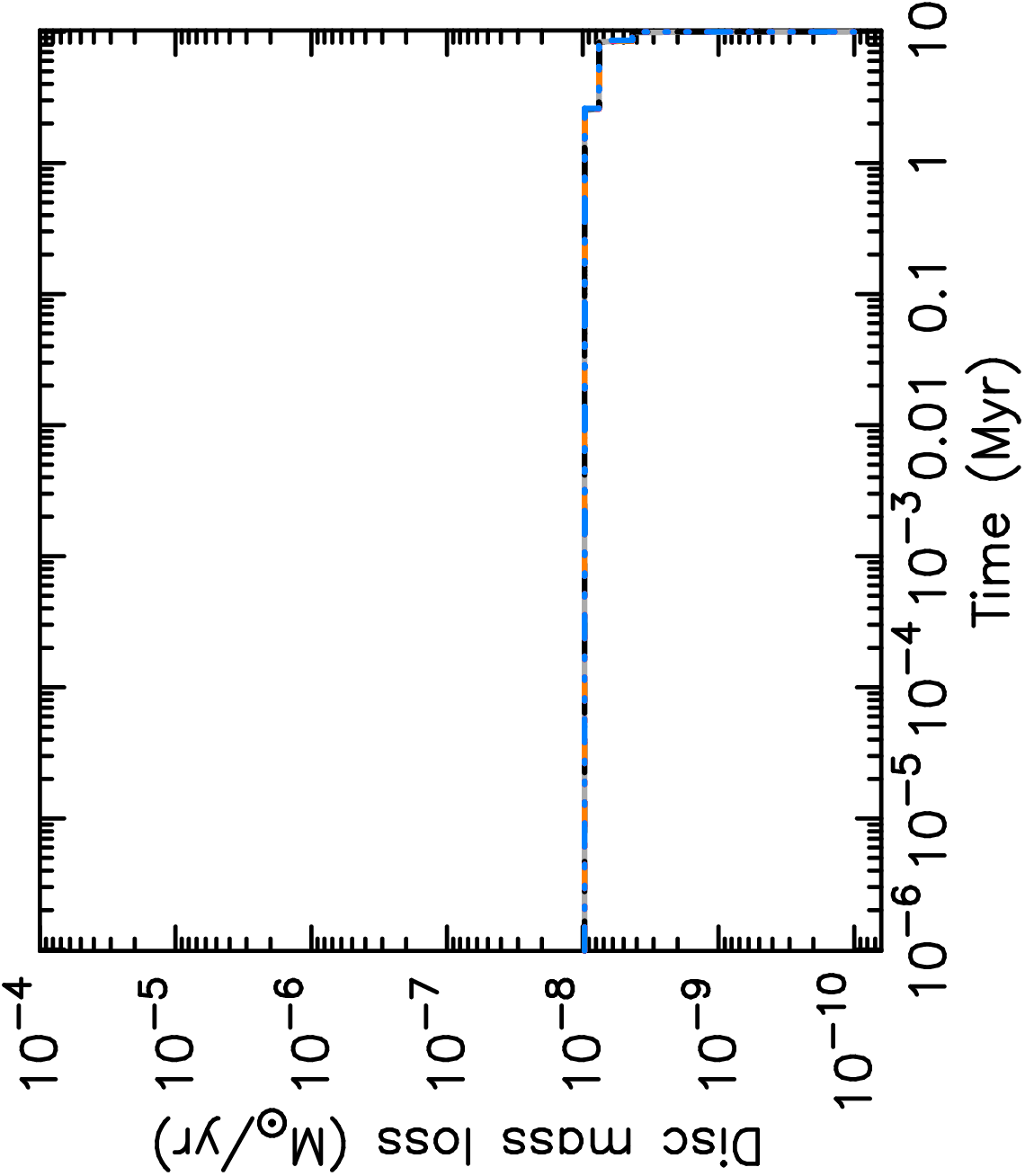}}}
\caption[bf]{Evolution of a single disc in a radiation field where the disc radius is allowed to evolve according to Eqn.~\ref{rescale_disc}. In this simulation the radiation field is $10^4 G_0$, the stellar mass is 1\,M$_\odot$, the initial disc mass is 0.1\,M$_\odot$ and the initial disc radius is 10\,au. We show the evolution of the disc mass in panel (a), the evolution of the disc radius in panel (b) and the mass-loss rate responsible for this evolution in panel (c). The solid grey lines are the results for a timestep of 0.1\,Myr, the dashed black lines are 0.01\,Myr, the dot-dashed orange lines are 0.001\,Myr, the dotted pink lines are $10^{-4}$\,Myr and the dot-dot-dot-dashed blue lines are $10^{-5}$\,Myr.}
\label{disc_comparison_resolution_10000G0_10_p1}
  \end{center}
\end{figure*}

\begin{figure*}
  \begin{center}
\setlength{\subfigcapskip}{10pt}
\hspace*{-1.0cm}\subfigure[Disc mass]{\label{disc_comparison_resolution_10000G0_10_p01-a}\rotatebox{270}{\includegraphics[scale=0.26]{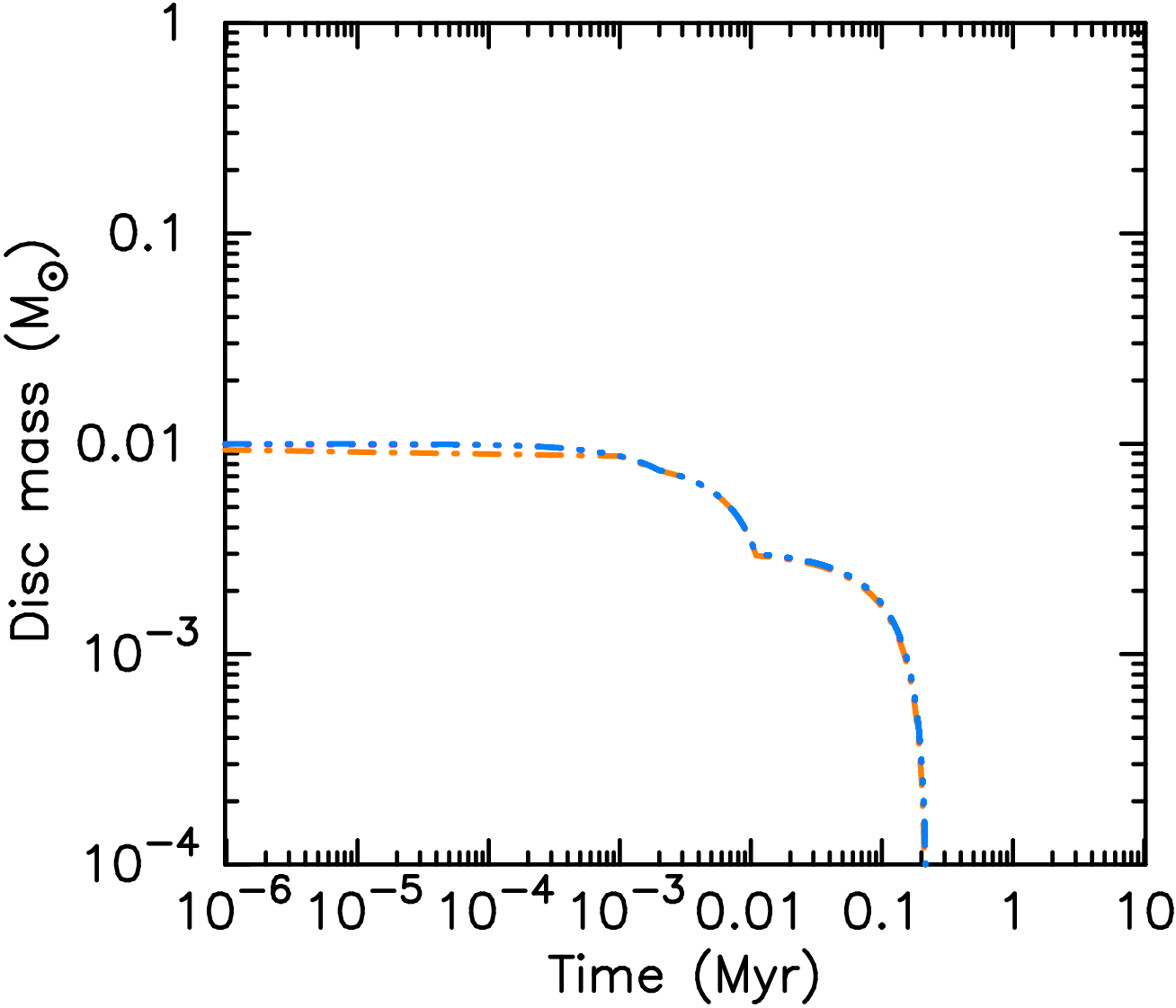}}}
\hspace*{0.3cm}\subfigure[Disc radius]{\label{disc_comparison_resolution_10000G0_10_p01-b}\rotatebox{270}{\includegraphics[scale=0.26]{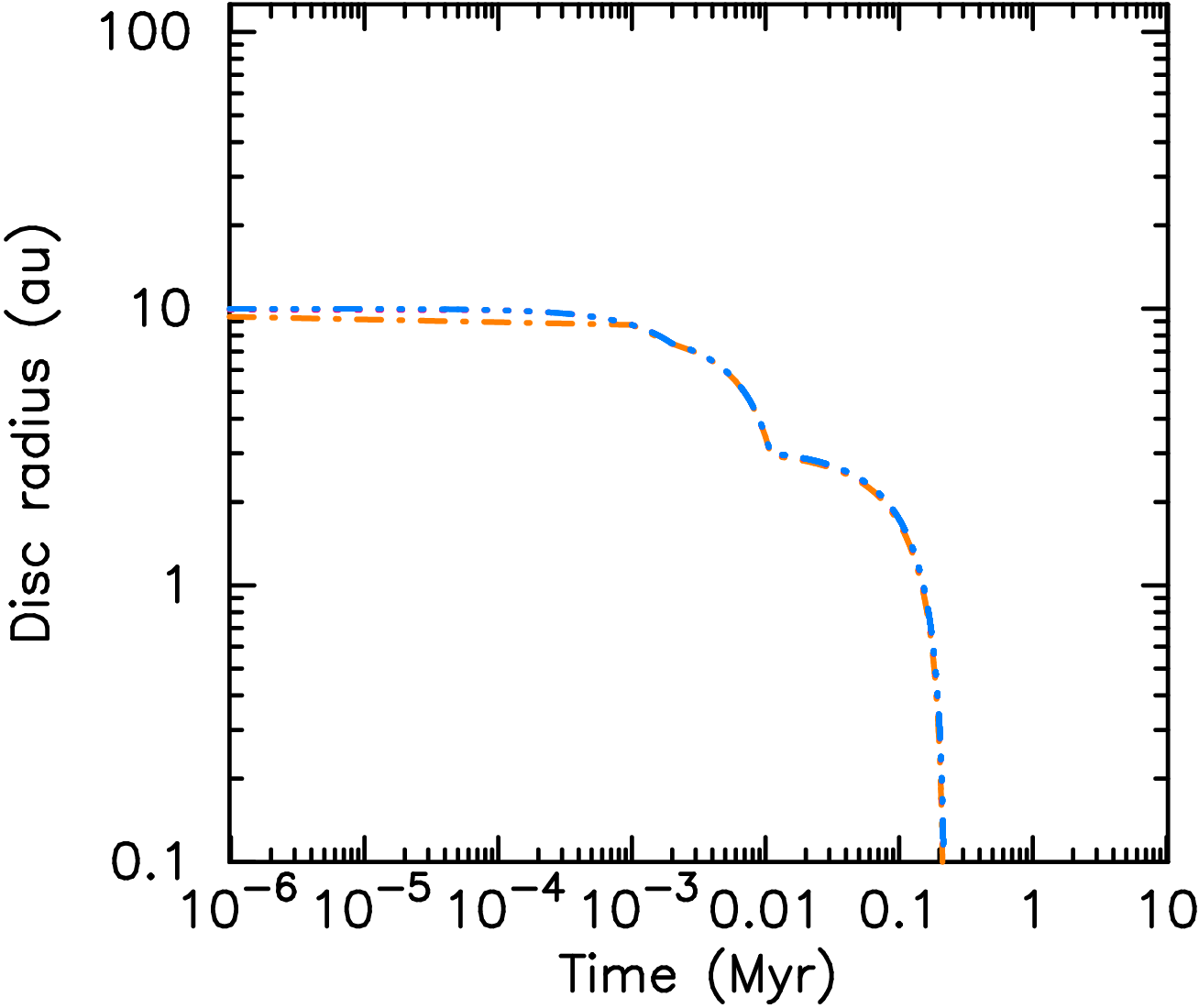}}}
\hspace*{0.3cm}\subfigure[Mass loss rate]{\label{disc_comparison_resolution_10000G0_10_p01-c}\rotatebox{270}{\includegraphics[scale=0.26]{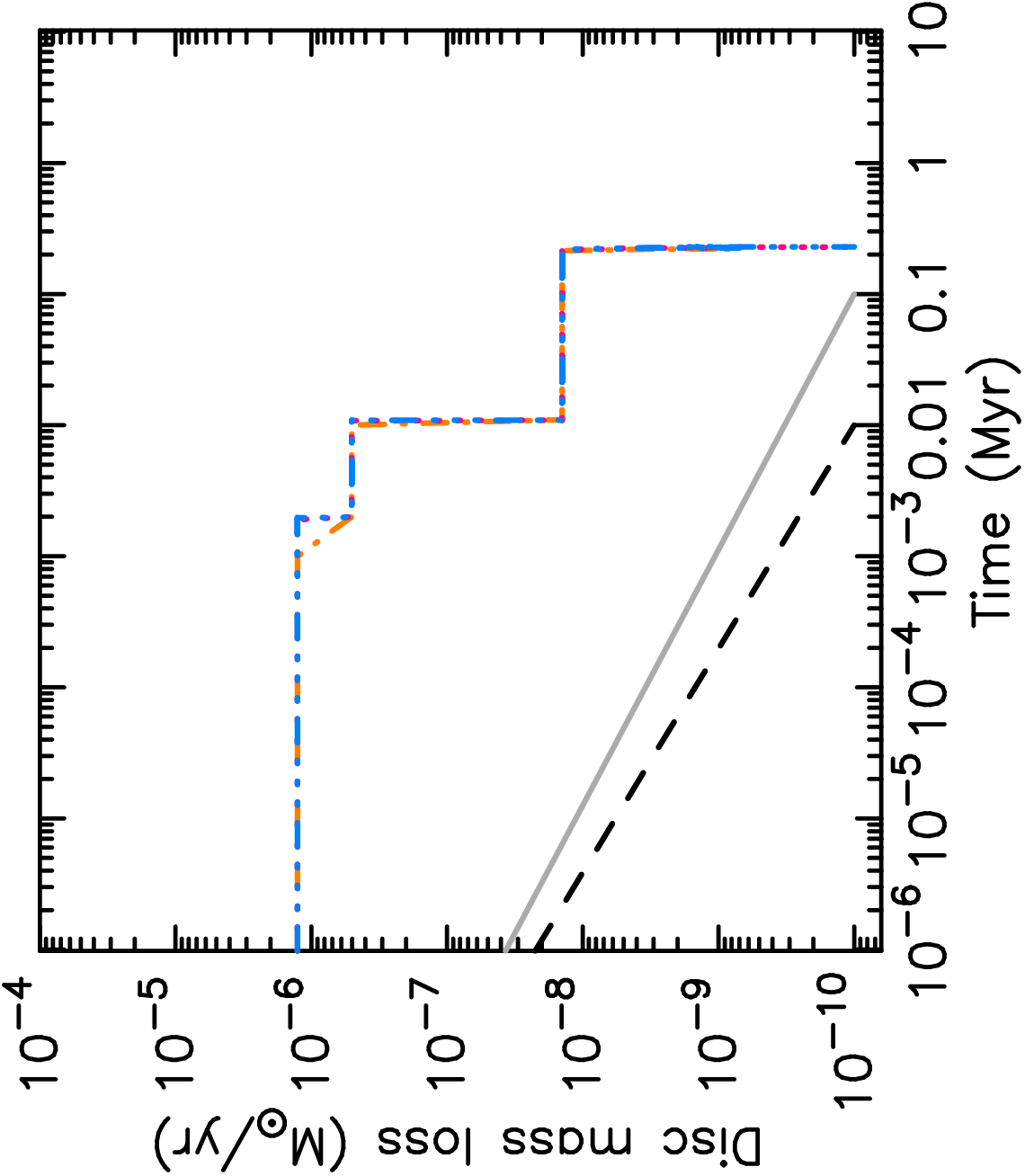}}}
\caption[bf]{Evolution of a single disc in a radiation field where the disc radius is allowed to evolve according to Eqn.~\ref{rescale_disc}. In this simulation the radiation field is $10^4 G_0$, the stellar mass is 0.1\,M$_\odot$, the initial disc mass is 0.01\,M$_\odot$ and the initial disc radius is 10\,au. We show the evolution of the disc mass in panel (a), the evolution of the disc radius in panel (b) and the mass-loss rate responsible for this evolution in panel (c). The solid grey lines are the results for a timestep of 0.1\,Myr, the dashed black lines are 0.01\,Myr, the dot-dashed orange lines are 0.001\,Myr, the dotted pink lines are $10^{-4}$\,Myr and the dot-dot-dot-dashed blue lines are $10^{-5}$\,Myr.}
\label{disc_comparison_resolution_10000G0_10_p01}
  \end{center}
\end{figure*}

To determine the optimum timestep for calculating the disc evolution, we fix the snapshot output for the dynamical information from the $N$-body simulation to be 0.1\,Myr (changing this timescale makes no discernible difference to the $G_0$ values in the simulation). We then reduce the timestep of the disc evolution calculation (both mass-loss due to photoevaporation, and internal viscous evolution).

We show examples of the evolution of single discs in a fixed radiation field in Figs.~\ref{disc_comparison_resolution_10000G0_100_p1}--\ref{disc_comparison_resolution_10000G0_10_p01}. In each figure, we show the evolution of the disc mass in panel (a), the evolution of the disc radius according to Eqn.~\ref{rescale_disc} in panel (b), and the mass-loss rate that induces this disc evolution in panel (c). The mass-loss rates are not constant, even in a constant radiation field, because the \texttt{FRIED} grid provides mass-loss rates that depend on the disc mass and radius, as well as the radiation field. In all panels the coloured lines indicate different timesteps in the algorithm; $10^{-1}$\,Myr (solid grey), $10^{-2}$\,Myr (dashed black), $10^{-3}$\,Myr (dot-dashed orange), $10^{-4}$\,Myr (dotted pink), and $10^{-5}$\,Myr (dot-dot-dot-dashed blue). In some plots of the disc mass and disc radius evolution (panels (a) and (b)), the black and grey lines are not shown because the values are immediately zero after the first timestep. %
In Fig.~\ref{disc_comparison_resolution_10000G0_100_p1} we show the evolution of a 0.1\,M$_\odot$ disc around a 1\,M$_\odot$ star in a very strong ($10^4G_0$) radiation field, where the initial disc radius is 100\,au. In this simulation, the timestep of 0.1\,Myr (the solid grey line) is far too coarse, and the disc is immediately destroyed. The versions of the simulation with the smaller timesteps converge, although there is some deviation when the timestep is 0.01\,Myr (compare the dashed black lines to the other lines). This disc is destroyed altogether (i.e.\,\,the disc mass drops to zero) after 3.5\,Myr in the convergent simulations.

If we keep the radiation field high ($10^4G_0$) and the radius of the disc at 100\,au, but decrease the mass of the host star to 0.1\,M$_\odot$ and the mass of the disc to 0.01\,M$_\odot$, then this disc is destroyed on much faster timescales (0.03\,Myr -- see Fig.~\ref{disc_comparison_resolution_10000G0_100_p01}). Here, timesteps of 0.1 and 0.01\,Myr are unsuitable, and there is some slight divergence between simulations where the timestep is $10^{-3}$\,Myr and the two simulations where the timestep is lower. 

Next, in Fig.~\ref{disc_comparison_resolution_1000G0_100_p01} we reduce the radiation field to $10^3G_0$, but keep all other parameters the same (the mass of the host star is 0.1\,M$_\odot$, the mass of the disc is 0.01\,M$_\odot$, and the disc radius is 100\,au. In this simulation, again timesteps of 0.1 and 0.01\,Myr are unsuitable, but there is better convergence between timesteps of $10^{-3}$\,Myr and the two simulations where the timestep is lower. There is a slight difference in the temporal evolution of the discs (compare the orange dot-dashed lines to the blue and pink lines), but the disc is destroyed at the same time (2.9\,Myr) for timesteps of $10^{-3}$\,Myr and lower.  

When we decrease the initial disc radius from 100\,au to 10\,au, we also see convergence of the simulations for timesteps of $10^{-3}$\,Myr and smaller, and these simulations are shown in Fig.~\ref{disc_comparison_resolution_10000G0_10_p1} (for a host star mass 1\,M$_\odot$ star and disc mass 0.1\,M$_\odot$) and in   Fig.~\ref{disc_comparison_resolution_10000G0_10_p01} (for a host star mass 0.1\,M$_\odot$ star and disc mass 0.01\,M$_\odot$).

In summary, the disc evolution aspect of our simulations reaches reasonable convergence for timesteps of  $10^{-3}$\,Myr, and whilst an even smaller timestep would further increase the accuracy of the calculations, we deem it an unnecessary extra computational expense to do this.

\begin{figure*}
  \begin{center}
\setlength{\subfigcapskip}{10pt}
\hspace*{0.3cm}\subfigure[\texttt{FRIED} grid, fixed disc radii, 100au]{\label{scally_comparison-a}\rotatebox{270}{\includegraphics[scale=0.35]{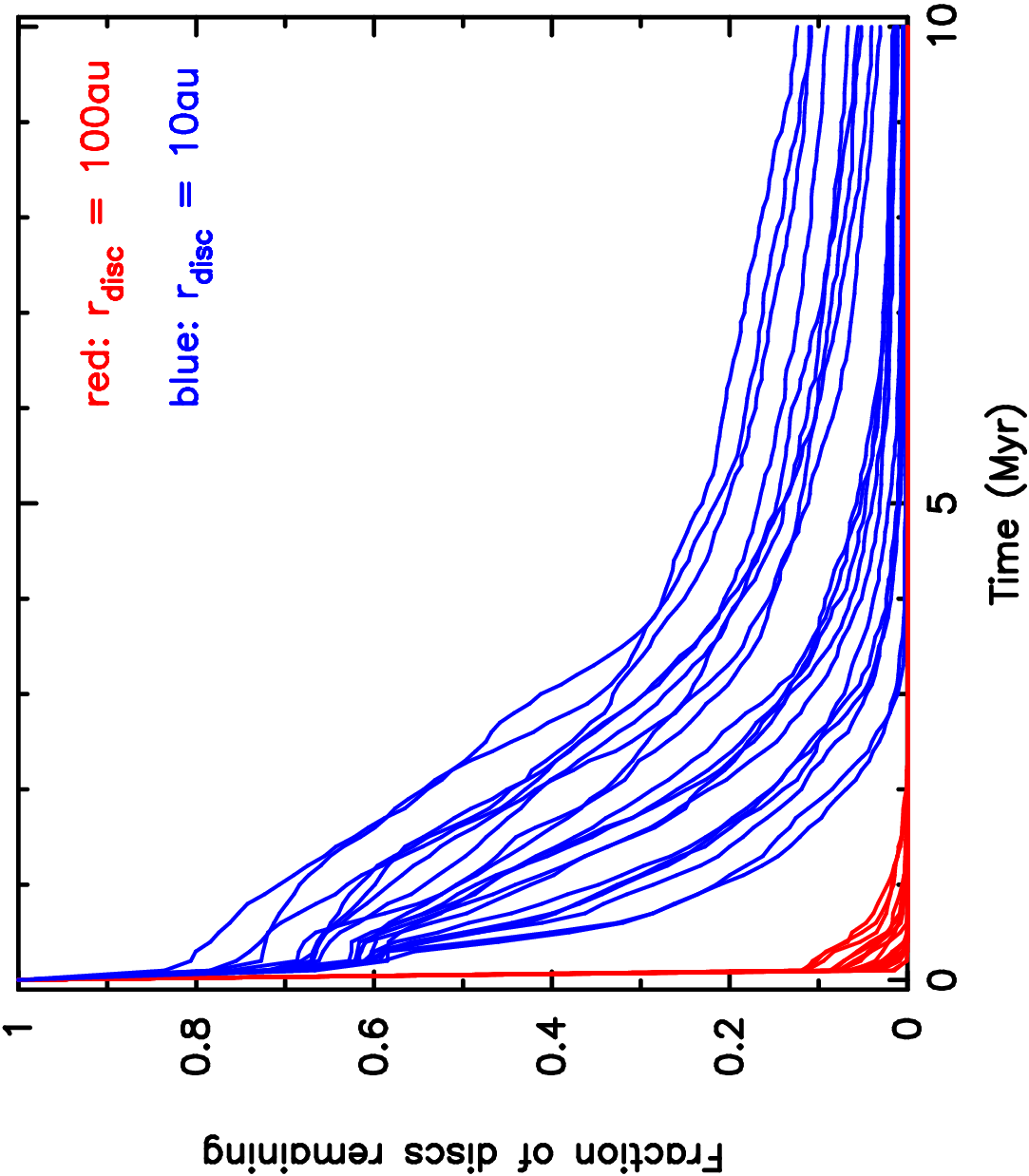}}}
\hspace*{0.5cm}\subfigure[\citet{Nicholson19a}, fixed disc radii]{\label{scally_comparison-b}\rotatebox{270}{\includegraphics[scale=0.35]{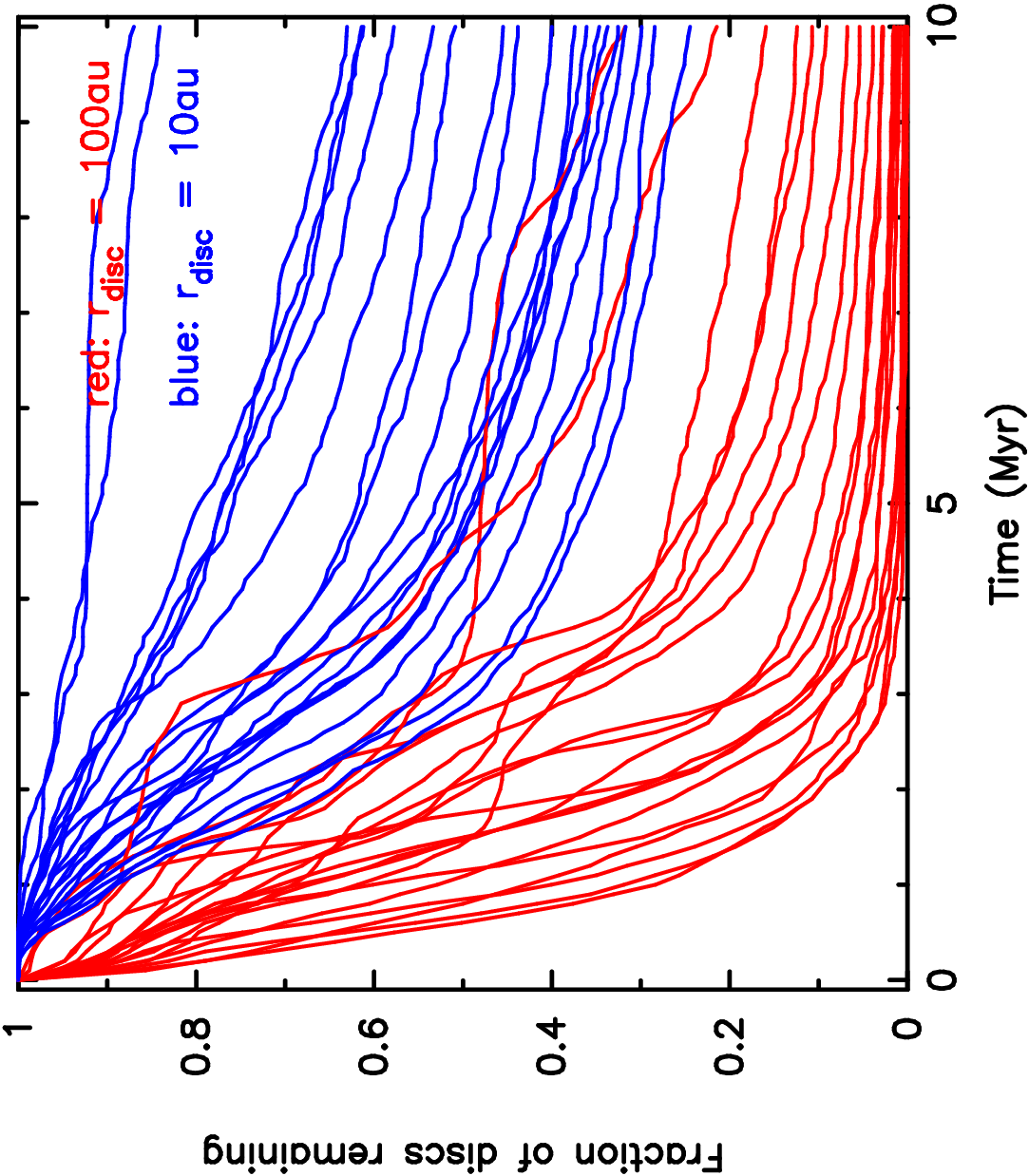}}}
\caption[bf]{Comparison of the evolution of disc fractions in our default simulation ($\tilde{\rho} = 100$\,M$_\odot$\,pc$^{-3}$, $D = 2.0$, $\alpha_{\rm vir} = 0.3$). with different prescriptions for the disc mass-loss due to photoevaporation. Both panels show the evolution of disc fractions in simulation where the disc s have initial radii of 10\, and 100\,au. The lefthand panel shows the disc fractions when the mass-loss rates from the \texttt{FRIED} grid are adopted, with the disc radii fixed to their initial values. The righthand panel shows the disc fractions when the FUV-induced mass-loss rates are determined from Eqn.~\ref{scally_massloss} \citep{Scally01,Nicholson19a}. } 
\label{scally_comparison}
  \end{center}
\end{figure*}

\section{Comparison with previous photoevaporation models}
\label{appendix:scally_comp}

Despite in some instances adopting very similar initial conditions for our star-forming regions to those in our previous work \citep{Nicholson19a}, we find that the mass-loss rates due to FUV photoevaporation in the \texttt{FRIED} grid models \citep{Haworth18b} are much higher. In \citet{Nicholson19a}, the mass-loss due to EUV radiation is the same as that adopted here and by other authors \citep[Eqn.~\ref{euv_equation}, e.g.\,\,][]{Johnstone98,Winter19b}, but we used the FUV photoevaporation mass-loss rate $\dot{M}_{\rm FUV}$ derived by \citet{Scally01}, which is \emph{independent} of the distance to the ionising star(s), $d$:
\begin{equation}
\dot{M}_{\rm FUV} \simeq 2 \times 10^{-9} r_{\rm disc} \,\,{\rm M_\odot \,yr}^{-1},
\label{scally_massloss}
\end{equation}
where $r_{\rm disc}$ is the radius of the disc in au, as before. This more simplistic prescription appears to have severely underestimated the FUV mass-loss rate, as shown in Fig.~\ref{scally_massloss}.  Here, we show the evolution of disc fractions in our default  star-forming regions ($\tilde{\rho} = 100$\,M$_\odot$\,pc$^{-3}$, $D = 2.0$, $\alpha_{\rm vir} = 0.3$) and where the discs have initial radii  of 10\,au and 100\,au. In the lefthand panel the mass-loss is calculated with the \texttt{FRIED} grid, and the disc radii are kept constant (for a fairer comparison with \citet{Nicholson19a}). In the righthand panel we implement the FUV mass-loss according to Eqn.~\ref{scally_massloss}, as in \citet{Scally01} and \citet{Nicholson19a}.

\label{lastpage}

\end{document}